\documentclass[letterpaper,journal]{IEEEtran}
\usepackage{amsmath,amsfonts,amssymb}
\usepackage{algorithm}
\usepackage{algpseudocode}
\usepackage{array}
\usepackage[english]{babel}
\usepackage{cite}
\usepackage{xcolor}
\usepackage{listings}
\usepackage{graphicx}
\usepackage[caption=false,font=scriptsize,labelfont=rm,textfont=rm]{subfig}
\usepackage{adjustbox}
\usepackage{threeparttable}
\usepackage{tikz}
\usepackage{xspace}
\usepackage[per-mode=symbol]{siunitx}
\usepackage{enumitem}
\usepackage{textcomp}
\usepackage{xurl}
\usepackage{hyperref}
\usepackage{cleveref}
\usepackage{amsthm}

\floatname{algorithm}{Pseudocode}
\newtheorem{lemma}{Lemma}

\makeatletter
\providecommand{\theHALG@line}{}
\renewcommand{\theHALG@line}{\thealgorithm.\arabic{ALG@line}}
\makeatother

\DeclareSIUnit{\slots}{sl}
\DeclareSIUnit{\bytes}{bytes}
\DeclareSIUnit{\bits}{bits}
\DeclareSIUnit{\rbs}{RBs}
\DeclareSIUnit{\mbps}{Mbps}
\DeclareSIUnit{\kbps}{kbps}

\newcommand{\ballnumber}[1]{\tikz[baseline=(myanchor.base)] \node[circle,fill=white,draw=black,inner sep=1pt] (myanchor) {\color{black}\bfseries\footnotesize #1};}

\addto\captionsenglish{

}

\renewcommand{\S}{\textsection\kern-0.15em}
\crefname{section}{\S}{\S}
\crefname{subsection}{\S}{\S}
\crefname{appendix}{\S}{\S}
\crefname{figure}{Fig.}{Fig.}
\crefname{table}{Tab.}{Tab.}
\crefname{equation}{Eq.}{Eq.}
\crefname{algorithm}{Pseudocode}{Pseudocode}

\newcommand{\thesystem}{{LatencyScope}\xspace}

\newcommand{\DDDU}{\textbf{DDDU}}

\title{\thesystem: A System-Level\\Mathematical Framework for 5G RAN Latency}

\author{Arman Maghsoudnia,
        Aoyu Gong,
        Raphael Cannatà,
        Dan Mihai Dumitriu,
        and Haitham Hassanieh
\thanks{Arman Maghsoudnia, Aoyu Gong, Raphael Cannatà, and Haitham Hassanieh are with EPFL, Lausanne, Switzerland.}
\thanks{Dan Mihai Dumitriu is with Pavonis LLC, Tolochenaz, Switzerland.}}

\begin{document}

\maketitle

\begin{abstract}
This paper presents \thesystem, a mathematical framework for computing one-way uplink and downlink latency in fifth-generation radio access networks across diverse system configurations. \thesystem models latency sources across the protocol stack, including radio interfaces, scheduling decisions, processing delays, frame structures, and hardware and software constraints, while capturing dependencies among configuration parameters and stochastic sources of delay. The framework also includes a configuration analyzer that uses these models to search billions of candidate settings and identify those that satisfy latency-reliability targets under user-specified constraints. We validate \thesystem on two open-source fifth-generation radio access network testbeds, as well as on measurements from a public commercial fifth-generation network. The results show that \thesystem closely matches empirical latency distributions, captures observed lower and upper latency bounds, and substantially outperforms prior analytical models and widely used fifth-generation network simulators. \thesystem can determine whether ultra-reliable low-latency communication targets are feasible for a given deployment and, when they are feasible, efficiently find satisfying configurations, helping network operators reason about latency modeling, configuration analysis, and system-level bottlenecks.

\end{abstract}


\section{Introduction}
\label{sec:introduction}

\IEEEPARstart{I}{ntroduced} in 5G New Radio (NR) to serve mission-critical applications, Ultra-Reliable Low-Latency Communications (URLLC) promises extremely low latency within the Radio Access Network (RAN) stack, aiming for \SI{0.5}{\milli\s} latency for both uplink (UL) and downlink (DL) channels (\SI{1}{\milli\s} round trip), with high reliability (above \SI{99.999}{\percent})~\cite{3GPP-TR-38.913-version-18}. While the specific requirements vary across applications~\cite{3GPP-TR-38.824-version-16}, URLLC will support a wide range of real-time applications~\cite{Benzaoui2020DeterministicLN,uitto2021evaluation}, including autonomous vehicles~\cite{autonomous-vehicles-5g}, industrial automation~\cite{brown2018ultra, EricssonRobotDemo}, smart grids~\cite{smart-grid-5g}, virtual $\&$ augmented reality (VR/AR)~\cite{s23073682}, professional live audio production~\cite{NokiaSennheiser}, public safety communications~\cite{public-safety-5g,public-safety-5g-applications}, and gaming~\cite{11360713}.
Even conventional applications such as web browsing can benefit from the improved responsiveness that low-latency 5G networks can provide~\cite{dchannel, virtualchannel}.

Despite the standardization of URLLC in late 2017~\cite{3GPP-first-NR-with-URLLC}, more than a decade of discussions~\cite{5g-first-responders}, and several years of commercial 5G rollouts~\cite{commercial-5g-launches}, practical 5G deployments, to date, cannot consistently achieve the latency and reliability requirements of URLLC~\cite{NokiaSennheiser,rischke20215g,rischke2022empirical,lackner2022measurement,EricssonRobotDemo, fezeu2023depth}. 
Moreover, discussions around 6G indicate even stricter latency goals of \SI{0.1}{\milli\s} uplink and downlink (\SI{0.2}{\milli\s} round trip)~\cite{She2024, GUPTA2021103521, 10.1109/COMST.2023.3243918, 10092856}. 
However, it is still unclear which cellular configurations and system parameters can achieve URLLC in practice. 
This is because we still lack a clear understanding of latency in the RAN: what are the bottlenecks, how different configuration parameters impact latency, and how they interplay with each other~\cite{10.1145/3696348.3696862}.

Prior work on analyzing latency in the 5G RAN relies on oversimplified models and simulations that do not capture the full system complexity~\cite{patriciello20185g, patriciello2019impact, rischke2022empirical, zhao2023physical, coll2023end, skocaj2023data, mostafavi2024edaf}.
Widely used simulators, such as MATLAB 5G Toolbox and the ns-3 based 5G-LENA~\cite{5Glena}, also fail to reproduce the latency distributions observed in real-world deployments and severely underestimate the latency as we show in \cref{sec:eval}.

Accurately estimating latency and identifying system configurations that minimize it, however, is challenging. The 5G RAN has hundreds of configuration parameters~\cite{chroma, auric}. Identifying which ones mainly affect latency and correctly modeling their impact is non-trivial. Intricate dependencies among parameters often yield counterintuitive behavior. For instance, prior work has frequently assumed that URLLC can be achieved by simply shortening transmission time slots~\cite{wirth20165g, 7247338,8636206, 8683972}. In practice, however, using very short slots might shift the bottleneck to processing and radio interface delays. Depending on the chosen TDD (Time Division Duplexing) pattern--which governs the allocation of uplink and downlink slots--packets may miss uplink opportunities or scheduling request windows, ultimately increasing latency rather than reducing it as we show in \cref{sec:config_analyzer}. Moreover, several variables like processing time, packet arrival time, delays inside the RF radios, etc. are nondeterministic.
Analyzing the latency distribution based on the distributions of these random variables quickly becomes intractable.

In this paper, we present \thesystem, a mathematical framework that can accurately estimate one-way latency distributions (for both uplink and downlink) inside the 5G RAN. \thesystem addresses the above challenges through extensive mathematical modeling of the latency sources at every layer in the RAN. We take a system-level approach where we account for various protocol specifications, hardware constraints, and system implementation details and model the dependencies between various configuration parameters. Using these analytical models, \thesystem can output the latency for an instantiation of the variables and configuration parameters. To account for nondeterministic components, we collect measurements of nondeterministic variables from an actual 5G system to generate empirical distributions.\footnote{In cases where we lack open-source access and cannot directly measure these variables (e.g., user equipment processing time), we learn an approximate distribution from latency measurements.} We then randomly sample these distributions to generate possible instantiations of the distribution.

We have also developed, within \thesystem, a configuration analyzer that identifies the set of configurations meeting specified latency requirements (e.g., minimum achievable latency or maximum latency under reliability constraints).
Through careful pruning, the analyzer can evaluate billions of configurations in a matter of hours, which would be computationally infeasible using simulators like MATLAB and ns-3.
\thesystem allows us to answer questions like: How can we configure our 5G system to achieve the lowest uplink latency if we restrict our system to operate in the sub-\SI{6}{\giga\hertz} bands with a TDD pattern and grant-based scheduling? What are the different configurations that can satisfy \SI{0.5}{\milli\s} latency with 99.99\% reliability? How reliably can we achieve \SI{1}{\milli\s} latency under a given configuration or traffic pattern?

We implement \thesystem and validate its results on two real-world private 5G testbeds with open-source RAN implementations, srsRAN~\cite{srsRAN} and OpenAirInterface (OAI)~\cite{OAI}, as well as a commercial 5G standalone network. We use synthetic and real application traffic, including video conferencing and online multiplayer gaming, to show that the latency distribution calculated by \thesystem accurately matches real-world latency, which we quantify using the Wasserstein metric~\cite{vaserstein1969markov,kantorovich1960mathematical,mallows1972note}. Our results reveal that:
\begin{itemize}[leftmargin=*, topsep=2pt]
    \item \thesystem achieves a low Wasserstein distance of \num{0.003} to \num{0.035} compared to real-world latency distributions.
    \item \thesystem calculates the P5 and P95 latency bounds with relative mean errors of \qty{5.2}{\percent} and \qty{2.4}{\percent}, respectively.
    \item Compared to prior analytical work~\cite{patriciello2019impact, zhao2023physical} and widely used 5G simulators such as MATLAB 5G Toolbox and 5G-LENA~\cite{5Glena}, \thesystem achieves over 40$\times$ better Wasserstein distance and over 21$\times$ and 35$\times$ lower mean absolute error in estimating the min and max latency, respectively.
    \item Using \thesystem's configuration analyzer, we find that among the \num{351.3}M analyzed possible configuration choices in the sub-\SI{6}{\giga\hertz} bands, none can meet the URLLC target of \SI{0.5}{\milli\s} one-way latency at \qty{99.99}{\percent} reliability. However, \qty{1.05}{\percent} of them can satisfy the relaxed target of \SI{1}{\milli\s} latency at the same reliability level using a grant-free configuration (see~\cref{table:percentage_reliability}).
\end{itemize}

To our knowledge, \thesystem is the first system-level analytical framework that provides a general model for 5G RAN one-way latency and validates it against real 5G measurements.
\thesystem can pinpoint bottlenecks currently preventing URLLC requirements from being met in practice and help configure RAN parameters to achieve the required latency under the given constraints of the system (e.g., hardware or regulatory limitations).
\noindent \textbf{Ethics:} This work does not raise any ethical considerations.

\section{Background}
\label{sec:background}

\noindent{\bf 5G NR (New Radio): }
Similar to 4G LTE, 5G NR uses Orthogonal Frequency-Division Multiplexing (OFDM) at the PHY layer, where the bandwidth is divided into frequency subcarriers on which data is modulated. However, unlike 4G which has a fixed subcarrier spacing (SCS), the SCS in 5G can be selected among seven numerologies ($\mu$).
Numerologies \numrange{0}{2} are available in low and mid-frequency bands, known as Frequency Range~1 (FR1), which spans \SI{410}{\mega\hertz} to \SI{7.125}{\giga\hertz}.
Numerologies \numrange{2}{6} are available in Frequency Range~2 (FR2), covering \SIrange{24.25}{52.6}{\giga\hertz} and, in recent releases, extending to \SI{71}{\giga\hertz}~\cite{3GPP-TR-38.913-version-18,3GPP-TS-138-211}.
The SCS can be derived as $\SI{15}{\kilo\hertz} \cdot 2^{\mu}$~\cite{3GPP-TS-138-211}. For all numerologies using normal cyclic prefix, 14 OFDM symbols are grouped into a time \textit{slot}, with duration $\SI{1}{\milli\s} / 2^{\mu}$.

Due to the shared nature of the wireless medium, the base station (gNB) allocates time-frequency resources to user equipment (UEs). This allocation process is known as resource scheduling.
After allocation, the gNB sends the scheduling decisions to UEs as part of downlink control information (DCI).
In dynamically scheduled, grant-based uplink access, UL resources are allocated by the gNB on demand.
In this case, a UE without available uplink resources sends a Scheduling Request (SR) via uplink control information (UCI), which the gNB uses to assign an uplink grant.

5G NR supports both frequency-division duplexing (FDD) and time-division duplexing (TDD).
In FDD, downlink (DL) and uplink (UL) transmissions use separate, non-overlapping frequency bands, typically in paired spectrum~\cite{3GPP-scs-fr1}.
In contrast, TDD uses the entire channel with different time-slots (and symbols) allocated for DL and UL.
A TDD period can be composed of one or two consecutive patterns.
A pattern consists of several DL slots \textbf{D}, followed by one mixed slot \textbf{M}, and several UL slots \textbf{U}.
The mixed slot symbols can be used for DL or UL.
TDD offers a more flexible allocation between DL and UL and allows changing the ratio of DL to UL resources.

\begin{figure}[!t]
  \includegraphics[width=\linewidth]{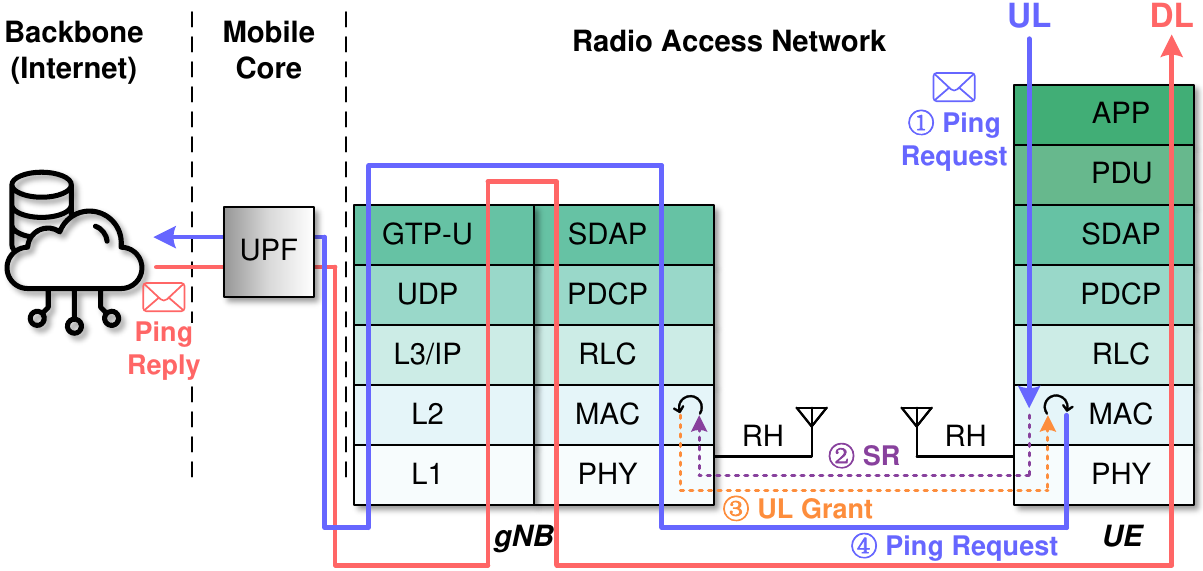}

  \caption{Path of a representative small packet through the 5G stack, illustrated using an ICMP echo request/reply.}

  \label{fig:ping-journey}

\end{figure}

\vskip 0.03in \noindent{\bf 5G Network Stack: }
To make the packet flow concrete, we use an ICMP echo request/reply (ping) as a representative small-packet example, since it exercises both uplink transmission for the request and downlink transmission for the reply.
The use of ping is only illustrative; the latency components identified along this path apply to general user-plane packets, and later sections extend the model to larger packets, bursty traffic and queuing delays, and more complex traffic patterns.
The path begins at the Application (APP) layer in the UE with the creation of a ping request.
The packet is carried as IP traffic within an already established PDU session, which provides the UE's user-plane connectivity to the data network through the 5G core.
It then enters the 5G user-plane protocol stack: the Service Data Adaptation Protocol (SDAP) layer maps the packet to a QoS flow and radio bearer, and the Packet Data Convergence Protocol (PDCP) layer performs functions such as header compression, ciphering, and integrity protection where configured.
Next, the data moves through the Radio Link Control (RLC) layer for segmentation and reassembly.
The MAC layer manages access to the shared medium.
The UE first sends a Scheduling Request (SR) (\ballnumber{2} in~\cref{fig:ping-journey}) and then sends the data after receiving a UL grant (\ballnumber{3} in~\cref{fig:ping-journey}).
In the PHY layer, the data is encoded and modulated into samples.
The Radio Head (RH) converts these samples into
Radio Frequency (RF) signals and sends them to the gNB over the air.
On the gNB side, the RH captures the signals and converts them into samples that are then demodulated and decoded into the data.
The gNB reconstructs the request from PHY to SDAP and encapsulates it into a General Packet Radio Service Tunneling Protocol User Plane (GTP-U) packet,
which is carried toward the UPF through the GTP-U, UDP, IP (L3), L2, and L1 layers.
The UPF decapsulates the payload and forwards it over IP.

The ping reply traces back the same route. However, it can be immediately scheduled for DL transmission at gNB's MAC layer.
The ping journey involves multiple steps contributing to the latency, which we explore in the following section.

\begin{figure*}[!t]
  \includegraphics[width=0.98\linewidth]{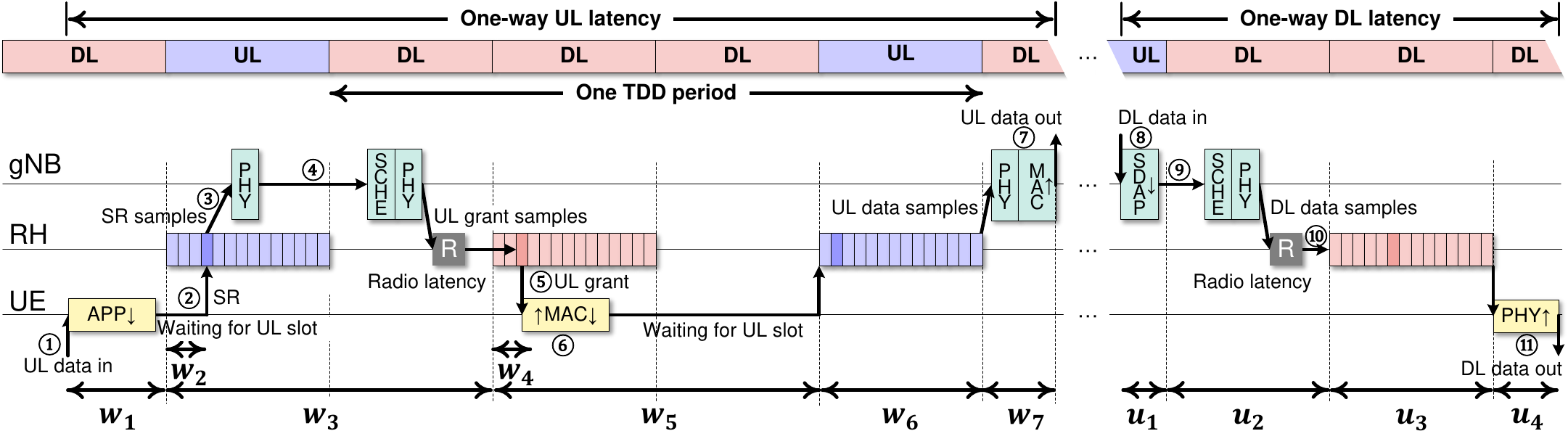}

  \caption{Overview of the system-level latency along the path of a packet with \textit{TDD Common Configuration} of \DDDU~ pattern.}

  \label{fig:overall-latency}

\end{figure*}

\section{Latency Sources}
\label{sec:latency-analysis}
To better understand the nature of latency in 5G systems, we classify each source of delay based on two dimensions: (i) the type of latency source and (ii) its inherent stochastic characteristics (deterministic or nondeterministic). We group latency sources into three main types:

\noindent
(1) \textbf{Processing latency}: representing the time required for decision-making and data processing. \\
(2) \textbf{Protocol latency}: introduced by protocol mechanisms and configurations. \\
(3) \textbf{Radio latency}: covering the time spent in the Radio Head (RH) and its interactions with the physical layer.

For clarity, we start with a high-level analysis, which we formalize in the following section. We detail the ping journey in \cref{fig:ping-journey} from a packet flow perspective and present a temporal breakdown in \cref{fig:overall-latency}.
While larger packets and buffer waiting times add complexity, requiring additional buffer reports and grants, the underlying high-level principles remain consistent across scenarios.
Consequently, we defer the formal analysis and modeling of these more complex cases to the appendix, though we keep their evaluation to demonstrate the generalizability of our analysis and model.

For a ping request, the UL transmission begins at the UE in \ballnumber{1}.
The UE prepares the scheduling request (SR), processing through the layers from APP down to PHY (APP$\downarrow$), to request network resources.
However, the UE must wait for the next UL slot to respect the TDD pattern.
\ballnumber{2} The UE transmits the SR in the next UL slot to the RH at the gNB side.
\ballnumber{3} The RH passes samples to the PHY layer, where they are demodulated and decoded.
The MAC layer receives the SR and schedules a UL grant for the UE.
The scheduling (SCHE) does not occur right away, as it is performed periodically in every slot.
\ballnumber{4} Consequently, the grant is scheduled in the next slot.
\ballnumber{5} Afterward, the gNB sends the grant to the UE as part of the control information\@.
\ballnumber{6} After the UE receives the grant, it waits for the next UL slot and sends the ping request to the gNB ($\uparrow$MAC$\downarrow$)\@.
\ballnumber{7} Finally, the gNB processes the UL data samples and passes the ping request to the UPF through the layers from MAC up to SDAP (MAC$\uparrow$)\@.

For the ping reply, the DL transmission begins at the gNB in \ballnumber{8}\@.
The gNB processes the DL data through the layers from SDAP down to RLC (SDAP$\downarrow$)\@.
\ballnumber{9} Again, as the scheduling is performed periodically in every slot, the data waits in the RLC layer and is scheduled in the next slot.
\ballnumber{10} Next, the scheduling result specifies the DL slot where the data will be transmitted to the UE\@.
\ballnumber{11} Finally, the UE receives the DL data in one or multiple symbols of the DL slot.
It then passes the data through the layers from PHY up to APP (PHY$\uparrow$)\@.

Given the breakdown, the three latency sources are summarized as follows.
(i) The \emph{processing latency} exists in both UE and gNB (cf. \ballnumber{1}, \ballnumber{4}, \ballnumber{6}--\ballnumber{9}, \ballnumber{11}).
This latency includes the time taken for processing data through the layers from APP down to PHY in the UE and from PHY up to SDAP in the gNB.
(ii) The \emph{radio latency} ($\mathbb{R}$) also appears in both the UE (cf. \ballnumber{2}, \ballnumber{5}, \ballnumber{6}, \ballnumber{11}) and gNB (cf. \ballnumber{3}, \ballnumber{7}, \ballnumber{10}).
This latency consists of the time spent in RF chains (e.g., analog-to-digital and digital-to-analog conversions), queuing delays on interface buses, and the bus transmission time.
(iii) The \emph{protocol latency} is the most significant, including the configurations in use (cf. \ballnumber{1}, \ballnumber{4}, \ballnumber{6}, \ballnumber{7}, \ballnumber{9}).
Specifically, the SR and grant procedure noticeably increases the latency of UL transmissions (cf. \ballnumber{2} \ballnumber{5}).
An alternative grant-free access mechanism allocates resources to UEs without SRs, reducing latency but facing scalability issues as the number of UEs increases~\cite{0ae4d1d5415845e980e046fe596c9907}.

It is crucial to note the following points:
\vskip 0.03in \noindent \emph{$\bullet$ These latency sources can be influenced by numerous factors specific to the system in use.}
For instance, in software-based 5G implementations, processing latency may increase or be random due to the non-real-time nature of operating system (OS) scheduling. Radio latency also varies with the type of interface--PCIe, QSFP, Ethernet, or USB--used to connect the radio head (RH) to the processor running the 5G stack.

\vskip 0.03in \noindent \emph{$\bullet$ These sources are also interdependent.}
For instance, the MAC scheduler must be designed to account for the total processing time in subsequent layers and radio latency. Failure to do so may result in the radio not being ready for transmission, leading to a corrupted signal. Since the sum of these delays is nondeterministic, practical implementations require the scheduler to include a margin to ensure the radio is ready on time, further increasing latency.

\vskip 0.03in \noindent \emph{$\bullet$ Any of these sources can bottleneck the system.} For instance, if the radio latency is \SI{0.5}{\milli\s}, selecting a shorter TDD pattern may not reduce latency and could even increase it. In particular, if the latency exceeds one TDD pattern in the \textbf{DU} configuration, an entire pattern is missed before the gNB can respond to the scheduling request.
Although counterintuitive, it is preferable to increase the TDD pattern duration to \textbf{DDUU}, allowing responses to scheduling requests without missing a full pattern as we show in detail in \cref{sec:config_analyzer}.

\section{\thesystem}
\subsection{Formal Models}
\label{sec:model-foundation}

In this section, we present our mathematical modeling of latency in 5G systems.
{\bf We focus on latency inside the 5G RAN. We model the DL latency as the time for a packet to travel from the beginning of the SDAP layer at the gNB, shown in Fig.~\ref{fig:ping-journey},  to the end of the SDAP layer at the UE, and the UL latency as the time for a packet to travel in the reverse direction.}
For clarity, we split the latency analysis and modeling into several cases.
While some cases could be combined, we present them separately to build up to the general model step by step and then address specific cases.

\begin{enumerate}[leftmargin=*, topsep=0pt]
    \item UL latency for size 1 packets in TDD
    \item UL latency for size 2 packets in TDD
    \item UL latency when radio delay $>$ TDD period
    \item UL latency under congestion with RLC-layer queuing
    \item UL latency with grant-free access
    \item DL latency in TDD
    \item UL and DL latency with mini-slot configuration
    \item UL and DL latency in FDD
    \item UL latency under multi-UE contention
\end{enumerate}
 We start by analyzing the uplink latency in TDD, similar to the ping request example shown in the previous section. We define two sizes of packets:
 \begin{itemize}[leftmargin=*, topsep=0pt]
 \item Size 1: Packets that can be fully transmitted in the initial resources granted by the gNB to the UE following the SR (scheduling request).
 \item Size 2: Larger packets that require additional resources.
 \end{itemize}

We further extend the model to account for congestion, where limited resources lead to buffering and queuing at the RLC layer, as well as the case of large radio latency and grant-free access, which allows uplink transmissions to skip the scheduling request and transmit immediately in preallocated resources.
We also analyze downlink latency as well as the mini-slot configuration and FDD. Finally, we analyze uplink latency with multi-UE contention. Due to limited space and the increased complexity of the mathematical models, we will present the formal model for the first case and defer the remaining cases to the Appendix~\ref{sec:appendix_c}. However, the analysis follows similar logic.
We will provide an evaluation of all these cases, including those whose models are deferred to the Appendix, in~\cref{sec:eval}.

\vskip 0.06in \noindent {\bf Modeling of the Uplink Latency Case 1:}
We break the total latency into distinct components, as illustrated in~\cref{fig:overall-latency} for uplink transmissions.
The total UL latency is the sum of the consecutive components (\cref{fig:overall-latency}) expressed as follows:
\setlength{\belowdisplayskip}{1pt}
\setlength{\abovedisplayskip}{1pt}
\setlength{\belowdisplayshortskip}{1pt}
\setlength{\abovedisplayshortskip}{1pt}
{
\begin{equation}
    \text{Total UL Latency} = w_1 + w_3 + w_5 + w_6 + w_7
    \label{eq:totalUL}
\end{equation}
}
In the following, we describe each component in detail.

\vskip 0.06in \noindent {\bf Modeling $w_1$:}
This variable represents the time from packet generation until the scheduling request is transmitted for that packet.
It includes both the UE’s preparation time for the scheduling request message and the wait time for the next available uplink slot. Two configurable parameters, the \textit{Scheduling Request Periodicity} (${SR}_P$) and the \textit{Scheduling Request Offset} ($SR_O$), determine when scheduling requests can be sent.
The offset indicates the first slot (relative to the gNB’s start) in which a scheduling request can be transmitted, and the periodicity specifies the interval (number of time slots) between consecutive scheduling requests.  Let $T$ be the number of time slots in the TDD pattern. Then, the scheduling request (SR) can only be sent in time slots: $\{SR_O+\{1,\cdots,T \}\cdot SR_P \}$. However, some of these slots might be DL, and the SR must be sent in a UL slot.

To find $w_1$, we must identify all possible UL slots for scheduling requests.
The earliest slot after the UE is ready determines $w_1$. Let  $d$ be the number of downlink slots in the TDD pattern. Hence, the UL slots in the TDD pattern are $\{d+1,\cdots,T-1\}$. The problem can be written as:

Find the largest set \( A = \{k'_0, k'_1, \ldots, k'_{T'-1}\} \), such that:

{\small
 \begin{equation}
     A \subseteq \{1, \ldots, T\}, \quad
    k'_0 < k'_1 < \cdots < k'_{T'-1},
     \label{eq:aSubsetOneT}
 \end{equation}

 \begin{equation}
    SR_O + A \cdot SR_P \equiv_T B
     \label{eq:offsetPeriodicity}
 \end{equation}

\begin{equation}
    B  \subseteq \{d+1, \ldots, T-1\},
    \label{eq:bSubsetDL}
\end{equation}
}

where $\{SR_O + A \cdot SR_P\}$ is the set of uplink slots in which we can send an SR, and $\equiv_T$ is the set equal operator modulo $T$. \cref{eq:aSubsetOneT} limits the solution to a finite set, \cref{eq:offsetPeriodicity} ensures that scheduling requests are sent in slots dictated by the offset and periodicity, and
\cref{eq:bSubsetDL} ensures that scheduling requests are sent in uplink slots.

\begin{lemma}
\label{lemma1}
Let $n = GCD(SR_P, T)$, $\varphi(\cdot)$ be Euler's totient function, and the set $D = \left\{\left\lceil \frac{d+1 - SR_O}{n} \right\rceil, \ldots, \left\lfloor \frac{T-1 - SR_O}{n} \right\rfloor \right\}$\footnote{$D =  \emptyset \text{ if } \left\lceil \frac{d+1 - SR_O}{n} \right\rceil \geq \left\lfloor \frac{T-1 - SR_O}{n} \right\rfloor$}, then:

{\small
\begin{equation*}
\begin{aligned}
A = \Bigg\{&
\left( j \cdot \left( \frac{SR_P}{n} \right)^{\varphi\left( \frac{T}{n} \right) - 1}
\bmod \frac{T}{n} \right) + \frac{T}{n} i \\
&\Bigg|\; j \in D,\; i \in \{0, \ldots, n-1\}
\Bigg\}
\end{aligned}
\end{equation*}
}
\end{lemma}

The proof of this lemma can be found in Appendix~\ref{sec:Proofs}.

Now that we have found $ A $, we know the slot numbers where the scheduling request can be sent.
We define $ o_1 $ as the time the packet is generated relative to the gNB's start (zero time reference). Assuming the UE needs $ l_1 $ seconds to prepare the request after data generation, and each slot is $ S $ seconds, we can calculate the first slot in which the UE is ready to send the request after $SR_O$ as:
{\small
\begin{equation}
    {SR}_{\text{ready}} = \left\lfloor \frac{o_1 + l_1}{S} \right\rfloor - SR_O
    \label{eq:srReady}
\end{equation}
}
Since the SR slots are periodic of period $T\cdot SR_P$ (with respect to the location they occupy in the TDD pattern), define quotient of the division of $ {SR}_{\text{ready}} $ by $ T \cdot SR_P $  as $ q $ and the remainder as $ r $.

{\small
\begin{equation}
    q = \left\lfloor \frac{{SR}_{\text{ready}}}{T \cdot SR_P} \right\rfloor, \quad
    r = {SR}_{\text{ready}} \mod {T \cdot SR_P}
    \label{eq:q_r_w1_combined}
\end{equation}
}

$ w_1 $ is calculated as follows.
We locate the slot where the UE is ready to send the scheduling request.
Then we identify the next slot in which the scheduling request can be sent.
For $i \in \{0, \ldots, T'-2\}$ we have the following cases:

{\footnotesize
\begin{align}
    w_1 =
    \begin{cases}
        S (SR_O + (k'_0 + q T) SR_P) - o_1 &\hspace{-5pt} \text{if } r < k'_0 \cdot SR_P\\
        S (SR_O + (T + k'_0 + q T) SR_P) - o_1 &\hspace{-5pt} \text{if } r > k'_{T'-1} \cdot SR_P \\
        S (SR_O + (k'_{i+1} + q T) SR_P) - o_1 &\hspace{-5pt} \text{if } k'_i \leq \frac{r}{SR_P} \leq k'_{i+1}
    \end{cases}
    \label{eq:w1}
\end{align}
}

\vskip 0.06in \noindent {\bf Modeling of $w_2$:}
This variable represents the time at which the scheduling request is sent relative to the slot's start.
The request is transmitted via the Physical Uplink Control Channel (PUCCH) within that slot.
In normal cyclic prefix, each 5G slot consists of 14 symbols, and the scheduling request is sent in one of these symbols.
A configured PUCCH resource occupies one or more contiguous OFDM symbols within the slot.
We call the first symbol ${UC_{st}}$ and the number of symbols $UC_{no}$.
Consequently, $w_2$ can be calculated as:

{\small
\begin{equation}
    w_2 = \frac{{UC_{st}} + UC_{no}}{14} \cdot S, \quad
    {UC_{st}} + UC_{no} \leq 14
    \label{eq:combined_w2_constraint}
\end{equation}
}

\vskip 0.06in \noindent {\bf Modeling of $w_3$:}
This variable represents the time between the beginning of the slot that sends the scheduling request and the DL slot delivering the UL grant.
The gNB processes the scheduling request and issues a grant to the UE, specifying resources and time slot for data transmission.
Like UL control information, the grant is sent via the Physical Downlink Control Channel (PDCCH) in the slot where it appears. Let:

\noindent $-$ $ p_1:$ The gNB's processing time for the scheduling request.

\noindent $-$ $ p_2:$ The MAC layer's processing time for allocating resources based on the UE's demands.

\noindent $-$ $ p_3:$ The physical layer’s processing time for generating radio samples for transmission.

Typically, the MAC layer processes the grant at the start of each slot, followed by the physical layer.
Therefore, we can express the time at which the MAC starts processing the grant as follows:

{\small
\begin{equation}
    \text{Start of MAC Processing} = o_1 + w_1 + \left\lceil \frac{w_2 + p_1}{S} \right\rceil \cdot S
    \label{eq:startOfMACProcessing-w3}
\end{equation}
}

The physical layer must timestamp the samples before submitting them to the radio for transmission.
The radio processes these samples and transmits them at the designated time.
If the gNB does not provide enough lead time, the radio may fail to process and transmit the samples in time, resulting in underflow and data loss.
Hence, the gNB must supply the samples to the radio in advance.
We define the number of slots the MAC schedules in advance as $a_1$.
The time the radio needs to prepare the samples before transmission is $r_1$.
Based on this, two constraints apply.
First, MAC and physical layer processing must finish before radio processing begins.
Second, the radio must have enough time to prepare the samples before transmission, expressed respectively as follows.

{\small
\begin{equation}
    \left\lceil \frac{p_2 + p_3}{S} \right\rceil \leq a_1 + 1, \quad
    r_1 < (a_1 + 1) \cdot S - (p_2 + p_3)
    \label{eq:combined_constraints}
\end{equation}
}

The scheduled slot for grant transmission is as follows:

{\small
\begin{equation}
    \text{Scheduled Slot for Grant} = \frac{o_1 + w_1}{S} + \left\lceil \frac{w_2 + p_1}{S} \right\rceil + a_1 + 1
    \label{eq:scheduledSlotForGrant}
\end{equation}
}

The value of $w_3$ can be calculated as follows.
The cases follow the logic that the scheduled slot for the grant should be a DL slot.
If the slot determined by \cref{eq:scheduledSlotForGrant} is a UL slot, the grant is sent in the first DL slot of the next TDD period (second case in \cref{eq:w3}).

{\small
\begin{equation*}
re_1 \equiv_T \frac{o_1 + w_1}{S} + \left\lceil \frac{w_2 + p_1}{S} \right\rceil + a_1 + 1
\end{equation*}
\begin{equation}
\label{eq:w3}
w_3 =
\begin{cases}
    \left( \left\lceil \frac{w_2 + p_1}{S} \right\rceil + a_1 + 1 \right) \cdot S; & \text{if } re_1 \leq d \\[10pt]
    \left( \left\lceil \frac{w_2 + p_1}{S} \right\rceil + a_1 + 1 + (T - re_1) \right) \cdot S; & \text{otherwise}.
\end{cases}
\end{equation}
}

\vskip 0.06in \noindent {\bf Modeling of $w_4$:}
This variable represents the time, with respect to the beginning of the slot, at which the PDCCH occasion carrying the uplink grant has been fully received.
The uplink grant is conveyed by DCI on the PDCCH, which the UE monitors over configured time-frequency resources.
Let $DC_{st}$ denote the zero-based index of the first OFDM symbol of the PDCCH occasion within the slot, and let $DC_{no}$ denote its duration in OFDM symbols.
The 3GPP standard specifies that the corresponding PDCCH resources occupy a contiguous set of OFDM symbols, with duration no larger than three OFDM symbols.
For normal cyclic prefix, $w_4$ is modeled as follows.

{\small
\begin{equation}
    w_4 = \frac{DC_{st} + DC_{no}}{14} \cdot S, \quad
    DC_{no} \in \{1, 2, 3\}.
    \label{eq:combined_w4_pdcchConstraint}
\end{equation}
}

\vskip 0.06in \noindent {\bf Modeling of $w_5$:}
This variable represents the time between the beginning of the slot in which the grant is sent and the UL slot where the UE sends data using the allocated resources.
Each grant in PDCCH includes information specifying the slot in which the UE should send data.
This slot is communicated using an offset value called $ k_2 $, which indicates how many slots after the grant the UE should transmit.
The value of $ k_2 $ must align with the UE's processing capability.
If the UE requires $ l_2 $ seconds for MAC and PHY processing, the minimum value of $ k_2 $ can be calculated as follows.
This equation determines the first slot after the grant where the UE is finished with the processing.

{\small
\begin{equation}
    k_{2_{\text{min}}} = \left\lceil \frac{w_4 + l_2}{S} \right\rceil, \quad
    k_{2_{\text{min}}} \leq k_2
    \label{eq:combined_k2min_constraint}
\end{equation}
}

If we assume that the grant aims to minimize latency by setting $ k_2 $ as low as possible,  $ w_5 $ can be calculated as:

{\small
\begin{equation}
    re_2 \equiv_T \frac{o_1 + w_1 + w_3}{S} + k_{2_{\text{min}}}
    \label{eq:re2}
\end{equation}
\begin{equation}
    k_2 =
    \begin{cases}
    k_{2_{\text{min}}}, & \text{if } re_2 \geq d; \\
    k_{2_{\text{min}}} + d - re_2, & \text{otherwise.}
    \end{cases}, \quad
    w_5 = k_2 \cdot S
    \label{eq:combined_k2_w5}
\end{equation}
}

\cref{eq:re2} calculates the slot number within a TDD period.
The second case in \cref{eq:combined_k2_w5} represents the scenario where the $k_{2_{\text{\text{min}}}}$-th slot after the grant is a DL slot.
In this case, the grant must be postponed until the next available UL slot, which is the first UL slot in the same TDD period.

\vskip 0.06in \noindent {\bf Modeling of $w_6$ and $w_7$:}
$w_6$ represents the time of the UL data transmission.
If we assume that the UL data can be transmitted within a single Physical Uplink Shared Channel (PUSCH) slot, then $w_6 = S$.
We present this case for clarity, while the general model supports transmissions spanning multiple slots, as detailed in the appendices.
Once the uplink data reaches the gNB radio, it must be forwarded and processed through all layers of the 5G stack before being forwarded to the core network.
This processing time is represented by $w_7$.
We aggregate the processing time across all 5G stack layers into a single variable, $p_4$. Hence, $w_7=p_4$.
While gNB processing could be improved to occur symbol by symbol (as with PDCCH and PUCCH), we assume it begins only after the entire slot is received (\cref{fig:overall-latency}).
This worst-case assumption accounts for uncertainty in the symbol carrying the data and the likelihood that large transmissions span most of the slot.

\subsection{Stochastic Framework}
\label{sec:probabilistic_approach}

In \cref{sec:model-foundation}, we introduced a model for system latency using several variables. However, some of these variables are nondeterministic and fluctuate over time. For example, the values of the \textit{UE Processing Time} (denoted by $l_1$) and the \textit{gNB Processing Time} (denoted by $p_4$) can substantially affect overall latency. To appreciate the need to model the stochastic nature of these variables, consider a 5G system where the scheduling request (SR) period is set to \SI{20}{\milli\s}.
If a UE is idle and suddenly has a new packet to send, it must wait for the next SR opportunity before transmission.
Suppose the next SR window is \SI{1}{\milli\s} away, and the UE's processing time averages \SI{1}{\milli\s} but has some variability.
In this case, the UE may miss the imminent SR window due to longer-than-average processing, incurring an extra \SI{20}{\milli\s} delay until the following SR opportunity. Similarly, the gNB processing time varies depending on the system load, contributing to nondeterministic latency.

Because of these uncertainties, latency must be examined from a probabilistic perspective. In particular, there are two types of variables defined in \cref{sec:model-foundation}: (1) Constants dictated by the chosen system configuration, such as $S$, the time slot duration, and $T$, the number of time slots in a TDD pattern. (2) Random variables like $o_1$ (the data packet arrival time), $l_1$ and $p_4$ mentioned above. To generate the output distribution for the latency, we sample these variables from their input distribution. For each sampling instance of the random variables, we calculate the latency. We then aggregate the latency measurements across samples to generate the latency distribution. The input distribution of these random variables can be either:

\vskip 0.06in \noindent {\it (i) Empirical distribution}: The empirical distribution is generated by collecting real measurements of the values of these random variables from an experimental setup and aggregating them to generate a probability density function (PDF).

\vskip 0.06in \noindent {\it (ii) Learned distribution}: In some cases, it is not feasible to directly measure these variables since the 5G UE chipsets are not opened (e.g. Qualcomm SDX65~\cite{qualcomm-sdx-65}).
To address this, we learn the distribution by assuming it follows a known distribution (e.g., Gaussian with mean $\mu$ and standard deviation $\sigma$).
We then collect the latency of \num{10000} packets sent with constant inter-arrival time as our ground truth on a given configuration and find the final output distribution of the latency.
We iterate over different parameters (e.g., $\mu$ and $\sigma$) of the input distribution and generate the output latency distribution using the models as described above. We then use the Wasserstein distance~\cite{vaserstein1969markov,kantorovich1960mathematical,mallows1972note} as our error function between the calculated and measurement distributions to find the best parameters that fit the real-world data. By doing so, we can indirectly estimate the input distribution of these random variables.
It is important to note that we only learn the distribution (train) from a single configuration. However, we evaluate it (test) on completely new configurations that generate very different output latency distributions.

In this paper, we focus on the three
important variables:
\vskip 0.06in \noindent {\bf UE Processing Time} ($l_1$ in \cref{sec:model-foundation}).
We estimate the distribution of this variable from measurements and approximate it with a Gaussian model based on the observed empirical distribution.
Note, however, that depending on the UE's model and load, the processing time can vary in the range of \SI{2}{\milli\s}~\cite{10.1145/3750718.3750743}.
The typical UE's processing time mean in our experiments is around \qtyrange[range-phrase=~to~,range-units=single]{1}{3}{\milli\s}, and the standard deviation is around \qtyrange[range-phrase=~to~,range-units=single]{0.1}{0.6}{\milli\s} for low and high loads, respectively.
We train the parameters on a certain model and load and use it to test the model with different configurations on similar loads.

\vskip 0.06in \noindent {\bf gNB Processing Time} ($p_4$ in \cref{sec:model-foundation}).
Modeling the gNB processing time can be easier than the UE processing time, due to our access to open-source gNBs.
However, for commercial gNBs, we can also use a learned distribution since we find that the empirical distributions can be approximated with a log-normal distribution.
Typically, the mean processing time of the gNB in our experiments ranges from \qtyrange[range-phrase=~--~,range-units=single]{0.5}{1.66}{\milli\s} under low and high loads, respectively; the corresponding standard deviation is approximately \SI{0.1}{\milli\s}.

\vskip 0.06in \noindent {\bf Packet Inter-Arrival Time} (related to $o_1$ in \cref{sec:model-foundation}). The random variable $o_1$ represents packet generation time, but its standalone distribution is not particularly informative.
Instead, the relevant quantity is the distribution of packet inter-arrival times.
These inter-arrival times can be measured by observing packet generation events at the UE and the gNB, corresponding to uplink and downlink traffic, respectively, together with timing information from the gNB. Since the model defines $o_1$ relative to the start of the gNB time reference, all packet generation times are expressed in this reference frame.
Using this information, we construct the inter-arrival time distribution, sample from it, and subsequently reconstruct $o_1$ values.
These generated $o_1$ samples are then used to calculate the latency distribution using \thesystem's models.

\subsection{Configuration Analyzer}\label{sec:analyzer}

Determining an appropriate configuration for a 5G gNB is a non-trivial task, as it requires tuning many parameters.
Some of these parameters are constrained by standards or regulatory requirements.
For example, in Switzerland, the n78 band is the only spectrum that can be licensed for campus networks, and it operates exclusively in TDD mode.
Similarly, according to the 3GPP standard, NR FR1 bands above \SI{2.69}{\giga\hertz} are not defined for FDD operation~\cite{3GPP-scs-fr1}.
Nevertheless, many other configuration parameters remain under the operator's control.

Even when a relaxed goal of minimizing the average latency is in mind, a naive approach, like setting the TDD pattern's period to the minimal supported value, might result in a worse latency, as we show in \cref{sec:config_analyzer}.
We design an analyzer that systematically explores the configuration space to identify the desired valid configuration for a gNB under a given set of constraints that can achieve a certain performance goal.
The objective for the analyzer can be user-defined, such as minimizing the average latency or the 99\% latency of the distribution.
Although we restrict the search to parameters that mainly affect latency, the configuration space still grows combinatorially, with around 32 billion configurations in the sub-\SI{6}{\giga\hertz} frequency bands for grant-based access alone.
This number arises from the range of possible values of each configuration parameter as detailed in Appendix~\ref{sec:eval_analyzer_config}.

To make the search tractable, we reduce the configuration space through pruning based on three principles.
First, each parameter value is validated against 5G standard constraints, ensuring that only standard-compliant configurations are considered.
For example, the slot duration is restricted to $1\,\mathrm{ms}/2^u$, where $u \in \{0,1,2\}$ in the sub-\SI{6}{\giga\hertz} bands and $u \in \{2,3,4,5,6\}$ in the millimeter-wave bands.
This step eliminates many invalid configurations.

Second, the analyzer accounts for cross-parameter dependencies that can make otherwise valid individual settings infeasible when combined.
For instance, if the SR period and SR offset are chosen independently of the TDD pattern, the SR may always fall in a downlink slot, preventing the UE from transmitting an SR.
Similarly, the number of slots in a TDD pattern and the slot duration cannot be set independently, since their product (the TDD period) is restricted to specific values by the standard.

After these pruning steps, which eliminate invalid and infeasible configurations, further pruning depends on the optimization objective.
If the goal is to minimize a latency metric (e.g., average, maximum, or minimum), the analyzer can discard suboptimal configurations when it can prove that better ones exist.
In contrast, if the objective is to identify all configurations that satisfy a given latency-reliability target, such pruning cannot be applied, as all configurations must be evaluated.

In the case of latency minimization, the analyzer further prunes suboptimal configurations as follows.
For parameters whose impact on latency is monotonic, it fixes them to their optimal values to further reduce the search space.
For example, placing UL control information at the beginning of a slot ensures that scheduling requests are sent and received earlier, and varying this parameter cannot improve latency.
After pruning, the remaining configurations are evaluated in two phases.
In the coarse phase, the analyzer runs \thesystem's model with a few packets to estimate latency distributions and retains the top \qty{10}{\percent} of configurations.
In the fine phase, it evaluates these configurations with more packets to identify the one that best meets the chosen objective.

In addition to identifying the single best configuration, the analyzer also supports finding all configurations that satisfy a given objective.
In this mode, it disables parameter fixing based on monotonic effects and the two-phase coarse-fine evaluation, since these mechanisms eliminate non-optimal configurations rather than enumerate all valid ones.
The configuration analyzer allows us to answer questions like:

\vskip 0.03in \noindent \emph{-- What configurations can achieve a \SI{1}{\milli\s} uplink latency with a reliability of \qty{95}{\percent} given that we use numerology 3?}

\vskip 0.03in \noindent \emph{-- Is it possible to achieve \SI{0.5}{\milli\s} downlink latency with a reliability of \qty{99.99}{\percent} if the radio latency is \SI{0.3}{\milli\s}?}

\vskip 0.03in \noindent \emph{-- What configurations can achieve URLLC specs without using grant-free scheduling?}

\vskip 0.03in \noindent \emph{-- What configuration achieves the lowest average latency given the constraint that we must use TDD with a pattern of 4 DL and 2 UL and no mixed-slots?}

For reliability, we follow the definition in 3GPP TR~38.913~\cite{3GPP-TR-38.913-version-18}, where reliability is evaluated as the success probability of transmitting $X$ bytes within a specified delay, measured from the radio protocol layer~2/3 SDU ingress point to the corresponding SDU egress point of the radio interface, under a given channel-quality condition.
For URLLC, TR~38.913 specifies a general reliability requirement of $1-10^{-5}$ for one transmission of a \SI{32}{\byte} packet with a user-plane latency of \SI{1}{\milli\s}.
\thesystem's models consider single-transmission operation, with HARQ retransmissions disabled.
Thus, packets that are not successfully delivered in the initial transmission are not recovered through HARQ and are treated as deadline violations in the latency-reliability evaluation.
Under this setting, the reported reliability quantifies the probability that a packet is delivered within the target latency using only the initial transmission, together with the stochastic components of \thesystem.
This provides a conservative indication of whether a configuration can satisfy the latency-reliability target without relying on retransmission-based recovery.
Accordingly, reliability can be inferred from the CDF of the output latency distribution: for example, if the \qty{99}{\percent} percentile latency is \SI{3}{\milli\s}, then the system achieves a latency of \SI{3}{\milli\s} with \qty{99}{\percent} reliability.

\section{Implementation and Setup}
\label{sec:experimental-setup}

\begin{figure}[t]
  \centering
  \includegraphics[width=\linewidth]{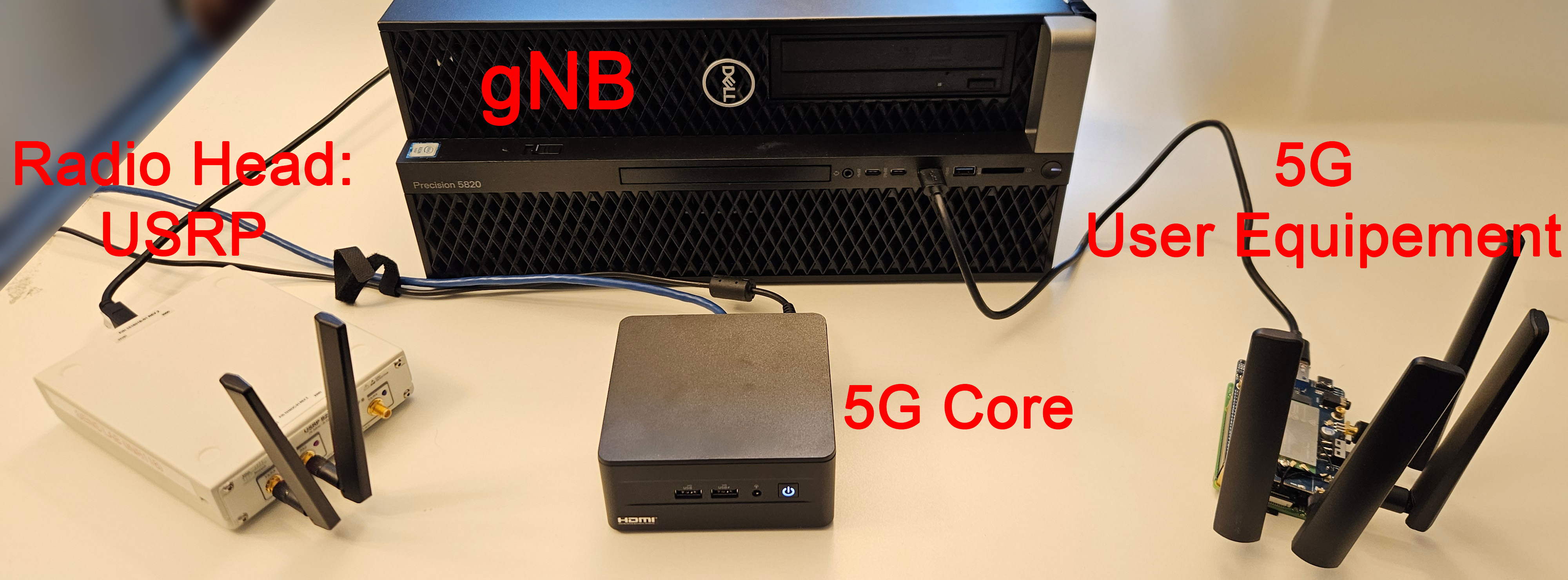}

  \caption{The private 5G testbed setup.}\label{fig:testbed}

\end{figure}
\begin{figure*}[t!]
  \centering
  \captionsetup[subfloat]{width=0.185\linewidth,justification=centering,singlelinecheck=false}
  \subfloat[UL, TDD (3\textbf{D}/1\textbf{U}), SR = 2 ms, $k_2=2$, $a_1=3$, Constant, srsRAN\label{fig:4211_sr2_bsr1_k2_mac1}]{
    \includegraphics[width=0.185\linewidth]{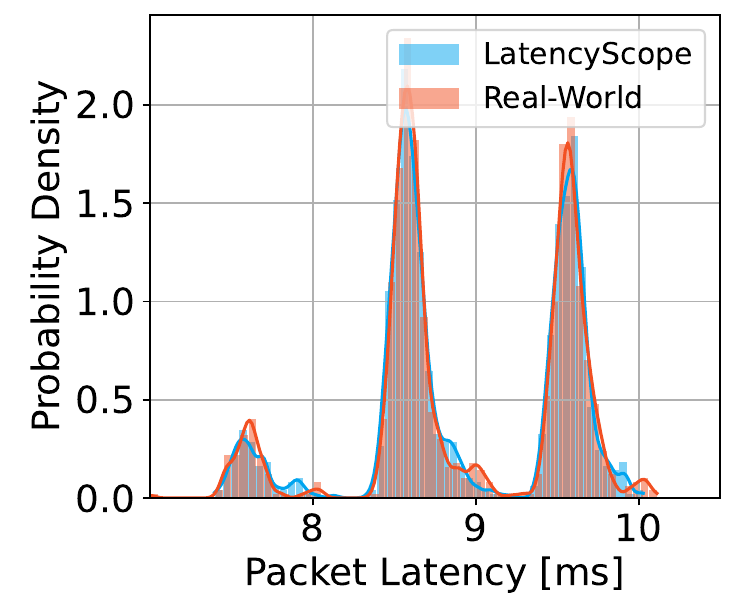}
  }
  \hfil
  \subfloat[UL, TDD (3\textbf{D}/1\textbf{U}), SR = 4 ms, $k_2=2$, $a_1=3$, Gaussian, srsRAN\label{fig:4211_sr4_bsr1_k2_mac1_gauss}]{
    \includegraphics[width=0.185\linewidth]{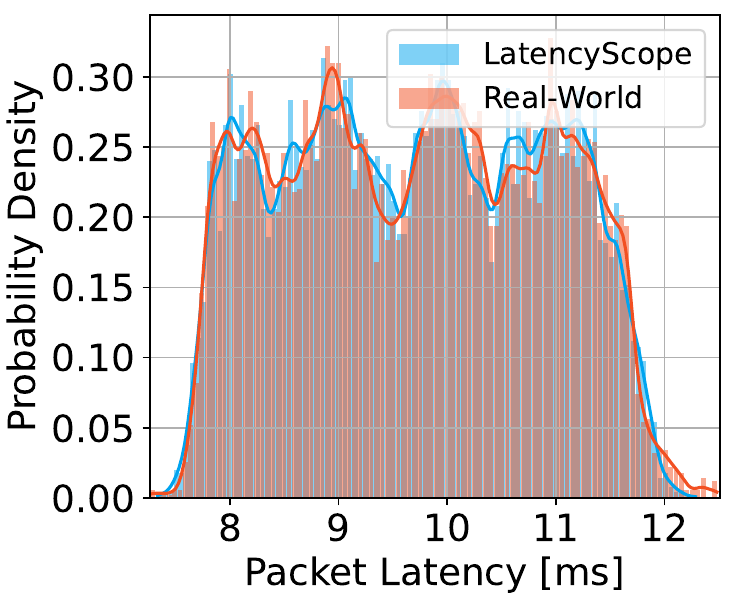}
  }
  \hfil
  \subfloat[UL, TDD (7\textbf{D}/3\textbf{U}), SR = \SI{10}{\milli\s}, $k_2=3$, $a_1=3$, Zoom, srsRAN\label{fig:10613_sr10_bsr1_k3_mac1_zoom}]{
    \includegraphics[width=0.185\linewidth]{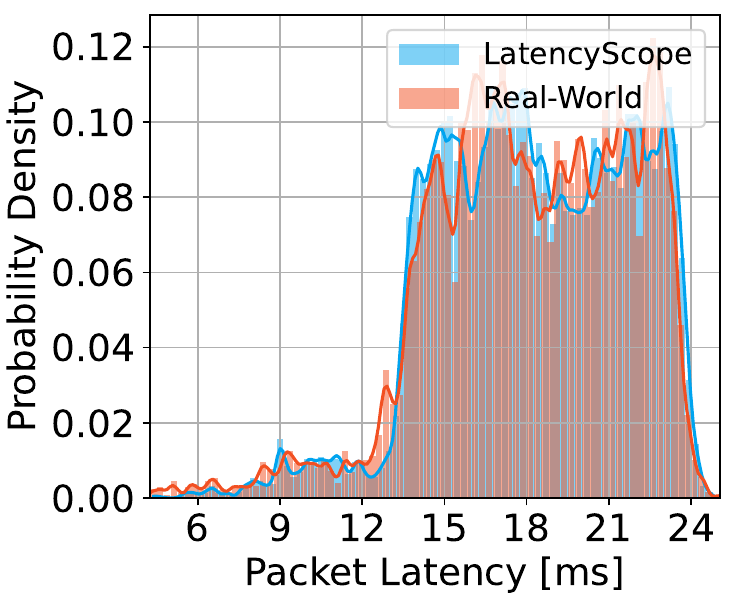}
  }
  \hfil
  \subfloat[UL, TDD (7\textbf{D}/3\textbf{U}), SR = \SI{20}{\milli\s}, $k_2=3$, $a_1=7$, Dota~2, srsRAN\label{fig:10613_sr20_bsr1_k3_mac5_dota2}]{
    \includegraphics[width=0.185\linewidth]{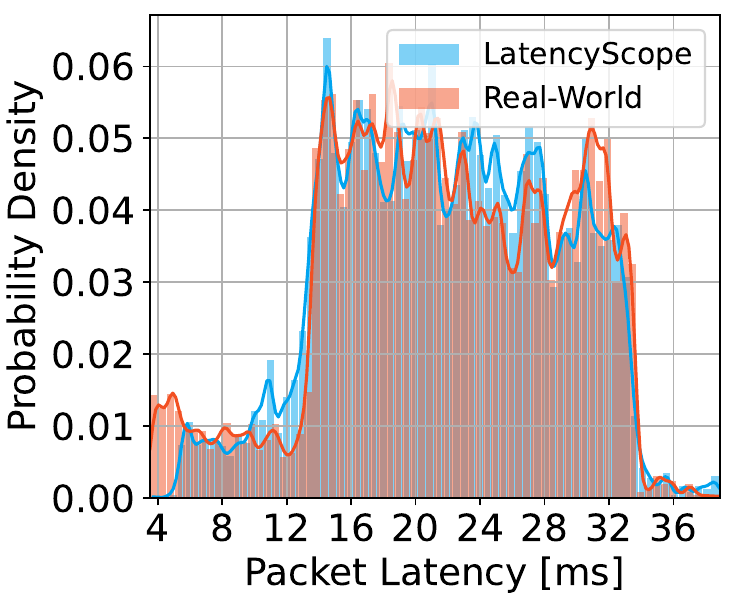}
  }
  \hfil
  \subfloat[UL, TDD (3\textbf{D}/1\textbf{U}), SR = 10 ms, $k_2=2$, $a_1=5$, Constant, \textbf{OAI}\label{fig:OAI}]{
    \includegraphics[width=0.185\linewidth]{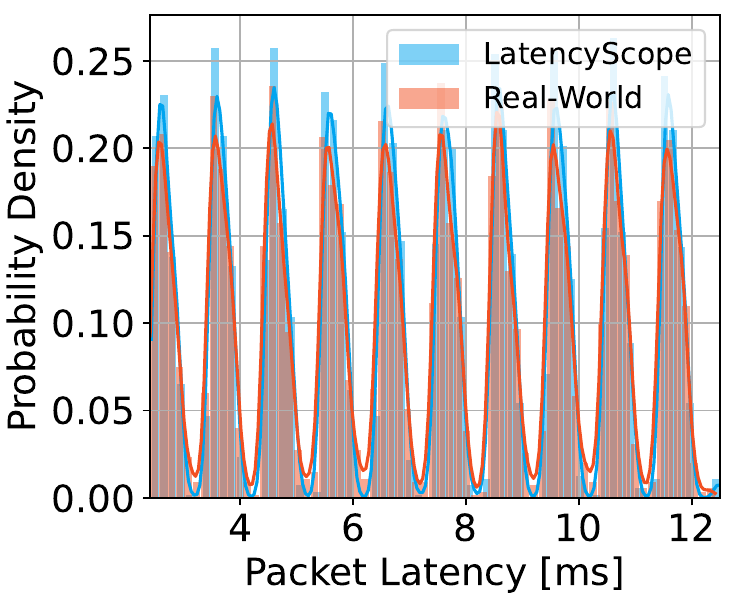}
  }
  \\
  \hfil
  \subfloat[UL, FDD, SR = 2 ms, $k_2=2$, $a_1=3$, Constant, srsRAN\label{fig:FDD}]{
    \includegraphics[width=0.185\linewidth]{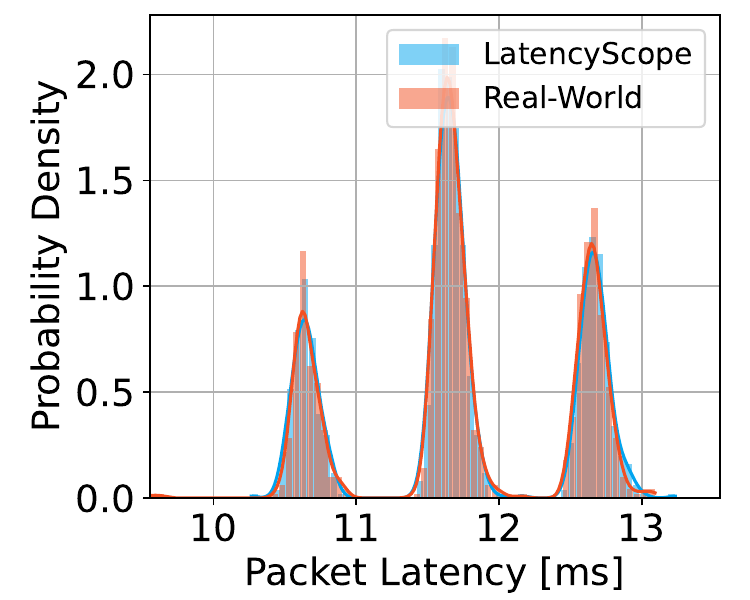}
  }
  \hfil
  \subfloat[UL, TDD (7\textbf{D}/3\textbf{U}), SR = \SI{10}{\milli\s}, $k_2=3$, $a_1=3$, large packets, srsRAN\label{fig:LargePacket}]{
    \includegraphics[width=0.185\linewidth]{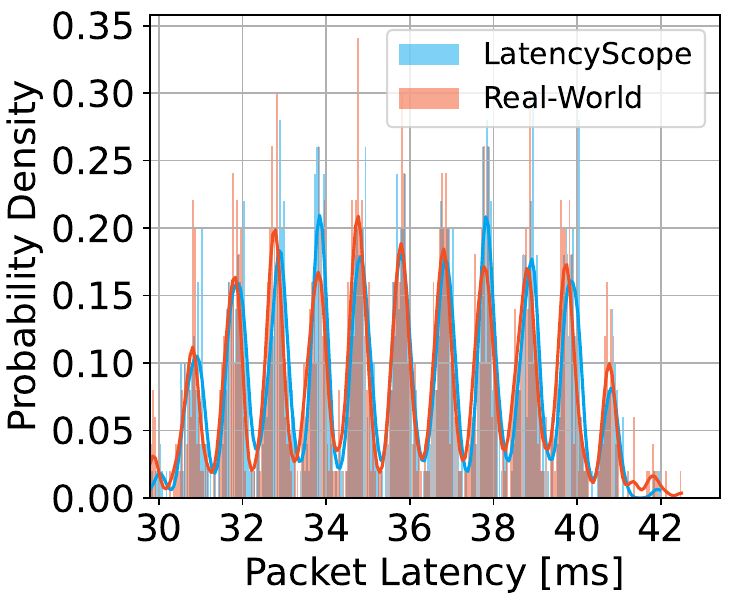}
  }
  \hfil
  \subfloat[UL, TDD (4\textbf{D}/1\textbf{U}), SR = \SI{40}{\milli\s}, $k_2=1$, $a_1=1$, constant, \textit{MNO}\label{fig:UL_mno1_1}]{
    \includegraphics[width=0.185\linewidth]{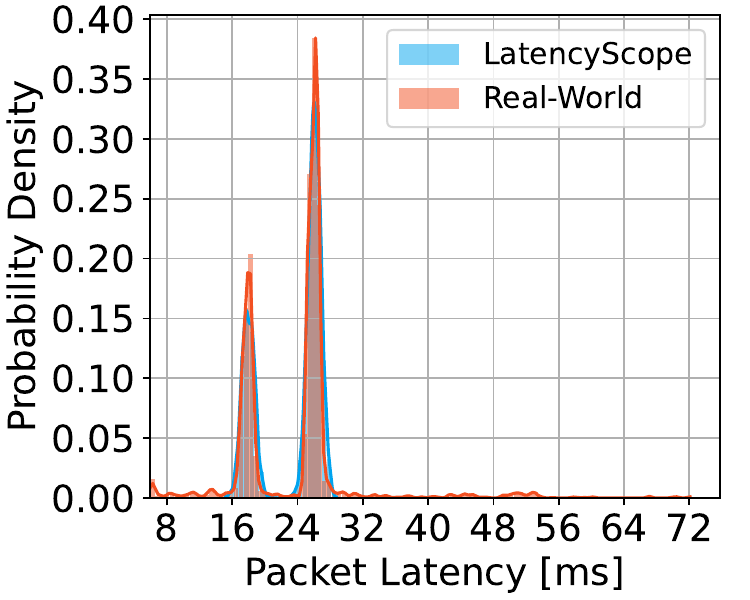}
  }
  \hfil
  \subfloat[UL, TDD (4\textbf{D}/1\textbf{U}), SR = \SI{40}{\milli\s}, $k_2=1$, $a_1=1$, constant, \textit{MNO}\label{fig:UL_mno1_2}]{
    \includegraphics[width=0.185\linewidth]{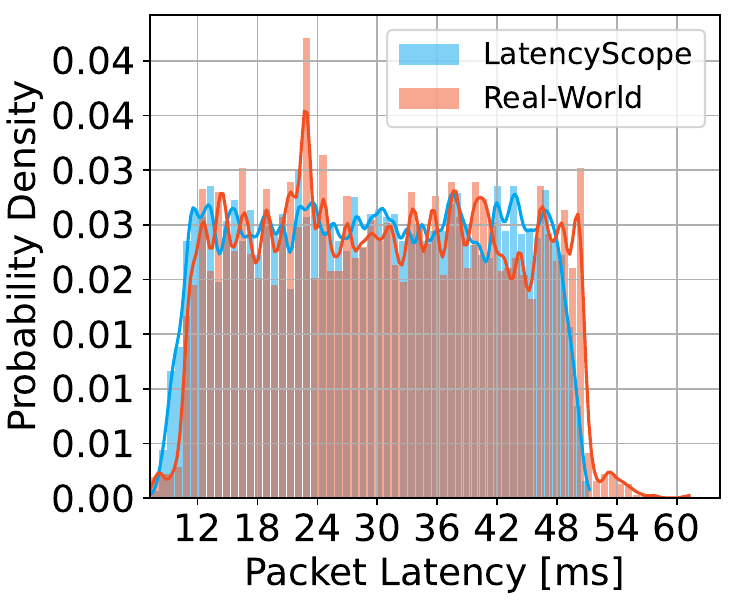}
  }
  \hfil
  \subfloat[DL, TDD (4\textbf{D}/1\textbf{U}), $a_1=1$, constant, \textit{MNO}\label{fig:DL_mno1_2}]{
    \includegraphics[width=0.185\linewidth]{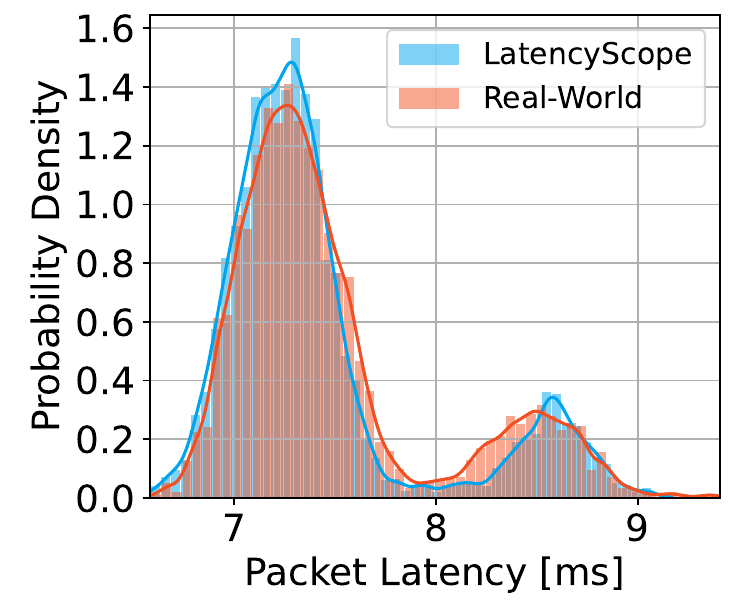}
  }

  \caption{Latency distributions over the 0.2nd-99th percentile latency range across scenarios on a commercial 5G operator (\textit{MNO}) and two private 5G testbeds (srsRAN, OAI).
  Traffic includes real applications (Zoom, Dota~2) and synthetic workloads:
  (1) Constant---\SI{64}{\bytes} packets with fixed inter-arrival time (\SI{101}-\SI{1000.01}{\milli\s});
  (2) Gaussian---\SI{64}{\bytes} packets with Gaussian inter-arrival times (mean \SI{105}{\milli\s}, std \SI{0.05}{\milli\s});
  (3) Large packets---\SI{40000}{\bytes} packets with fixed \SI{201}{\milli\s} inter-arrival time.
  Panels (a--e,g--j) use numerology~1; panel~(f) uses numerology~0.}

  \label{fig:latency-distribution-2}

\end{figure*}

\subsection{Private 5G Testbed}

We evaluate \thesystem on two private 5G networks deployed in our lab. We use modified versions of commonly used open-source 5G RAN implementations, namely srsRAN~\cite{srsRAN} and OpenAirInterface (OAI)~\cite{OAI}, and the open-source core network implementation Open5GS~\cite{open5gs}.
For our radio platform, we use the USRP B210 and BladeRF xA9.
We run the gNB code on an Intel Xeon W-2225 CPU.
Furthermore, we use Telit FN990A40 5G modems as UEs.
The testbed is shown in \cref{fig:testbed}.
Due to the constraints of the open-source 5G implementations, for TDD, we obtain ground truth latency measurements using band n78 with a SCS of \SI{30}{\kilo\hertz}, and for FDD, we conduct measurement using band n3 with a SCS of \SI{15}{\kilo\hertz}.
We evaluate different FDD and TDD configurations, such as \SI{2}{\milli\s}, \SI{2.5}{\milli\s}, and \SI{5}{\milli\s} TDD periods, with different TDD patterns and different SR periods.

\subsection{Commercial 5G Network}

We also evaluate \thesystem using latency measurements collected from a public commercial mobile network operator (\textit{MNO}).
As of January 2026, the anonymized \textit{MNO} is the sole operator in Switzerland offering 5G Standalone (SA) services to consumers.
Measurements were conducted using a Quectel RM500Q-GL 5G modem with a consumer-grade 5G data plan.
To measure one-way latency, packets were transmitted from the UE to our synchronized server (colocated with the UE) over the public Internet as we do not have access to MNO's RAN.
Packets therefore traverse the 5G RAN, the mobile core network, and the public Internet.
For uplink measurements, sending both ICMP or UDP packets from the UE to the server is feasible.
For downlink measurements, due to NAT and firewall restrictions, the server cannot directly initiate connection to the UE; therefore the UE first sends ICMP packets to the server, which then replies to the UE.

The measured end-to-end latency includes additional wired-network delays from the core and the Internet, which cannot be isolated from the 5G RAN in commercial deployments. Moreover, \thesystem requires network configuration parameters to compute latency distributions.
We obtain these parameters by decoding RRC messages captured with QCSuper~\cite{qcsuper}, which interfaces with Qualcomm-based devices to record RRC/NAS/protocol messages.
This allows us to extract key settings such as the numerology, TDD pattern, and SR period.
However, some parameters (e.g., $a_1$) remain unobservable. Through extensive measurements, we find that for our traces, the residual wired/core delay was well approximated as an additive Gaussian component.
Therefore, we perform a parameter sweep over the unknown model parameters (e.g., $a_1$ and the Gaussian mean and variance) using one measurement trace, selecting the best match to the observed latency distribution. The learned parameters are then reused for new measurements in \cref{sec:eval}.
Parameters of \textit{MNO}, either learned or directly obtained by decoding RRC messages, are summarized in Appendix~\ref{sec:mno1-params}.
\subsection{\thesystem Implementation}

To evaluate \thesystem, we implemented all the models summarized in~\cref{sec:model-foundation} in Python.
The implementation can operate in two modes: (1) Synthetic-traffic: generated with specific traffic-arrival distribution and packet sizes, and (2) Real-traffic: replayed from previously captured application traffic. We use a custom-built C-based traffic generator and replayer to achieve precise inter-arrival times, outperforming conventional baseline generators~\cite{ping, fping, tcpreplay}, as shown in Appendix~\ref{sec:traffic_generator_evaluation_app}. We also implemented a separate module for the configuration analyzer. Since each configuration can be evaluated independently, the analyzer lends itself naturally to parallel and distributed execution. We employ the Ray framework~\cite{ray} to run the code across 14 machines simultaneously to obtain the results in \cref{sec:possible_urllc_configurations}.

\section{Evaluation}\label{sec:eval}

\begin{figure*}[t]
  \centering
  \begin{minipage}[t]{0.65\linewidth}
    \centering
    \includegraphics[width=\linewidth]{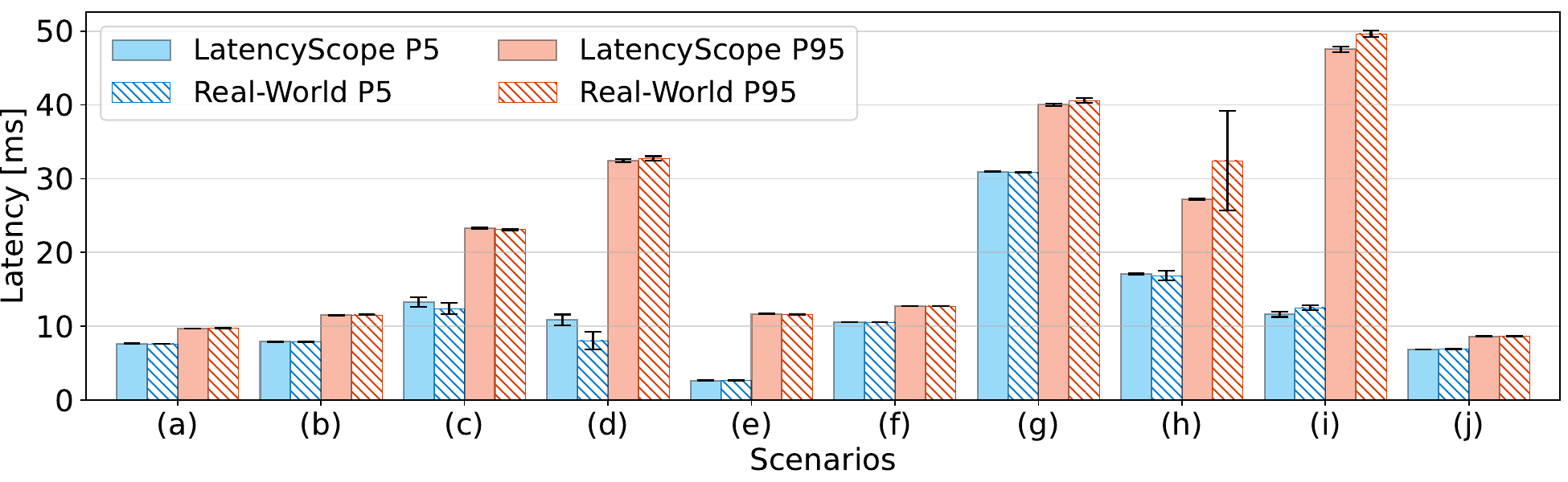}

    \caption{Latency bounds within the 5th-95th percentile range comparing \thesystem and real-world measurements under the same scenarios as \cref{fig:latency-distribution-2}. Error bars show 4th-96th percentiles.}\label{fig:latency_bar}
  \end{minipage}
  \hfil
  \begin{minipage}[t]{0.33\linewidth}
    \centering
    \includegraphics[width=\linewidth]{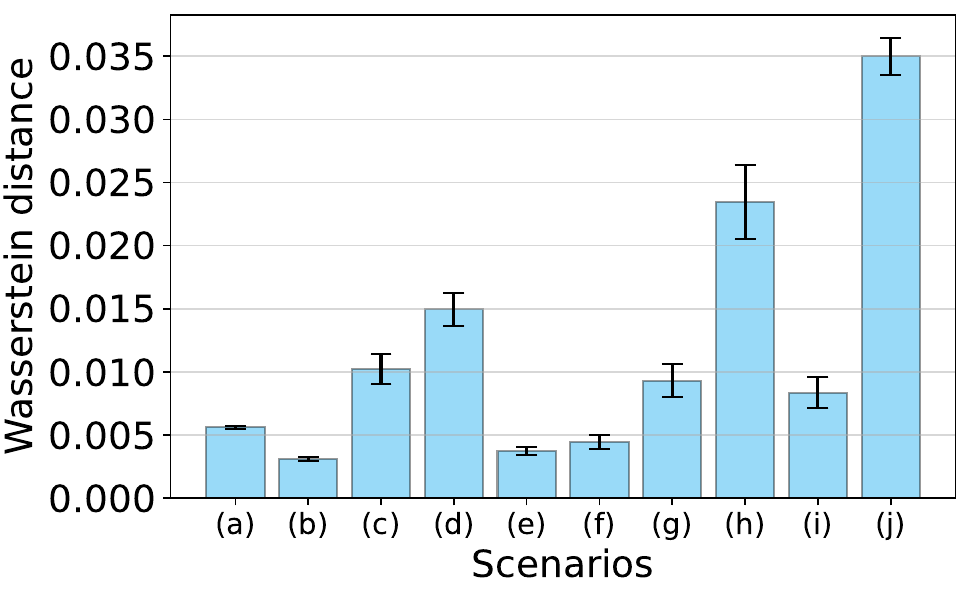}

    \caption{Wasserstein distances between \thesystem and real-world measurements over the 5th-95th percentile latency range. Error bars show 4th-96th percentiles.}\label{fig:wasserstein_distance}
  \end{minipage}

\end{figure*}

\subsection{\thesystem Accuracy}

We first demonstrate how our model can accurately determine latency distributions and bounds.
In \cref{fig:latency-distribution-2}, we show the comparison between the latency distribution estimated by \thesystem and the latency distribution measured in a public commercial 5G operator (\textit{MNO}) and our real-world 5G testbed based on srsRAN or OAI under various system conditions (all denoted by real-world measurements).
Using our 5G testbed, we have the flexibility to evaluate the model across different frame structures, SR periods, $k_2$ offsets, MAC's in-advance scheduling slots ($a_1$), packet arrival patterns, implementations (srsRAN and OAI), UL and DL channels, duplexing (FDD and TDD configurations), and varying packet sizes.
For the commercial 5G network, we run experiments in an urban outdoor non-line-of-sight environment during peak hours, approximately 5 p.m.--9 p.m.~\cite{8125190}, with the device located approximately \SI{1}{\kilo\meter} from the serving cell tower.\footnote{We have noticed that during peak hours, \textit{MNO} uses dynamic SR periods that likely adapt to network conditions (e.g., load).
Since we have access to the RRC messages, we constantly extract the SR period and update the corresponding value in \thesystem for each packet transmission.}
We evaluate \thesystem under both our own generated traffic (e.g., with constant and Gaussian inter-arrival times) and traffic from operational applications (Zoom video calls and Dota 2 online gaming data).\footnote{For such applications, most traffic arrives in bursts/trains rather than as a steady stream.
Therefore, capturing this bursty behavior requires modeling buffering effects at the RLC layer.
The corresponding model is presented in Appendix~\ref{sec:multi-packet} and Fig.~\ref{fig:multi-packet-states}, while its evaluation is included in Fig.~\ref{fig:latency-distribution-2}c--d.}
We can observe from all the cases in \cref{fig:latency-distribution-2} that our model can closely match the real-world latency distributions, accurately capturing the distribution range and the overall shape.
This consistency demonstrates our model's capability of replicating both the spread and the structural characteristics of the measured latency distributions. We present ten additional scenarios in Appendix~\ref{sec:model_evaluation}.
We also verify \thesystem's capability to determine latency lower and upper bounds. Fig.~\ref{fig:latency_bar} shows that \thesystem is able to accurately match these bounds across various 5G configurations.

Finally, we use the \emph{Wasserstein distance}~\cite{vaserstein1969markov,kantorovich1960mathematical,mallows1972note} to quantify the difference between the estimated and measured latency distribution, and we present the results in \cref{fig:wasserstein_distance}.
This metric represents the minimum ``cost'' of turning one distribution into another.
We can see from \cref{fig:wasserstein_distance} that the average values on the normalized distributions range from 0.003 to 0.035 which is considered accurate in practice~\cite{panaretos2019statistical}.

\subsection{Multi-UE Scenario}

\begin{figure}[t]
  \centering
  \subfloat[Low Contention.\label{fig:contention_low}]{
    \includegraphics[width=0.47\linewidth]{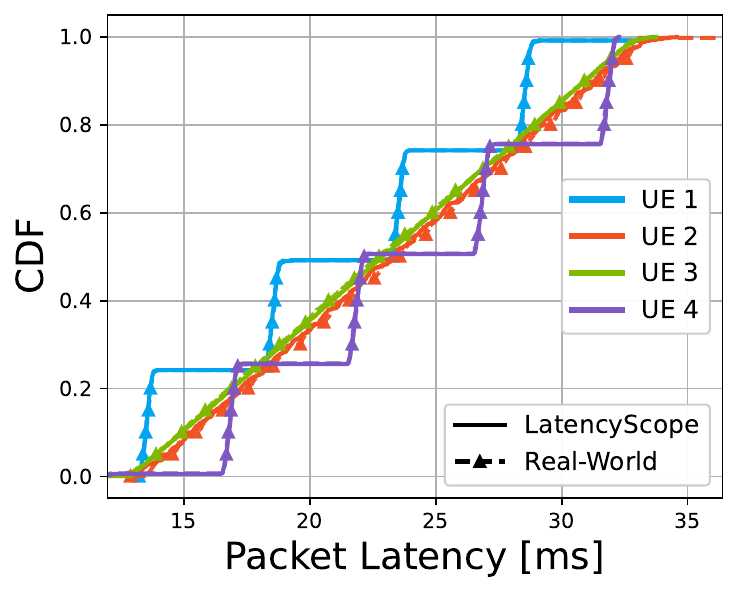}
  }
  \subfloat[High Contention.\label{fig:contention_high}]{
    \includegraphics[width=0.47\linewidth]{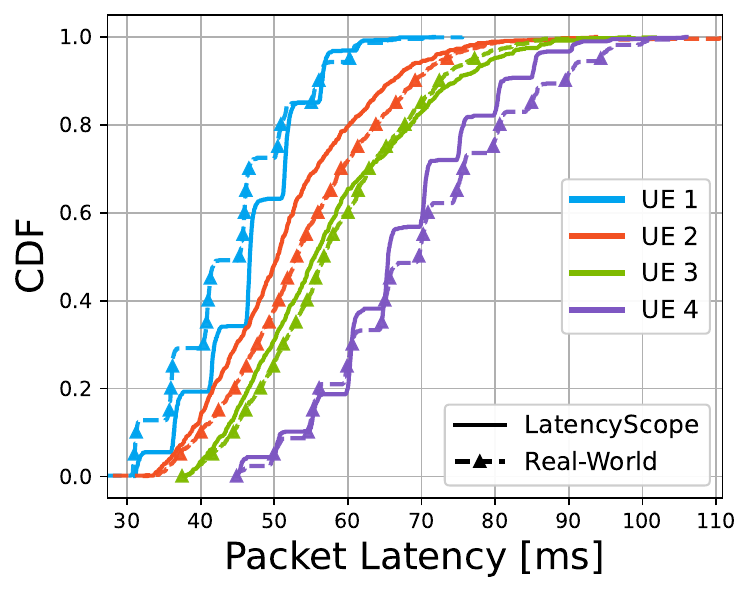}
  }
  \caption{Latency CDFs under contention from 4 UEs.}
  \label{fig:multi-ue-contention}
\end{figure}

\thesystem also models latency under multi-UE contention (see Appendix~\ref{sec:contention-model}).
Since contention dynamics depend on the scheduling policy at the gNB, we model the round-robin scheduler used in our testbed.
The same framework can be readily extended to incorporate other scheduling strategies.
\cref{fig:multi-ue-contention} presents latency distributions when four UEs connect to a single gNB and contend for uplink access.
We evaluate two operating regimes: low contention, where each UE generates traffic at an average rate of \SI{70}{\kbps} (\cref{fig:contention_low}), and high contention, where the average per-UE rate is \SI{5.6}{\mbps} (\cref{fig:contention_high}).
In the high contention case, the total offered load (\SI{22.4}{\mbps}) saturates the full capacity of our \SI{20}{\mega\hertz} testbed.
Under low contention, representative of many URLLC scenarios~\cite{khan2022urllc}, \thesystem closely matches the measured latency distributions.
Under high contention, it continues to track the overall distribution trends, with minor deviations as contention effects become more pronounced.

\subsection{Comparison to Prior Analytical Models}

\begin{figure}[t]
  \centering
  \begin{minipage}[t]{0.49\linewidth}
    \centering
    \includegraphics[width=\linewidth]{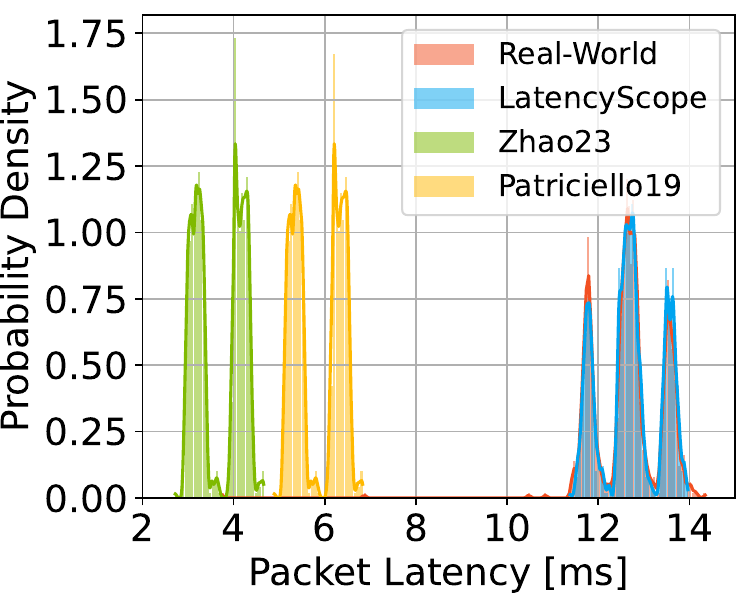}

    \caption{Distributions of \thesystem and baselines.}\label{fig:related_bar}
  \end{minipage}
  \hfil
  \begin{minipage}[t]{0.49\linewidth}
    \centering
    \includegraphics[width=\linewidth]{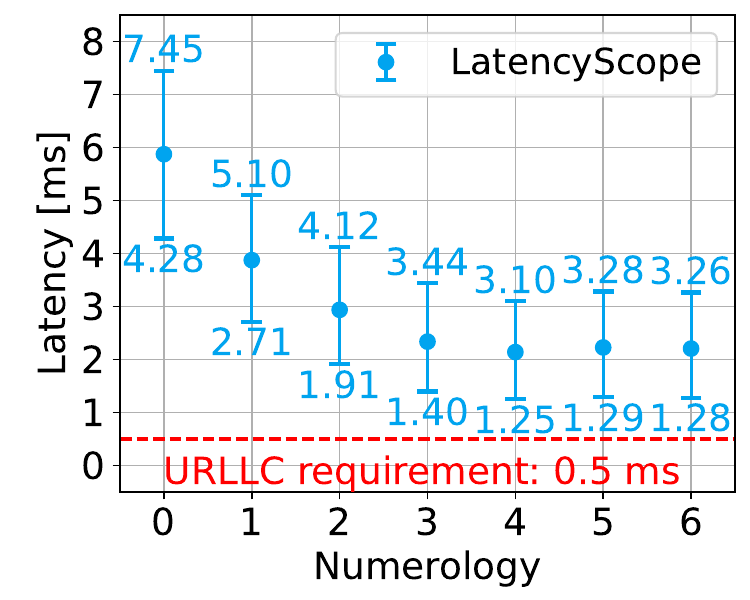}

    \caption{UL latency values (min, avg, and max) vs $\mu$.}\label{fig:numerologyVSlatency}
  \end{minipage}

\end{figure}

 In Fig.~\ref{fig:related_bar}, we compare \thesystem against two baselines--Patriciello19 \cite{patriciello2019impact} and Zhao23 \cite{zhao2023physical}--under the same configuration.
Patriciello19 assumed a fixed two slots gNB processing time and focused solely on the FDD configuration, while Zhao23 examined different duplexing but neglected hardware processing time.
For a fair comparison, we modify Patriciello to support TDD configurations and enable both models to process recorded traffic.
Both baselines significantly underestimate latency, resulting in distributions that deviate considerably from the real-world distribution, with average Wasserstein distances of 0.62 and 0.82.
The errors primarily arise from:
1) The baselines failed to identify system bottlenecks, leading to an underestimation of the required TDD periods for finishing UL transmissions.
2) They did not account for nondeterministic components, resulting in substantial distortions in the distribution.
In contrast, our model matches the measured results with a Wasserstein distance of 0.01.

\begin{figure}[t]
  \centering
  \subfloat[Uplink, TDD (3\textbf{D}/1\textbf{U}), SR = \SI{4}{\milli\s}, $k_2=2$, $a_1=3$, Constant, srsRAN\label{fig:4211_sr4_bsr1_k2_mac1_cdf}]{
    \includegraphics[width=0.47\linewidth]{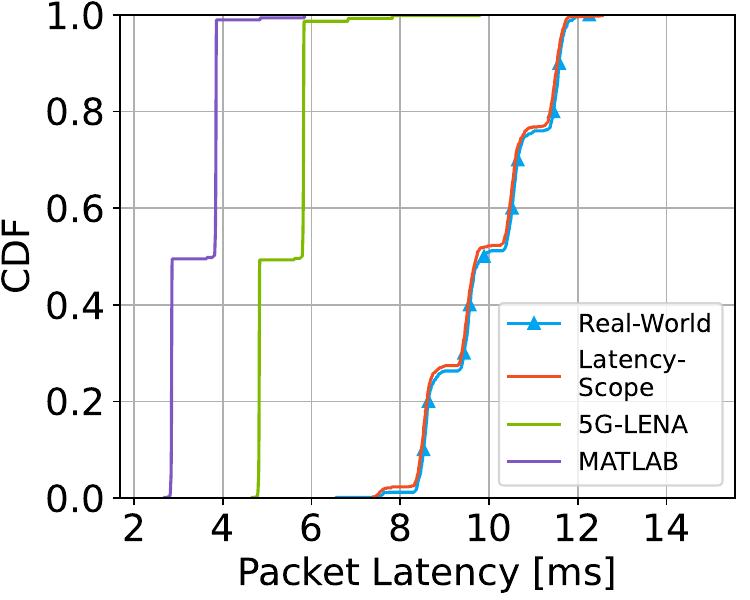}
  }
  \hfill
  \subfloat[Uplink, TDD (7\textbf{D}/3\textbf{U}), SR = \SI{10}{\milli\s}, $k_2=3$, $a_1=3$, Constant, srsRAN\label{fig:10613_sr10_bsr1_k3_mac1_cdf}]{
    \includegraphics[width=0.47\linewidth]{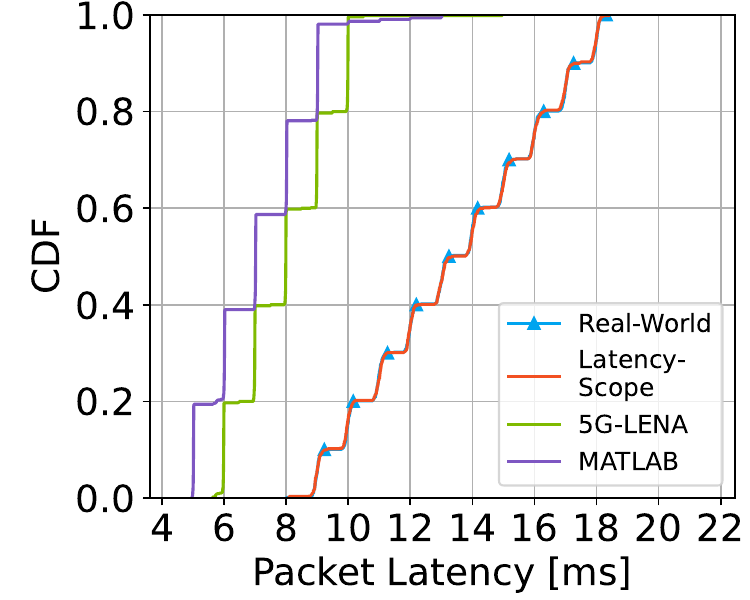}
  }
  \caption{Comparison of latency distributions between \thesystem and two 5G simulators for two scenarios.}
  \label{fig:latency-comparison-simulation}
\end{figure}

\begin{figure}[t]
    \centering
    \begin{minipage}{0.49\linewidth}
        \includegraphics[width=\linewidth]{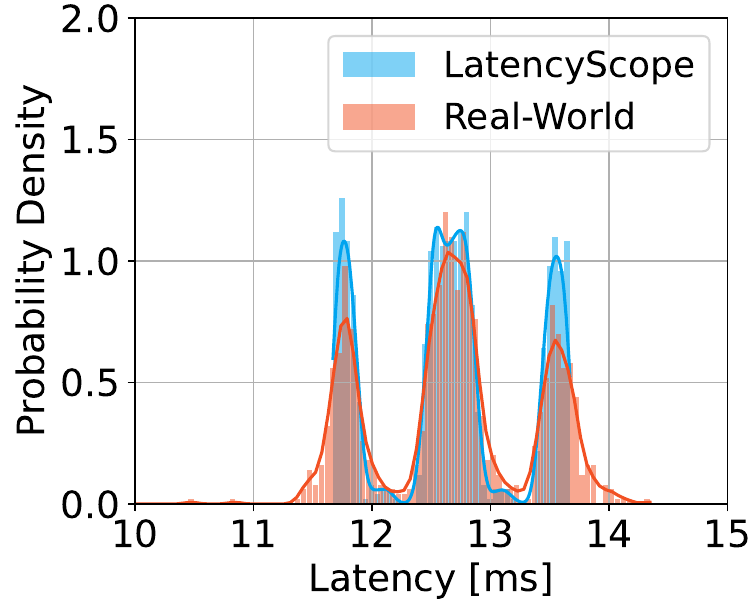}
    \end{minipage}
    \hfill
    \begin{minipage}{0.49\linewidth}
        \includegraphics[width=\linewidth]{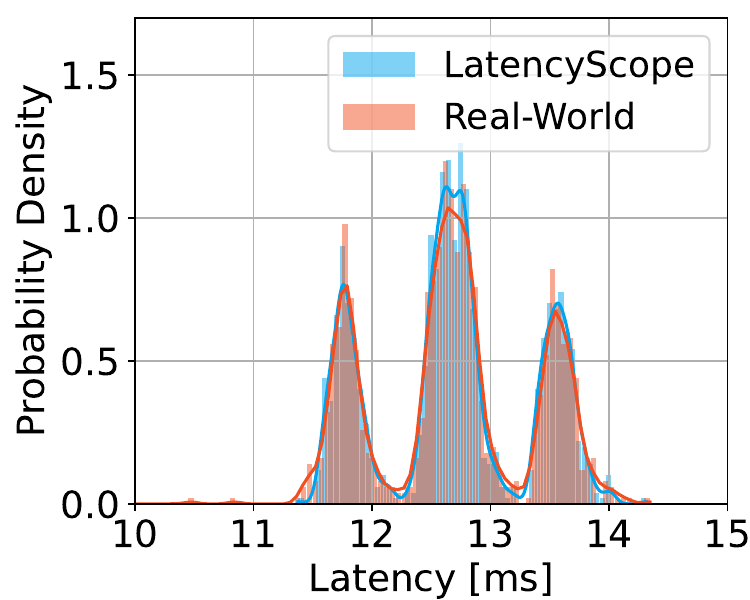}
    \end{minipage}

    \caption{Comparison of modelling $l_1$ and $p_4$ as random variables (right) versus constant values (left)}\label{fig:random_variables}

\end{figure}

\subsection{Comparison with 5G Simulators}

One might ask: \textit{Why develop a new analytical model for latency estimation instead of relying on existing 5G simulators?}
To address this, we evaluated \thesystem against two 5G simulators: the MATLAB 5G Toolbox~\cite{matlab}, and 5G-LENA~\cite{5Glena}, an open-source module built on top of ns-3~\cite{ns3}.
\thesystem demonstrates superior performance in both accuracy and computational efficiency compared to these simulators.
As illustrated in \cref{fig:latency-comparison-simulation}, \thesystem more accurately captures the latency distribution across two representative scenarios. In contrast, both MATLAB and 5G-LENA significantly underestimate the latency range and fail to replicate the actual distribution shape.
Beyond accuracy, the computational cost of traditional simulators is prohibitive.
For example, simulating the transmission of \num{10000} packets for a single configuration required a few minutes in 5G-LENA and MATLAB. In contrast, \thesystem computes the full latency distribution in just \SI{2}{\milli\s}--making it three orders of magnitude faster than 5G-LENA and MATLAB.

\subsection{Need for a Stochastic Framework}

We evaluate the importance of random variables (RVs) for the determination of latency. We do so by showing the difference in the latency of the packets when we model the UE preparation time ($l_1$) and gNB processing time ($p_4$) with and without RVs. In \cref{fig:random_variables}, we observe that the distribution of the latency of the packets is different when we model the $l_1$ and $p_4$ with RVs. The Wasserstein distance between the model and real-world measurements is \num{0.056} when we do not model the RVs and \num{0.027} when we do, which is a significant improvement.

\subsection{Configuration Analyzer: Minimizing Average Latency}
\label{sec:config_analyzer}
In this section, we evaluate the capability of the configuration analyzer to identify the configuration that minimizes average latency under system constraints.
Specifically, we consider a representative setting where the system is restricted to operate in the sub-\SI{6}{\giga\hertz} band with grant-based access.
Within these constraints, the analyzer explores a search space of 32 billion possible combinations, as discussed in~\cref{sec:analyzer}.
After pruning, the search space is reduced to \num{698.8} thousand combinations, enabling the analyzer to identify the optimal configuration in \SI{45}{\second} on a single Intel W-2225 CPU.

\begin{figure}[t]
  \centering
  \includegraphics[width=.8\linewidth]{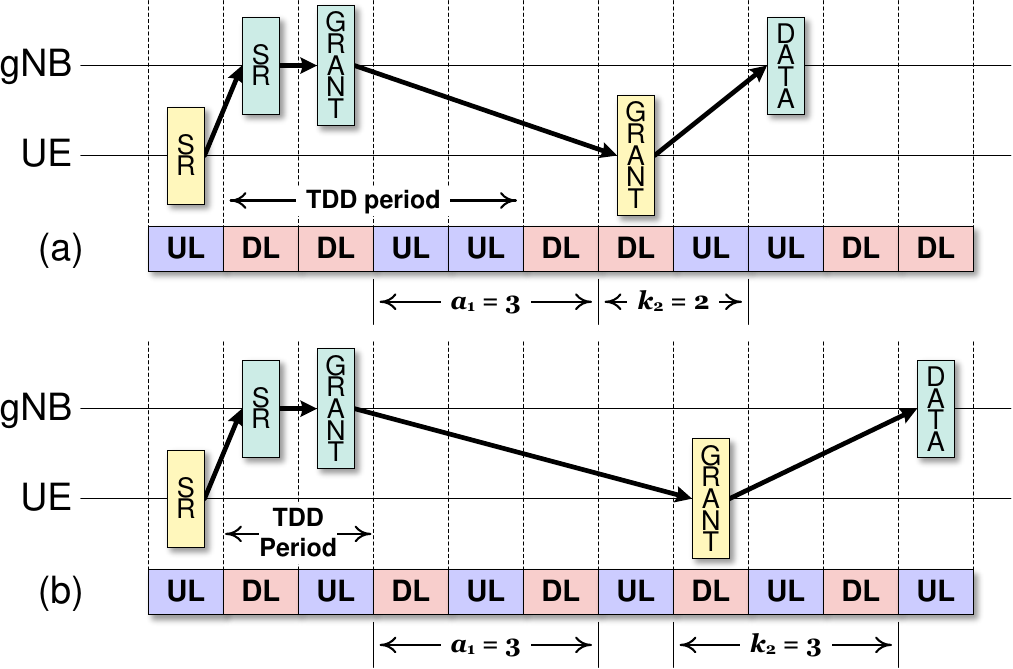}

  \caption{Example of how a shorter TDD pattern does not imply a lower latency. TDD pattern (a) \textbf{D}\textbf{D}\textbf{U}\textbf{U}, and (b) \textbf{D}\textbf{U}.}\label{fig:counter-intuition}

\end{figure}

The analyzer also provides insight into the non-monotonic nature of the problem.
For example, we run the analyzer under numerology \num{2} (\SI{0.25}{\milli\s} slot duration) and a radio latency of \SI{450}{\micro\s} to identify the configuration that minimizes average latency.
Intuitively, one might expect that the shortest TDD period with two slots (\textbf{DU}) yields the lowest uplink latency, as it offers more frequent uplink transmission opportunities.
However, contrary to this intuition, the analyzer selects a four-slot TDD period (\textbf{DDUU}) as the optimal configuration.
For this configuration, the average latency is \SI{2.01}{\milli\s}, with minimum and maximum latencies of \SI{1.67}{\milli\s} and \SI{2.47}{\milli\s}, respectively.
In contrast, the two-slot TDD period yields \SI{2.7}{\milli\s}, \SI{2.95}{\milli\s}, and \SI{3.2}{\milli\s} for minimum, average, and maximum latency, respectively, all higher than the optimal case.

To understand this behavior, consider the example shown in~\cref{fig:counter-intuition}.
Both the gNB and UE must submit RF samples several slots in advance (parameters $a_1$ and $k_2$ in the figure) to allow sufficient time for radio processing.
With a shorter TDD pattern (\textbf{DU}), this requirement causes both the grant and the data transmission to miss immediate transmission opportunities and wait for the next available downlink and uplink slots respectively.
As a result, additional delay is introduced.
In contrast, with the \textbf{DDUU} pattern, the grant and data transmissions can be accommodated within the same TDD period, reducing the overall latency.

\subsection{Configuration Analyzer: Feasible URLLC Configurations}
\label{sec:possible_urllc_configurations}

Leveraging the configuration analyzer, we can obtain the latency variation across possible configurations.
This enables us to evaluate 1) whether a given latency can be achieved with specific system conditions and 2) the set of configurations that can achieve a given latency with a specified reliability.
The definition of reliability follows the 3GPP standard~\cite{3GPP-TR-38.913-version-18} and is described in~\cref{sec:analyzer}.
Such capabilities empower network operators to assess the system's inherent bottlenecks while selecting the most suitable hardware combinations and support service providers in designing URLLC-oriented devices to meet the latency requirements.

We present the number of configurations able to achieve a given latency-reliability target when using a radio unit with \SI{50}{\micro\s} latency in~\cref{table:percentage_reliability} and a radio unit with \SI{450}{\micro\s} latency, similar to our testbed, in~\cref{table:percentage_reliability2}.
Since the goal of this analysis is to evaluate all possible configurations, we leverage the Ray framework to parallelize the analyzer across 14 machines, significantly increasing the available computational power.
Based on \cref{table:percentage_reliability}, out of all configurations, none can reliably go below \SI{1}{\milli\s} in FR1 using grant-based access.
In FR2 (mmWave), no configuration can achieve sub-\SI{0.5}{\milli\s} latencies without resorting to grant-free access.
Grant-based transmission struggles to meet stringent latency targets, with no configurations achieving \SI{0.5}{\milli\s} and some meeting \SI{1}{\milli\s}.
In contrast, grant-free access offers improved latency performance, with up to \qty{11.58}{\percent} of configurations in FR1 and \qty{19.37}{\percent} in FR2 achieving \SI{2}{\milli\s} latency at \qty{50}{\percent} reliability.
Nevertheless, increasing reliability constraints significantly reduces the number of feasible configurations, particularly for grant-based schemes.

Based on~\cref{table:percentage_reliability2}, we observe that significantly fewer configurations satisfy the URLLC requirements when higher radio latency is assumed.
Overall, these results highlight the trade-offs among latency, reliability, and access schemes.

\begin{table}[t]
\caption{Number of configurations achieving a given latency reliability target.}
\label{table:percentage_reliability}
\centering
\begin{threeparttable}
\resizebox{\linewidth}{!}{
\begin{tabular}{cc|ccc|ccc} \hline
&& \multicolumn{3}{c|}{\textbf{Grant-based}} & \multicolumn{3}{c}{\textbf{Grant-free}} \\
\multicolumn{2}{c|}{Reliability [\si{\percent}]} & \textbf{50} & \textbf{90} & \textbf{99.99} & \textbf{50} & \textbf{90} & \textbf{99.99} \\ \hline
&\SI{0.5}{\milli\s} &  {0} &  {0} &  {0} &  {0} &  {0} &  {0}  \\
FR1 &\SI{1}{\milli\s} &  {0}  & {0}  & {0} & {15.3M} & {3.7M} & {3.7M}  \\
&\SI{2}{\milli\s} &  60.3k & {0} & {0} & {40.7M} &  {18M} & {13.3M}  \\ \hline \hline
& \SI{0.5}{\milli\s} &  {0} &  {0} &  {0} &  {356.6M} &  {9.8M}  & {9.8M}  \\
FR2 & \SI{1}{\milli\s} &  {19.7M} &  {0} &  {0} &  {1.6B} &  {469.2M} &  {396M}  \\
& \SI{2}{\milli\s} &  {1.1B} &  {496.0M} &  {409.0M} &  {3.7B} &  {2.0B} & {1.4B}  \\ \hline
\end{tabular}
}
\par\vspace{0.5em}\noindent
\begin{minipage}{\columnwidth}
\footnotesize
\emph{Note:} The post-pruning search space contains \num{351.3}M configurations for FR1 and \num{19.1}B configurations for FR2. A radio latency of \SI{50}{\micro\s} is assumed.
\end{minipage}
\end{threeparttable}

\end{table}

\begin{table}[t]
\caption{Number of configurations achieving a given latency reliability target.}
\label{table:percentage_reliability2}
\centering
\begin{threeparttable}
\begin{adjustbox}{max width=\columnwidth}
\begin{tabular}{cc|ccc|ccc} \hline
&& \multicolumn{3}{c|}{\textbf{Grant-based}} & \multicolumn{3}{c}{\textbf{Grant-free}} \\
\multicolumn{2}{c|}{Reliability [\si{\percent}]} & \textbf{50} & \textbf{90} & \textbf{99.99} & \textbf{50} & \textbf{90} & \textbf{99.99} \\ \hline
&\SI{0.5}{\milli\s} &  {0} &  {0} &  {0} &  {0} &  {0} &  {0}  \\
FR1 &\SI{1}{\milli\s} &  {0}  & {0}  & {0} & {13.7M} & {3.2M} & {3.2M}  \\
&\SI{2}{\milli\s} &  9.3k & {0} & {0} & {37.2M} &  {16.4M} & {12.2M}  \\ \hline \hline
& \SI{0.5}{\milli\s} &  {0} &  {0} &  {0} &  {165.0M} &  {8.9M}  & {8.9M}  \\
FR2 & \SI{1}{\milli\s} &  {0} &  {0} &  {0} &  {829.4M} &  {255.1M} &  {203.7M}  \\
& \SI{2}{\milli\s} &  {183.9M} &  {60.3M} &  {43.8M} &  {2.0B} &  {983.4M} &  {692.1M}  \\ \hline
\end{tabular}
\end{adjustbox}
\par\vspace{0.5em}\noindent
\begin{minipage}{\columnwidth}
\footnotesize
\emph{Note:} The post-pruning search space contains \num{326.6}M configurations for FR1 and \num{11.8}B configurations for FR2. A radio latency of \SI{450}{\micro\s} is assumed.
\end{minipage}
\end{threeparttable}
\end{table}

Finally, an important insight that \thesystem gives us is that simply reducing the numerology (slot duration) is not enough to achieve URLLC, an assumption made by several prior works~\cite{wirth20165g, 7247338,8636206, 8683972}. \cref{fig:numerologyVSlatency} shows how the uplink latency varies versus numerologies, given an average radio preparation time of 0.5 ms and average UE and gNB processing times of 1.46 ms and 0.4 ms.
While the results indicate that latency decreases as the numerology increases, other sources of latency quickly become the bottleneck, and increasing the numerology does not help. Using this setup, even with the highest numerology, the URLLC requirement (\SI{0.5}{\milli\s} for UL transmission) remains unachievable.

\section{Related Work}\label{sec:related}

Latency in 5G networks has been studied from multiple perspectives, including analytical modeling, protocol and system-level optimization, and empirical measurement. This section reviews prior work along these three dimensions and highlights the key limitations that motivate the need for a comprehensive and realistic latency model.

\subsection{Latency Analysis}

Latency modeling and analysis in 5G have been explored in recent years \cite{patriciello20185g, patriciello2019impact, rischke2022empirical, zhao2023physical, coll2023end, skocaj2023data, mostafavi2024edaf, 265003}. However, these works either use oversimplified models or simply present an analysis based on observed measurements or simulations.
For example, \cite{coll2023end} and \cite{skocaj2023data} excluded critical system components such as frame structures and hardware overheads. \cite{patriciello2019impact} focused exclusively on the 5G uplink scheduling mechanism and was limited to simulation-based evaluations.
\cite{zhao2023physical} ignores the system’s inherent bottlenecks and simply assumes that the latency decreases linearly with the time slot duration.
Unlike previous works, we propose a comprehensive mathematical model that accurately characterizes latency distributions and captures the complex interactions among critical system components. We also compare experimentally with two of these past works~\cite{patriciello2019impact, zhao2023physical} to show that they severely underestimate the latency in~\cref{sec:eval}.

\subsection{Latency Minimization}

Previous studies address the challenges of minimizing latency and its associated trade-offs through modifying protocols and slot configurations.
Nokia~\cite{Benzaoui2020DeterministicLN} emphasizes the need for slotted MAC, central scheduling, and synchronization for deterministic latency.
Additionally, \cite{7247338} discusses avoiding retransmissions to minimize latency, assuming inherent end-to-end latency within a few milliseconds.
Comprehensive reviews such as \cite{8367785,9406015,dfgh342h,8636206,8683972,8329619, EricssonPresentation} survey enabling technologies for URLLC but often overlook real-world constraints, focusing instead on idealized scenarios.
For instance, either negligible processing~\cite{8683972, EricssonPresentation} or protocol-based latencies~\cite{8683972} are assumed. Finally, research on scheduling algorithms for URLLC~\cite{jiURLLC2018, 10092856,10159393,9099267,9247169,8486430,9779514} focuses on managing URLLC packets alongside other services, assuming low-latency communication for a UE and addressing scalability. Finally,~\cite{265003} proposes sending scheduling requests in advance to reduce latency, an approach that is complementary to our work.

\subsection{Latency Measurements}

Fezeu et al.~\cite{fezeu2023depth} evaluate the latency of several commercial mmWave implementations, achieving sub-millisecond round-trip latency under optimal conditions.
However, they note that sub-millisecond latency is only achieved in \SI{4.4}{\percent} of packets, severely violating the reliability constraint. In the sub-\SI{6}{\giga\hertz} bands, Wirth et al.~\cite{wirth20165g} propose a PHY layer solution for 5G that achieves low latency.
However, since this work predates the 5G standards, it does not incorporate the standard specifications, particularly those concerning scheduling and protocol latency, which can significantly increase the overall latency. Joint work by Nokia and Sennheiser focusing on professional audio applications~\cite{NokiaSennheiser} achieves a minimum DL latency of approximately \SI{0.8}{\milli\s} for a single UE, going higher in steps of \SI{0.5}{\milli\s} in case of retransmission.
This work, however, only supports single-user point-to-point communication using a hardware-accelerated platform, of which the scalability is limited.

Additional empirical works conduct latency evaluations in campus networks~\cite{rischke20215g,rischke2022empirical,lackner2022measurement} or testbeds~\cite{EricssonRobotDemo,qualcomm-making-5g}:
Rischke et al.~\cite{rischke20215g,rischke2022empirical} report round-trip times (RTTs) between 12 and \SI{40}{\milli\s}, and evaluate one-way latencies ranging from \qtyrange[range-units=single]{2}{8}{\milli\s}.
Lackner et al.~\cite{lackner2022measurement} find RTT latencies of \qtyrange[range-units=single]{6}{12}{\milli\s}, varying significantly with different UEs.
Work from Qualcomm~\cite{qualcomm-making-5g} indicates mmWave URLLC latencies of \SI{1.9}{\milli\s} for DL and \SI{4.0}{\milli\s} for UL.
Finally, Ericsson~\cite{EricssonRobotDemo} demonstrates an industrial automation use case, achieving \SI{5}{\milli\s} latency.

\section{Discussion \& Future Work}\label{sec:discussion}
\thesystem is a framework that formally analyzes 5G RAN latency and provides a generalized mathematical model for it.
It further enables systematic exploration of the 5G configuration space, allowing operators to identify right configurations based on their specific requirements and rigorously assess the feasibility of meeting URLLC latency-reliability targets under practical constraints.
We outline several directions for future work and extensions.

\vskip 0.05in \noindent $\bullet$ \textit{Scheduling Algorithms:}
We model contention using a round-robin scheduling algorithm as a representative case.
The framework can be extended to other scheduling policies by adapting the scheduling logic as outlined in Appendix~\ref{sec:contention-model}.
Exploring additional schedulers is left for future work.

\vskip 0.05in \noindent $\bullet$ \textit{Wireless Channel:}
\thesystem focuses on latency bottlenecks and configuration trade-offs under favorable channel conditions, and does not explicitly model wireless channel quality or packet retransmissions at the MAC or RLC layers.
The impact of the wireless channel on URLLC reliability has been studied in prior work~\cite{9887913, 7880930}.
The framework can be extended to incorporate channel models and retransmission dynamics, which we leave for future work.

\vskip 0.05in \noindent $\bullet$ \textit{Power Saving Modes:}
Power saving modes (e.g., DRX) can introduce significant initial latency (up to \SI{300}{\milli\s} in our tests on \textit{MNO}) after long inactivity periods.
\thesystem focuses on active transmission behavior and does not explicitly model these modes.
The framework can be extended to incorporate them by tracking on/off durations, similarly to how SR periods are handled, and accounting for the resulting waiting time in latency calculations.
In practice, power saving modes are typically disabled for URLLC, and in our experiments on \textit{MNO}, the UE entered power saving mode only after at least \SI{10}{\second} of inactivity.

\bibliographystyle{IEEEtran}

\begingroup
\emergencystretch=6em
\Urlmuskip=0mu plus 2mu\relax
\sloppy
\bibliography{paper}

@inproceedings{7247338,
  author    = {Johansson, Niklas A. and Wang, Y.-P. Eric and Eriksson, Erik and Hessler, Martin},
  booktitle = {2015 IEEE International Conference on Communication Workshop (ICCW)},
  title     = {Radio access for ultra-reliable and low-latency 5G communications},
  year      = {2015},
  volume    = {},
  number    = {},
  pages     = {1184-1189},
  publisher = {IEEE},
  address   = {London, UK},
  doi       = {10.1109/ICCW.2015.7247338}
}

@article{dfgh342h,
  author   = {Feng, Daquan
              and Lai, Lifeng
              and Luo, Jingjing
              and Zhong, Yi
              and Zheng, Canjian
              and Ying, Kai},
  title    = {Ultra-reliable and low-latency communications: applications, opportunities and challenges},
  journal  = {Science China Information Sciences},
  year     = {2021},
  month    = {Jan},
  day      = {20},
  volume   = {64},
  number   = {2},
  pages    = {120301},
  issn     = {1869-1919},
  doi      = {10.1007/s11432-020-2852-1},
}

@article{8636206,
  author   = {Sutton, Gordon J. and Zeng, Jie and Liu, Ren Ping and Ni, Wei and Nguyen, Diep N. and Jayawickrama, Beeshanga A. and Huang, Xiaojing and Abolhasan, Mehran and Zhang, Zhang and Dutkiewicz, Eryk and Lv, Tiejun},
  journal  = {IEEE Communications Surveys \& Tutorials},
  title    = {Enabling Technologies for Ultra-Reliable and Low Latency Communications: From PHY and MAC Layer Perspectives},
  year     = {2019},
  volume   = {21},
  number   = {3},
  pages    = {2488-2524},
  doi      = {10.1109/COMST.2019.2897800}
}

@article{8683972,
  author   = {Feng, Daquan and She, Changyang and Ying, Kai and Lai, Lifeng and Hou, Zhanwei and Quek, Tony Q. S. and Li, Yonghui and Vucetic, Branka},
  journal  = {IEEE Vehicular Technology Magazine},
  title    = {Toward Ultrareliable Low-Latency Communications: Typical Scenarios, Possible Solutions, and Open Issues},
  year     = {2019},
  volume   = {14},
  number   = {2},
  pages    = {94-102},
  doi      = {10.1109/MVT.2019.2903657}
}

@article{8329619,
  author   = {Popovski, Petar and Nielsen, Jimmy J. and Stefanovic, Cedomir and Carvalho, Elisabeth de and Strom, Erik and Trillingsgaard, Kasper F. and Bana, Alexandru-Sabin and Kim, Dong Min and Kotaba, Radoslaw and Park, Jihong and Sorensen, Rene B.},
  journal  = {IEEE Network},
  title    = {Wireless Access for Ultra-Reliable Low-Latency Communication: Principles and Building Blocks},
  year     = {2018},
  volume   = {32},
  number   = {2},
  pages    = {16-23},
  doi      = {10.1109/MNET.2018.1700258}
}

@article{10092856,
  author   = {Haque, Md. Emdadul and Tariq, Faisal and Khandaker, Muhammad R. A. and Wong, Kai-Kit and Zhang, Yangyang},
  journal  = {IEEE Access},
  title    = {A Survey of Scheduling in 5G URLLC and Outlook for Emerging 6G Systems},
  year     = {2023},
  volume   = {11},
  number   = {},
  pages    = {34372-34396},
  doi      = {10.1109/ACCESS.2023.3264592}
}

@article{8367785,
  author   = {Parvez, Imtiaz and Rahmati, Ali and Guvenc, Ismail and Sarwat, Arif I. and Dai, Huaiyu},
  journal  = {IEEE Communications Surveys \& Tutorials},
  title    = {A Survey on Low Latency Towards 5G: RAN, Core Network and Caching Solutions},
  year     = {2018},
  volume   = {20},
  number   = {4},
  pages    = {3098-3130},
  doi      = {10.1109/COMST.2018.2841349}
}

@article{9406015,
  author   = {Ali, Rashid and Zikria, Yousaf Bin and Bashir, Ali Kashif and Garg, Sahil and Kim, Hyung Seok},
  journal  = {IEEE Access},
  title    = {URLLC for 5G and Beyond: Requirements, Enabling Incumbent Technologies and Network Intelligence},
  year     = {2021},
  volume   = {9},
  number   = {},
  pages    = {67064-67095},
  doi      = {10.1109/ACCESS.2021.3073806}
}

@article{10159393,
  author   = {Sohaib, Rana M. and Onireti, Oluwakayode and Sambo, Yusuf and Swash, Rafiq and Ansari, Shuja and Imran, Muhammad A.},
  journal  = {IEEE Access},
  title    = {Intelligent Resource Management for eMBB and URLLC in 5G and Beyond Wireless Networks},
  year     = {2023},
  volume   = {11},
  number   = {},
  pages    = {65205-65221},
  doi      = {10.1109/ACCESS.2023.3288698}
}

@article{9099267,
  author   = {Li, Jing and Zhang, Xing},
  journal  = {IEEE Wireless Communications Letters},
  title    = {Deep Reinforcement Learning-Based Joint Scheduling of eMBB and URLLC in 5G Networks},
  year     = {2020},
  volume   = {9},
  number   = {9},
  pages    = {1543-1546},
  doi      = {10.1109/LWC.2020.2997036}
}

@article{9247169,
  author   = {Yin, Hao and Zhang, Lyutianyang and Roy, Sumit},
  journal  = {IEEE Transactions on Communications},
  title    = {Multiplexing URLLC Traffic Within eMBB Services in 5G NR: Fair Scheduling},
  year     = {2021},
  volume   = {69},
  number   = {2},
  pages    = {1080-1093},
  doi      = {10.1109/TCOMM.2020.3035582}
}

@article{8486430,
  author   = {Anand, Arjun and de Veciana, Gustavo and Shakkottai, Sanjay},
  journal  = {IEEE/ACM Transactions on Networking},
  title    = {Joint Scheduling of URLLC and eMBB Traffic in 5G Wireless Networks},
  year     = {2020},
  volume   = {28},
  number   = {2},
  pages    = {477-490},
  doi      = {10.1109/TNET.2020.2968373}
}

@article{9779514,
  author   = {Prathyusha, Yerra and Sheu, Tsang-Ling},
  journal  = {IEEE Transactions on Vehicular Technology},
  title    = {Coordinated Resource Allocations for eMBB and URLLC in 5G Communication Networks},
  year     = {2022},
  volume   = {71},
  number   = {8},
  pages    = {8717-8728},
  doi      = {10.1109/TVT.2022.3176018}
}

@techreport{3GPP-TR-38.824-version-16,
  author      = {3GPP},
  title       = {Study on physical layer enhancements for NR ultra-reliable and low latency case (URLLC)},
  institution = {3GPP},
  year        = {2019},
  month       = {3},
  number      = {3GPP TR 38.824 version 16.0.0 Release 16},
  publisher   = {3GPP},
  url         = {https://www.3gpp.org/ftp//Specs/archive/38_series/38.824/38824-g00.zip}
}

@techreport{3GPP-TR-38.913-version-18,
  author      = {3GPP},
  title       = {5G; Study on scenarios and requirements for next generation access technologies},
  institution = {3GPP},
  year        = {2024},
  month       = {5},
  number      = {3GPP TR 38.913 version 18.0.0 Release 18},
  publisher   = {ETSI},
  url         = {https://www.etsi.org/deliver/etsi_tr/138900_138999/138913/18.00.00_60/tr_138913v180000p.pdf}
}

@techreport{3GPP-first-NR-with-URLLC,
  author      = {3GPP},
  title       = {NR; NR and NG-RAN Overall description; Stage-2},
  institution = {3GPP},
  year        = {2017},
  month       = {10},
  day         = {27},
  number      = {3GPP TS 38.300 version 1.1.1 Release 15},
  publisher   = {3GPP},
  url         = {https://www.3gpp.org/ftp/Specs/archive/38_series/38.300/38300-111.zip}
}

@techreport{3GPP-TS-138-211,
  author      = {3GPP},
  title       = {5G; NR; Physical channels and modulation},
  institution = {3GPP},
  year        = {2024},
  month       = {05},
  number      = {3GPP TS 38.211 version 18.2.0 Release 18},
  publisher   = {ETSI},
  url         = {https://www.etsi.org/deliver/etsi_ts/138200_138299/138211/18.02.00_60/ts_138211v180200p.pdf}
}

@techreport{3GPP-scs-fr1,
  author      = {3GPP},
  title       = {5G; NR; User Equipment (UE) radio transmission and reception; Part 1: Range 1 Standalone},
  institution = {3GPP},
  year        = {2022},
  month       = {5},
  number      = {3GPP TS 38.101-1 version 17.5.0 Release 17},
  publisher   = {ETSI},
  url         = {https://www.etsi.org/deliver/etsi_ts/138100_138199/13810101/17.05.00_60/ts_13810101v170500p.pdf}
}

@inproceedings{uitto2021evaluation,
  title        = {Evaluation of live video streaming performance for low latency use cases in 5g},
  author       = {Uitto, Mikko and Heikkinen, Antti},
  booktitle    = {2021 Joint European Conference on Networks and Communications \& 6G Summit (EuCNC/6G Summit)},
  publisher    = {IEEE},
  address      = {Porto, Portugal},
  pages        = {431--436},
  year         = {2021},
  organization = {IEEE}
}

@inproceedings{fezeu2023depth,
  author    = {Fezeu, Rostand A. K. and Ramadan, Eman and Ye, Wei and Minneci, Benjamin and Xie, Jack and Narayanan, Arvind and Hassan, Ahmad and Qian, Feng and Zhang, Zhi-Li and Chandrashekar, Jaideep and Lee, Myungjin},
  title     = {An In-Depth Measurement Analysis of 5G mmWave PHY Latency and Its Impact on End-to-End Delay},
  year      = {2023},
  publisher = {Springer-Verlag},
  address   = {Berlin, Heidelberg},
  booktitle = {Passive and Active Measurement: 24th International Conference, PAM 2023, Virtual Event, March 21–23, 2023, Proceedings},
  pages     = {284–312},
  numpages  = {29}
}

@article{jiURLLC2018,
  author   = {Ji, Hyoungju and Park, Sunho and Yeo, Jeongho and Kim, Younsun and Lee, Juho and Shim, Byonghyo},
  journal  = {IEEE Wireless Communications},
  title    = {Ultra-Reliable and Low-Latency Communications in 5G Downlink: Physical Layer Aspects},
  year     = {2018},
  volume   = {25},
  number   = {3},
  pages    = {124-130},
  doi      = {10.1109/MWC.2018.1700294}
}

@article{lackner2022measurement,
  title     = {Measurement and comparison of data rate and time delay of end-devices in licensed sub-6 GHz 5G standalone non-public networks},
  author    = {Lackner, Thorge and Hermann, Julian and Dietrich, Fabian and Kuhn, Christian and Angos, Mario and Jooste, Johannes L and Palm, Daniel},
  journal   = {Procedia CIRP},
  volume    = {107},
  pages     = {1132--1137},
  year      = {2022},
  publisher = {Elsevier}
}

@inproceedings{wirth20165g,
  author    = {Wirth, Thomas and Mehlhose, Matthias and Pilz, Jens and Holfeld, Bernd and Wieruch, Dennis},
  booktitle = {2016 50th Asilomar Conference on Signals, Systems and Computers},
  title     = {5G new radio and ultra low latency applications: A PHY implementation perspective},
  year      = {2016},
  volume    = {},
  number    = {},
  publisher = {IEEE},
  address   = {Pacific Grove, CA, USA},
  pages     = {1409-1413},
  doi       = {10.1109/ACSSC.2016.7869608}
}

@inproceedings{virtualchannel,
  title     = {Boosting Application Performance using Heterogeneous Virtual Channels: Challenges and Opportunities},
  author    = {Talal Touseef and William Sentosa and Milind Kumar Vaddiraju and Debopam Bhattacherjee and Bala Chandrasekaran and P. Brighten Godfrey and Shubham Tiwari},
  booktitle = {22nd ACM Workshop on Hot Topics in Networks (HotNets)},
  year      = {2023},
  pages     = {139 - 146},
  doi       = {10.1145/3626111.3628193},
  publisher = {ACM},
  address   = {Cambridge MA USA}
}

@inproceedings{dchannel,
  author    = {William Sentosa and Balakrishnan Chandrasekaran and P. Brighten Godfrey and Haitham Hassanieh and Bruce Maggs},
  title     = {{DChannel}: Accelerating Mobile Applications With Parallel High-bandwidth and Low-latency Channels},
  booktitle = {20th USENIX Symposium on Networked Systems Design and Implementation (NSDI 23)},
  year      = {2023},
  isbn      = {978-1-939133-33-5},
  address   = {Boston, MA},
  pages     = {419--436},
  url       = {https://www.usenix.org/conference/nsdi23/presentation/sentosa},
  publisher = {USENIX Association},
  month     = apr
}

@article{rischke20215g,
  title     = {5G campus networks: A first measurement study},
  author    = {Rischke, Justus and Sossalla, Peter and Itting, Sebastian and Fitzek, Frank HP and Reisslein, Martin},
  journal   = {IEEE Access},
  volume    = {9},
  pages     = {121786--121803},
  year      = {2021},
  publisher = {IEEE}
}

@article{brown2018ultra,
  title   = {Ultra-reliable low-latency 5G for industrial automation},
  author  = {Brown, Gabriel and Analyst, P and Reading, H},
  journal = {Technol. Rep. Qualcomm},
  url     = {https://www.qualcomm.com/content/dam/qcomm-martech/dm-assets/documents/ultra-reliable-low-latency-5g-for-industrial-automation.pdf},
  volume  = {2},
  pages   = {52065394},
  year    = {2018}
}

@inproceedings{Benzaoui2020DeterministicLN,
  author    = {Benzaoui, Nihel},
  booktitle = {2020 European Conference on Optical Communications (ECOC)},
  title     = {Deterministic Latency Networks for 5G Applications},
  year      = {2020},
  volume    = {},
  number    = {},
  pages     = {1-4},
  doi       = {10.1109/ECOC48923.2020.9333411},
  address   = {Brussels, Belgium},
  publisher = {Institute of Electrical and Electronics Engineers (IEEE)}
}

@online{NokiaSennheiser,
  author       = {Nokia and Sennheiser},
  title        = {Low Latency 5G for Professional Audio Transmission},
  year         = {2020},
  organization = {Nokia and Sennheiser},
  url          = {https://d1p0gxnqcu0lvz.cloudfront.net/documents/Nokia_Low_Latency_5G_for_Professional_Audio_Transmission_White_Paper_EN.pdf},
  note         = {Accessed on May 14th 2026}
}

@online{EricssonPresentation,
  author       = {Ericsson},
  title        = {5G Techniques for Ultra Reliable Low Latency Communication},
  year         = {2017},
  url          = {https://wp-files.comsoc.org/cscn-2017/files/2017/08/Janne_Peisa_Ericsson_CSCN2017.pdf},
  organization = {Ericsson}
}

@online{EricssonRobotDemo,
  author       = {Ericsson},
  title        = {You Need to See Our Dancing Hexapod Demo from MWC},
  year         = {2019},
  url          = {https://www.ericsson.com/en/blog/2019/3/dancing-hexapod-demo-a-success-at-mwc},
  organization = {Ericsson}
}

@article{autonomous-vehicles-5g,
  title    = {Autonomous vehicles in 5G and beyond: A survey},
  journal  = {Vehicular Communications},
  volume   = {39},
  pages    = {100551},
  year     = {2023},
  issn     = {2214-2096},
  doi      = {10.1016/j.vehcom.2022.100551},
  author   = {Saqib Hakak and Thippa Reddy Gadekallu and Praveen Kumar Reddy Maddikunta and Swarna Priya Ramu and Parimala M and Chamitha {De Alwis} and Madhusanka Liyanage}
}

@article{smart-grid-5g,
  title    = {5G network-based Internet of Things for demand response in smart grid: A survey on application potential},
  journal  = {Applied Energy},
  volume   = {257},
  pages    = {113972},
  year     = {2020},
  issn     = {0306-2619},
  doi      = {10.1016/j.apenergy.2019.113972},
  author   = {Hongxun Hui and Yi Ding and Qingxin Shi and Fangxing Li and Yonghua Song and Jinyue Yan}
}

@article{public-safety-5g,
  author   = {Volk, Mojca and Sterle, Janez},
  journal  = {IEEE Access},
  title    = {5G Experimentation for Public Safety: Technologies, Facilities and Use Cases},
  year     = {2021},
  volume   = {9},
  number   = {},
  pages    = {41184-41217},
  doi      = {10.1109/ACCESS.2021.3064405}
}

@article{public-safety-5g-applications,
  author   = {Li, Jingya and Nagalapur, Keerthi Kumar and Stare, Erik and Dwivedi, Satyam and Ashraf, Shehzad Ali and Eriksson, Per-Erik and Engström, Ulrika and Lee, Woong-Hee and Lohmar, Thorsten},
  journal  = {IEEE Communications Standards Magazine},
  title    = {5G New Radio for Public Safety Mission Critical Communications},
  year     = {2022},
  volume   = {6},
  number   = {4},
  pages    = {48-55},
  doi      = {10.1109/MCOMSTD.0002.2100036}
}

@inproceedings{5g-first-responders,
  author    = {Shunqing Zhang and
               Xiuqiang Xu and
               Yiqun Wu and
               Lei Lu},
  title     = {5G: Towards energy-efficient, low-latency and high-reliable communications
               networks},
  booktitle = {{IEEE} International Conference on Communication Systems, {ICCS} 2014,
               Macau, China, November 19-21, 2014},
  pages     = {197--201},
  publisher = {{IEEE}},
  address   = {Macau, China},
  year      = {2014},
  doi       = {10.1109/ICCS.2014.7024793},
  bibsource = {dblp computer science bibliography, https://dblp.org}
}

@misc{commercial-5g-launches,
  title  = {{Commercial 5G launches – 5G Observatory}},
  author = {Europian 5G Observatory},
  year   = {2018},
  note   = {Accessed on May 14th 2026},
  url    = {https://digital-strategy.ec.europa.eu/en/policies/5g-observatory}
}

@techreport{qualcomm-making-5g,
  author      = {Qualcomm},
  title       = {Making 5G NR a Reality},
  year        = {2020},
  institution = {Qualcomm},
  url         = {https://www.qualcomm.com/content/dam/qcomm-martech/dm-assets/documents/powerpoint_presentation_-_making_5g_nr_a_reality_february_2020_web.pdf}
}

@misc{srsRAN,
  title        = {{srsRAN Project: Open source RAN}},
  author       = {{Software Radio Systems}},
  howpublished = {\url{https://www.srsran.com/}},
  year         = {2022},
  note         = {[Online; Last accessed: 17-Jan-2025]}
}

@misc{open5gs,
  title        = {{Open5GS Project: Open source 5G core network}},
  author       = {{Open5GS}},
  howpublished = {\url{https://open5gs.org/}},
  year         = {2017},
  note         = {[Online; Last accessed: 17-Jan-2025]}
}

@article{OAI,
  author     = {Nikaein, Navid and Marina, Mahesh K. and Manickam, Saravana and Dawson, Alex and Knopp, Raymond and Bonnet, Christian},
  title      = {OpenAirInterface: A Flexible Platform for 5G Research},
  year       = {2014},
  issue_date = {October 2014},
  publisher  = {Association for Computing Machinery},
  address    = {New York, NY, USA},
  volume     = {44},
  number     = {5},
  issn       = {0146-4833},
  doi        = {10.1145/2677046.2677053},
  journal    = {SIGCOMM Comput. Commun. Rev.},
  month      = {oct},
  pages      = {33–38}
}

@misc{ping,
  title        = {{Ping}},
  author       = {{Mike Muuss}},
  howpublished = {\url{https://github.com/iputils/iputils/tree/master/ping}},
  year         = {1983},
  note         = {[Online; Last accessed: 19-Jan-2025]}
}

@misc{fping,
  title        = {{fping}},
  author       = {{Roland Schemers}},
  howpublished = {\url{https://fping.org/}},
  year         = {1992},
  note         = {[Online; Last accessed: 19-Jan-2025]}
}

@misc{tcpreplay,
  title        = {{tcpreplay}},
  author       = {{Aaron Turner}},
  howpublished = {\url{https://tcpreplay.appneta.com/}},
  year         = {1999},
  note         = {[Online; Last accessed: 19-Jan-2025]}
}

@misc{0ae4d1d5415845e980e046fe596c9907,
  title     = {Uplink Grant-free Access for Ultra-Reliable Low-Latency Communications in 5G: Radio Access and Resource Management Solutions},
  author    = {Abreu, {Renato Barbosa}},
  note      = {PhD supervisor: Prof. Preben Mogensen, Aalborg University Assistant PhD supervisors: Assoc. Prof. Gilberto Berardinelli, Aalborg University Prof. Klaus Pedersen, Aalborg University},
  year      = {2019},
  language  = {English},
  series    = {Ph.d.-serien for Det Tekniske Fakultet for IT og Design, Aalborg Universitet},
  publisher = {Aalborg Universitetsforlag}
}

@article{9792172,
  author   = {Moglia, Andrea and Georgiou, Konstantinos and Marinov, Blagoi and Georgiou, Evangelos and Berchiolli, Raffaella Nice and Satava, Richard M. and Cuschieri, Alfred},
  journal  = {IEEE Journal of Biomedical and Health Informatics},
  title    = {5G in Healthcare: From COVID-19 to Future Challenges},
  year     = {2022},
  volume   = {26},
  number   = {8},
  pages    = {4187-4196},
  doi      = {10.1109/JBHI.2022.3181205}
}

@article{s23073682,
  author         = {Hazarika, Ananya and Rahmati, Mehdi},
  title          = {Towards an Evolved Immersive Experience: Exploring 5G- and Beyond-Enabled Ultra-Low-Latency Communications for Augmented and Virtual Reality},
  journal        = {Sensors},
  volume         = {23},
  year           = {2023},
  number         = {7},
  article-number = {3682},
  numpages       = {25},
  url            = {https://www.mdpi.com/1424-8220/23/7/3682},
  pubmedid       = {37050742},
  issn           = {1424-8220},
  doi            = {10.3390/s23073682}
}

@article{GUPTA2021103521,
  title    = {6G-enabled Edge Intelligence for Ultra -Reliable Low Latency Applications: Vision and Mission},
  journal  = {Computer Standards \& Interfaces},
  volume   = {77},
  pages    = {103521},
  year     = {2021},
  issn     = {0920-5489},
  doi      = {10.1016/j.csi.2021.103521},
  author   = {Rajesh Gupta and Dakshita Reebadiya and Sudeep Tanwar}
}

@inbook{She2024,
  author    = {She, Changyang
               and Li, Yonghui},
  title     = {Ultra-Reliable and Low-Latency Communications in 6G: Challenges, Solutions, and Future Directions},
  booktitle = {Fundamentals of 6G Communications and Networking},
  year      = {2024},
  publisher = {Springer International Publishing},
  address   = {Cham},
  pages     = {27},
  isbn      = {978-3-031-37920-8},
  doi       = {10.1007/978-3-031-37920-8_24}
}

@article{10.1109/COMST.2023.3243918,
  author     = {Chafii, Marwa and Bariah, Lina and Muhaidat, Sami and Debbah, Merouane},
  title      = {Twelve Scientific Challenges for 6G: Rethinking the Foundations of Communications Theory},
  year       = {2023},
  issue_date = {Secondquarter 2023},
  publisher  = {IEEE Press},
  volume     = {25},
  number     = {2},
  issn       = {1553-877X},
  doi        = {10.1109/COMST.2023.3243918},
  journal    = {Commun. Surveys Tuts.},
  month      = {apr},
  pages      = {868–904},
  numpages   = {37}
}

@article{9887913,
  author   = {Varatharaajan, Sutharshun and Grossmann, Marcus and Del Galdo, Giovanni},
  journal  = {IEEE Access},
  title    = {5G New Radio Physical Downlink Control Channel Reliability Enhancements for Multiple Transmission-Reception-Point Communications},
  year     = {2022},
  volume   = {10},
  number   = {},
  pages    = {97394-97407},
  doi      = {10.1109/ACCESS.2022.3206027}
}

@inproceedings{7880930,
  author    = {Sybis, Michal and Wesolowski, Krzysztof and Jayasinghe, Keeth and Venkatasubramanian, Venkatkumar and Vukadinovic, Vladimir},
  booktitle = {2016 IEEE 84th Vehicular Technology Conference (VTC-Fall)},
  title     = {Channel Coding for Ultra-Reliable Low-Latency Communication in 5G Systems},
  year      = {2016},
  volume    = {},
  number    = {},
  pages     = {1-5},
  publisher = {IEEE},
  address   = {Montreal, QC, Canada},
  doi       = {10.1109/VTCFall.2016.7880930}
}

@inbook{Dudley_2009,
  title      = {Euler's theorem and function},
  booktitle  = {A Guide to elementary number theory},
  publisher  = {Mathematical Association of America},
  author     = {Dudley, Underwood},
  year       = {2009},
  pages      = {37-40},
  doi        = {10.5948/UPO9780883859186},
  isbn       = {978-0-88385-347-4},
  series     = {Dolciani Mathematical Expositions},
  collection = {Dolciani Mathematical Expositions},
  address    = {Washington, DC}
}

@techreport{3GPP-TS-38.101-1-version-17.8.0,
  author      = {3GPP},
  title       = {User Equipment (UE) radio transmission and reception; Part 1: Range 1 Standalone},
  institution = {3GPP},
  year        = {2023},
  month       = {1},
  number      = {3GPP TS 38.101-1 version 17.8.0 Release 17},
  publisher   = {3GPP},
  url         = {https://www.etsi.org/deliver/etsi_ts/138100_138199/13810101/17.08.00_60/ts_13810101v170800p.pdf}
}

@techreport{3GPP-TS-38.214-version-16.2.0,
  author      = {3GPP},
  title       = {Physical layer procedures for data},
  institution = {3GPP},
  year        = {2020},
  month       = {7},
  number      = {3GPP TS 38.214 16.2.0 Release 16},
  publisher   = {3GPP},
  url         = {https://www.etsi.org/deliver/etsi_ts/138200_138299/138214/16.02.00_60/ts_138214v160200p.pdf}
}

@inproceedings{patriciello20185g,
  author    = {Patriciello, Natale and Lagen, Sandra and Giupponi, Lorenza and Bojovic, Biljana},
  booktitle = {2018 IEEE 23rd International Workshop on Computer Aided Modeling and Design of Communication Links and Networks (CAMAD)},
  title     = {5G New Radio Numerologies and their Impact on the End-To-End Latency},
  year      = {2018},
  volume    = {},
  number    = {},
  pages     = {1-6},
  doi       = {10.1109/CAMAD.2018.8514979},
  address   = {Barcelona, Spain},
  publisher = {Institute of Electrical and Electronics Engineers (IEEE)}
}

@inproceedings{skocaj2023data,
  author    = {Skocaj, Marco and Conserva, Francesca and Grande, Nicol Sarcone and Orsi, Andrea and Micheli, Davide and Ghinamo, Giorgio and Bizzarri, Simone and Verdone, Roberto},
  booktitle = {2023 IEEE 34th Annual International Symposium on Personal, Indoor and Mobile Radio Communications (PIMRC)},
  title     = {Data-driven Predictive Latency for 5G: A Theoretical and Experimental Analysis Using Network Measurements},
  year      = {2023},
  volume    = {},
  number    = {},
  pages     = {1-6},
  doi       = {10.1109/PIMRC56721.2023.10293861},
  address   = {London, UK},
  publisher = {Institute of Electrical and Electronics Engineers (IEEE)}
}

@inproceedings{mostafavi2024edaf,
  author    = {Mostafavi, Samie and Tillner, Marius and Sharma, Gourav Prateek and Gross, James},
  booktitle = {IEEE INFOCOM 2024 - IEEE Conference on Computer Communications Workshops (INFOCOM WKSHPS)},
  title     = {EDAF: An End-to-End Delay Analytics Framework for 5G-and-Beyond Networks},
  year      = {2024},
  volume    = {},
  number    = {},
  pages     = {1-6},
  doi       = {10.1109/INFOCOMWKSHPS61880.2024.10620853},
  address   = {Vancouver, BC, Canada},
  publisher = {Institute of Electrical and Electronics Engineers (IEEE)}
}

@inproceedings{rischke2022empirical,
  author    = {Rischke, Justus and Vielhaus, Christian and Sossalla, Peter and Itting, Sebastian and Nguyen, Giang T. and Fitzek, Frank H. P.},
  booktitle = {2022 IEEE 23rd International Symposium on a World of Wireless, Mobile and Multimedia Networks (WoWMoM)},
  title     = {Empirical Study of 5G Downlink \& Uplink Scheduling and its Effects on Latency},
  year      = {2022},
  volume    = {},
  number    = {},
  pages     = {11-19},
  doi       = {10.1109/WoWMoM54355.2022.00017},
  address   = {Belfast, United Kingdom},
  publisher = {Institute of Electrical and Electronics Engineers (IEEE)}
}

@article{coll2023end,
  author    = {Coll-Perales, Baldomero and Lucas-Estañ, M. Carmen and Shimizu, Takayuki and Gozalvez, Javier and Higuchi, Takamasa and Avedisov, Sergei and Altintas, Onur and Sepulcre, Miguel},
  journal   = {IEEE Transactions on Vehicular Technology},
  title     = {End-to-End V2X Latency Modeling and Analysis in 5G Networks},
  year      = {2023},
  volume    = {72},
  number    = {4},
  pages     = {5094-5109},
  doi       = {10.1109/TVT.2022.3224614},
  publisher = {Institute of Electrical and Electronics Engineers (IEEE)}
}

@inproceedings{patriciello2019impact,
  author    = {Patriciello, Natale and Lagen, Sandra and Giupponi, Lorenza and Bojovic, Biljana},
  booktitle = {2019 IEEE Global Communications Conference (GLOBECOM)},
  title     = {The Impact of NR Scheduling Timings on End-to-End Delay for Uplink Traffic},
  year      = {2019},
  volume    = {},
  number    = {},
  pages     = {1-6},
  doi       = {10.1109/GLOBECOM38437.2019.9013231},
  address   = {Waikoloa, HI, USA},
  publisher = {Institute of Electrical and Electronics Engineers (IEEE)}
}

@inproceedings{zhao2023physical,
  author    = {Zhao, Yong and Xie, Weiliang},
  booktitle = {2023 International Wireless Communications and Mobile Computing (IWCMC)},
  title     = {Physical Layer Round Trip Latency Analysis and Estimation for 5G NR},
  year      = {2023},
  volume    = {},
  number    = {},
  pages     = {971-976},
  doi       = {10.1109/IWCMC58020.2023.10183295},
  address   = {Marrakesh, Morocco},
  publisher = {Institute of Electrical and Electronics Engineers (IEEE)}
}

@article{vaserstein1969markov,
  title     = {Markov processes over denumerable products of spaces, describing large systems of automata},
  author    = {Vaserstein, Leonid Nisonovich},
  journal   = {Problemy Peredachi Informatsii},
  volume    = {5},
  number    = {3},
  pages     = {64--72},
  year      = {1969},
  publisher = {Russian Academy of Sciences, Branch of Informatics, Computer Equipment and~…}
}

@article{kantorovich1960mathematical,
  title     = {Mathematical methods of organizing and planning production},
  author    = {Kantorovich, Leonid V},
  journal   = {Management science},
  volume    = {6},
  number    = {4},
  pages     = {366--422},
  year      = {1960},
  publisher = {INFORMS}
}

@article{mallows1972note,
  title     = {A note on asymptotic joint normality},
  author    = {Mallows, Colin L},
  journal   = {The Annals of Mathematical Statistics},
  volume    = {43},
  number    = {2},
  pages     = {508--515},
  year      = {1972},
  publisher = {JSTOR}
}

@misc{qualcomm-sdx-65,
  author = {Qualcomm},
  title  = {Qualcomm Snapdragon X65 5G Modem-RF System},
  year   = {2021},
  url    = {https://www.qualcomm.com/modems/products/snapdragon-x65-5g-modem-rf-system},
  note   = {[Online; Last accessed: May 14th 2026]}
}

@inproceedings{chroma,
  author    = {Ge, Changhan and Ge, Zihui and Liu, Xuan and Mahimkar, Ajay and Shaqalle, Yusef and Xiang, Yu and Pathak, Shomik},
  title     = {Chroma: Learning and Using Network Contexts to Reinforce Performance Improving Configurations},
  year      = {2023},
  isbn      = {9781450399906},
  publisher = {Association for Computing Machinery},
  address   = {New York, NY, USA},
  doi       = {10.1145/3570361.3613256},
  booktitle = {Proceedings of the 29th Annual International Conference on Mobile Computing and Networking},
  articleno = {42},
  numpages  = {16},
  location  = {Madrid, Spain},
  series    = {ACM MobiCom '23}
}

@inproceedings{auric,
  author    = {Mahimkar, Ajay and Sivakumar, Ashiwan and Ge, Zihui and Pathak, Shomik and Biswas, Karunasish},
  title     = {Auric: using data-driven recommendation to automatically generate cellular configuration},
  year      = {2021},
  isbn      = {9781450383837},
  publisher = {Association for Computing Machinery},
  address   = {New York, NY, USA},
  doi       = {10.1145/3452296.3472906},
  booktitle = {Proceedings of the 2021 ACM SIGCOMM 2021 Conference},
  pages     = {807–820},
  numpages  = {14},
  location  = {Virtual Event, USA},
  series    = {SIGCOMM '21}
}

@article{panaretos2019statistical,
  title     = {Statistical aspects of Wasserstein distances},
  author    = {Panaretos, Victor M and Zemel, Yoav},
  journal   = {Annual review of statistics and its application},
  volume    = {6},
  number    = {1},
  pages     = {405--431},
  year      = {2019},
  publisher = {Annual Reviews}
}

@misc{matlab,
  author       = {MathWorks},
  title        = {5G Toolbox version: R2025a},
  year         = {2025},
  publisher    = {MATLAB},
  organization = {MATLAB},
  address      = {Natick, MA, USA},
  url          = {https://www.mathworks.com/products/5g.html}
}

@inproceedings{ns3,
  title     = {Network simulations with the ns-3 simulator},
  author    = {Henderson, Thomas R and Lacage, Mathieu and Riley, George F},
  publisher = {Association for Computing Machinery},
  address   = {New York, NY, USA},
  url       = {https://conferences.sigcomm.org/sigcomm/2008/papers/p527-hendersonA.pdf},
  booktitle = {Proceedings of the ACM SIGCOMM 2008 Conference on Data Communication},
  pages     = {527},
  numpages  = {1},
  location  = {Seattle, WA, USA},
  series    = {SIGCOMM '08},
  year      = {2008},
  isbn      = {9781605581750}
}

@article{5Glena,
  title    = {An E2E simulator for 5G NR networks},
  journal  = {Simulation Modelling Practice and Theory},
  volume   = {96},
  pages    = {101933},
  year     = {2019},
  issn     = {1569-190X},
  doi      = {10.1016/j.simpat.2019.101933},
  author   = {Natale Patriciello and Sandra Lagen and Biljana Bojovic and Lorenza Giupponi},
}

@inproceedings{10.1145/3750718.3750743,
author = {Gong, Aoyu and Maghsoudnia, Arman and Cannat\`{a}, Raphael and Vlad, Eduard and Lomba, N\'{e}stor Lomba and Dumitriu, Dan Mihai and Hassanieh, Haitham},
title = {Towards URLLC with Open-Source 5G Software},
year = {2025},
isbn = {9798400721083},
publisher = {Association for Computing Machinery},
address = {New York, NY, USA},
doi = {10.1145/3750718.3750743},
booktitle = {Proceedings of the 1st Workshop on Open Research Infrastructures and Toolkits for 6G},
pages = {7–14},
numpages = {8},
location = {Coimbra, Portugal},
series = {OpenRIT6G '25}
}

@inproceedings{10.1145/3696348.3696862,
author = {Maghsoudnia, Arman and Vlad, Eduard and Gong, Aoyu and Dumitriu, Dan Mihai and Hassanieh, Haitham},
title = {Ultra-Reliable Low-Latency in 5G: A Close Reality or a Distant Goal?},
year = {2024},
isbn = {9798400712722},
publisher = {Association for Computing Machinery},
address = {New York, NY, USA},
doi = {10.1145/3696348.3696862},
booktitle = {Proceedings of the 23rd ACM Workshop on Hot Topics in Networks},
pages = {111–120},
numpages = {10},
location = {Irvine, CA, USA},
series = {HotNets '24}
}

@misc{ray,
  author       = {Anyscale},
  title        = {Ray},
  year         = {2020},
  publisher    = {Ray Project},
  organization = {Ray Project},
  address      = {San Francisco, CA, USA},
  url          = {https://docs.ray.io/}
}

@article{khan2022urllc,
  title     = {{URLLC and eMBB in 5G industrial IoT: A survey}},
  author    = {Khan, Benish Sharfeen and Jangsher, Sobia and Ahmed, Ashfaq and Al-Dweik, Arafat},
  journal   = {IEEE Open Journal of the Communications Society},
  volume    = {3},
  pages     = {1134--1163},
  year      = {2022},
  publisher = {IEEE}
}

@misc{qcsuper,
  author = {P1 Security},
  title = {QCSuper release 2.0.1},
  year = {2024},
  publisher = {GitHub},
  journal = {GitHub repository},
  howpublished = {\url{https://github.com/P1sec/QCSuper}},
  commit = {f5f1501c7ce09f6c167ae623233f674be09cdf87}
}

@ARTICLE{8125190,
  author={Zhang, Mingyang and Fu, Haohao and Li, Yong and Chen, Sheng},
  journal={IEEE Transactions on Big Data},
  title={Understanding Urban Dynamics From Massive Mobile Traffic Data},
  year={2019},
  volume={5},
  number={2},
  pages={266-278},
  doi={10.1109/TBDATA.2017.2778721}}

@inproceedings {265003,
author = {Zhaowei Tan and Jinghao Zhao and Yuanjie Li and Yifei Xu and Songwu Lu},
title = {{Device-Based} {LTE} Latency Reduction at the Application Layer},
booktitle = {18th USENIX Symposium on Networked Systems Design and Implementation (NSDI 21)},
year = {2021},
isbn = {978-1-939133-21-2},
pages = {471--486},
url = {https://www.usenix.org/conference/nsdi21/presentation/tan},
publisher = {USENIX Association},
month = apr
}

@INPROCEEDINGS{11360713,
  author={Nawaz, Faiza and Mughal, Hamna},
  booktitle={2025 5th International Conference on Digital Futures and Transformative Technologies (ICoDT2)},
  title={Enabling Low-Latency Real-Time Gaming through Edge Computing and Wireless Optimization: Trends, Challenges, and Future Directions},
  year={2025},
  volume={},
  number={},
  pages={1-6},
  doi={10.1109/ICoDT269104.2025.11360713}}
\endgroup

\appendices

\section{Proofs}
\label{sec:Proofs}

\subsection{Proof of lemma \ref{lemma1}}

\begin{proof}

We solve this problem using Euler’s theorem~\cite{Dudley_2009} in number theory.

Based on \cref{eq:offsetPeriodicity}, for $ x \in A, \, y \in B $, we can write:

{\small
\begin{equation}
    SR_O + x \cdot SR_P \equiv_T y
    \implies
    x \cdot \frac{SR_P}{n} \equiv_{\frac{T}{n}} \frac{y - SR_O}{n}
    \label{eq:xAndy}
\end{equation}
}

Based on Euler's theorem we can deduce (\(\varphi\) is Euler's totient function):

{\small
\begin{equation}
    n \mid y - SR_O
    \label{eq:nDividesyMinSR0}
\end{equation}
\begin{equation}
    x \equiv_{\frac{T}{n}} \frac{y - SR_O}{n} \cdot \left({\frac{SR_P}{n}}\right)^{\varphi(\frac{T}{n})-1}
    \label{eq:xBasedOnEuler}
\end{equation}
}

From \cref{eq:bSubsetDL}, we know that if $ y \in B $, then:

{\small
\begin{equation}
    y \subseteq \{d+1, \dots, T-1\}
    \label{eq:yInDL}
\end{equation}
}

We define set $ C $ as the largest set of integers satisfying \cref{eq:nDividesyMinSR0} and \cref{eq:yInDL}.
For any $ z \in C $, we can write:

{\small
\begin{equation}
    n \mid (z - SR_O) \iff  \exists a \in \mathbb{N}, \quad y = n \cdot a + SR_O
\end{equation}
\begin{equation}
    \implies d+1 \leq n \cdot a + SR_O \leq T-1
\end{equation}
\begin{equation}
    \iff \frac{d+1 - SR_O}{n} \leq a \leq \frac{T-1 - SR_O}{n}.
\end{equation}
}

Consequently, let
$L=\left\lceil \frac{d+1 - SR_O}{n} \right\rceil$ and
$U=\left\lfloor \frac{T-1 - SR_O}{n} \right\rfloor$.
If we define set $D$ as follows:

\begin{equation*}
D =
\begin{cases}
\{L, L+1, \ldots, U\}, & \text{if } L < U,\\
\emptyset, & \text{otherwise.}
\end{cases}
\end{equation*}

Then, $C$ can be written as:

\begin{equation}
    C = \{ n \cdot j + SR_O \mid j \in D \}.
\end{equation}

Next, we aim to prove that  $B = C$.
We must show $ B \subseteq C $ and $ C \subseteq B $.
We already know $ B \subseteq C $ because any $ y \in B $ satisfies \cref{eq:nDividesyMinSR0} and \cref{eq:yInDL}, and $ C $ is the largest set of integers meeting these conditions.
To show $ C \subseteq B $, we must demonstrate that for every $ z \in C $, there exists an $ x \in A $ such that \cref{eq:xAndy} holds.
We can find this $ x $ by defining it as in \cref{eq:xBasedOnEuler}; hence, we have proved that:

{\small
\begin{equation}
    B = C = \{ n \cdot j + SR_O \mid j \in D \}
\end{equation}
}

Now that we have found $ B $, we can find $ A $ by substituting all members of $ B $ as $ y $ in \cref{eq:xBasedOnEuler}, and finding all corresponding $ x $.
For each $ y \in B $ there are exactly $ n $ corresponding $ x \in \{1, \ldots, T\}$, which can be written as follows after substitution:

\begin{equation*}
\begin{aligned}
A = \Bigg\{&
\left(
j \cdot
\left( \frac{SR_P}{n} \right)^{\varphi\left( \frac{T}{n} \right) - 1}
\bmod \frac{T}{n}
\right)
+ \frac{T}{n} i \;\Bigg|\\
& j \in D,\; i \in \{0, \ldots, n-1\}
\Bigg\}.
\end{aligned}
\end{equation*}
\end{proof}

\section{Additional Results}
\label{sec:additional_results}

\subsection{Model and Configuration Analyzer Evaluation} \label{sec:model_evaluation}
\begin{figure*}[t]
    \renewcommand{\thesubfigure}{\roman{subfigure}}
  \captionsetup[subfloat]{width=0.185\linewidth,justification=centering,singlelinecheck=false}
  \centering
  \subfloat[Uplink, TDD (3\textbf{D}/1\textbf{U}), SR = 2 ms, $k_2=4$, $a_1=7$, Constant, srsRAN\label{fig:4211_sr2_bsr1_k4_mac5}]{
    \includegraphics[width=0.185\linewidth]{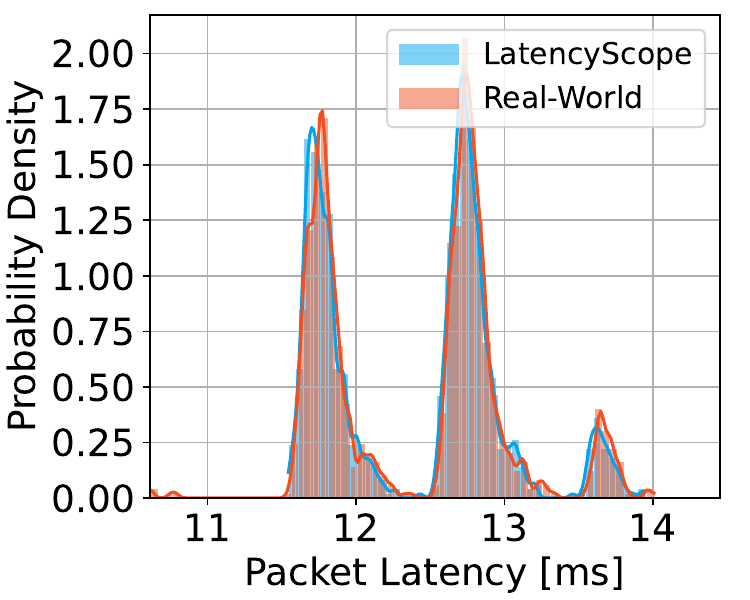}
  }
  \hfil
  \subfloat[Uplink, TDD (3\textbf{D}/1\textbf{U}), SR = 4 ms, $k_2=2$, $a_1=3$, Constant, srsRAN\label{fig:4211_sr4_bsr1_k2_mac1}]{
    \includegraphics[width=0.185\linewidth]{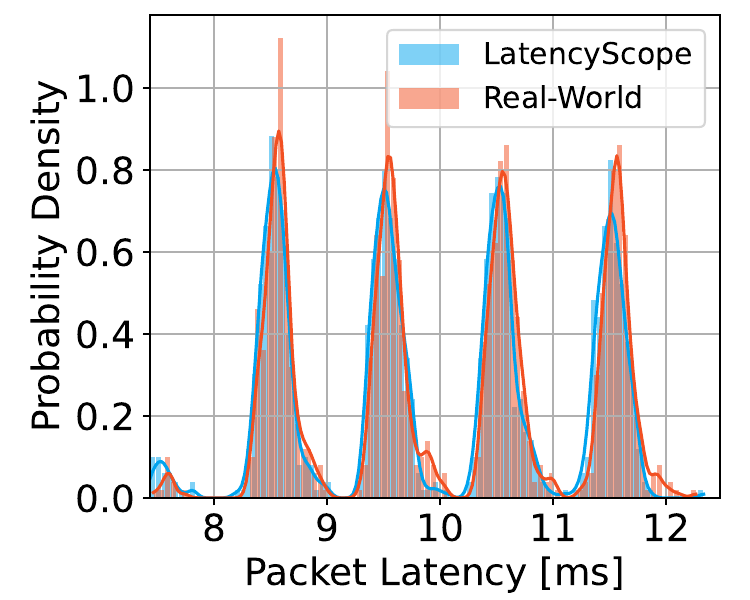}
  }
  \hfil
  \subfloat[Uplink, TDD (3\textbf{D}/2\textbf{U}), SR = 10 ms, $k_2=2$, $a_1=3$, Constant, srsRAN\label{fig:5212_sr10_bsr1_k2_mac1}]{
    \includegraphics[width=0.185\linewidth]{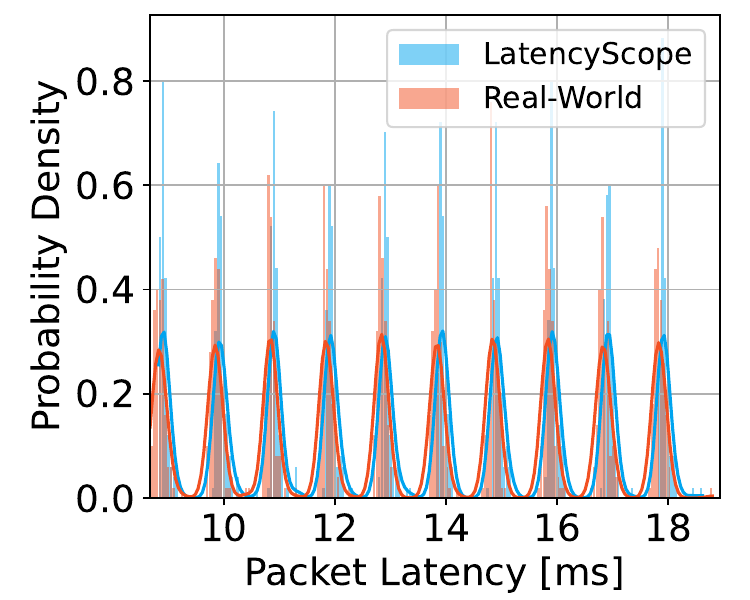}
  }
  \hfil
  \subfloat[Uplink, TDD (3\textbf{D}/2\textbf{U}), SR = 10 ms, $k_2=4$, $a_1=7$, Constant, srsRAN\label{fig:5212_sr10_bsr1_k4_mac5}]{
    \includegraphics[width=0.185\linewidth]{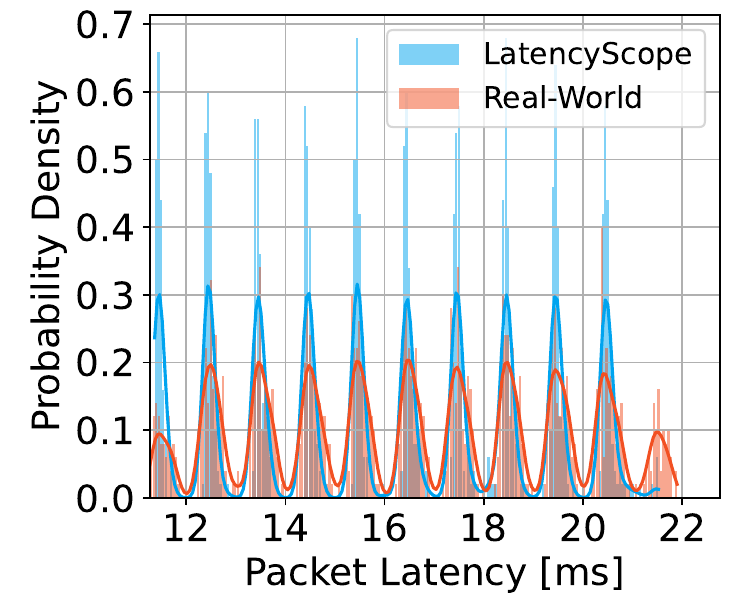}
  }
  \hfil
  \subfloat[Uplink, TDD (3\textbf{D}/2\textbf{U}), SR = 20 ms, $k_2=4$, $a_1=5$, Constant, srsRAN\label{fig:5212_sr20_bsr1_k4_mac3}]{
    \includegraphics[width=0.185\linewidth]{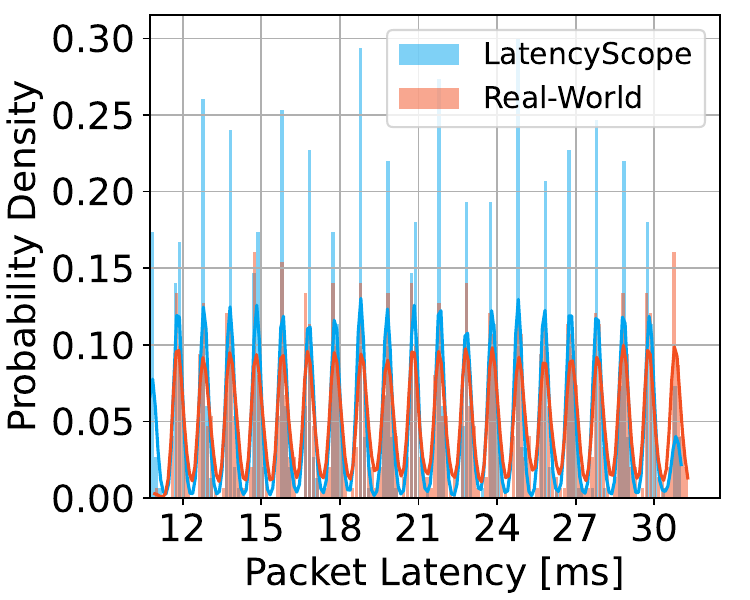}
  }
  \hfil
  \subfloat[Uplink, TDD (3\textbf{D}/2\textbf{U}), SR = 20 ms, $k_2=4$, $a_1=5$, Gaussian, srsRAN\label{fig:5212_sr20_bsr1_k4_mac3_gauss}]{
    \includegraphics[width=0.185\linewidth]{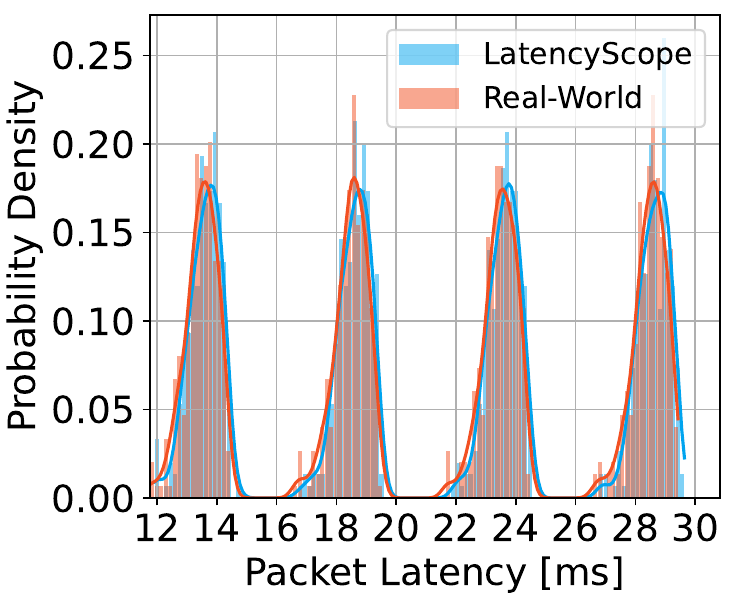}
  }
  \hfil
  \subfloat[Uplink, TDD (7\textbf{D}/3\textbf{U}), SR = 10 ms, $k_2=3$, $a_1=3$, Constant, srsRAN\label{fig:10613_sr10_bsr1_k3_mac1}]{
    \includegraphics[width=0.185\linewidth]{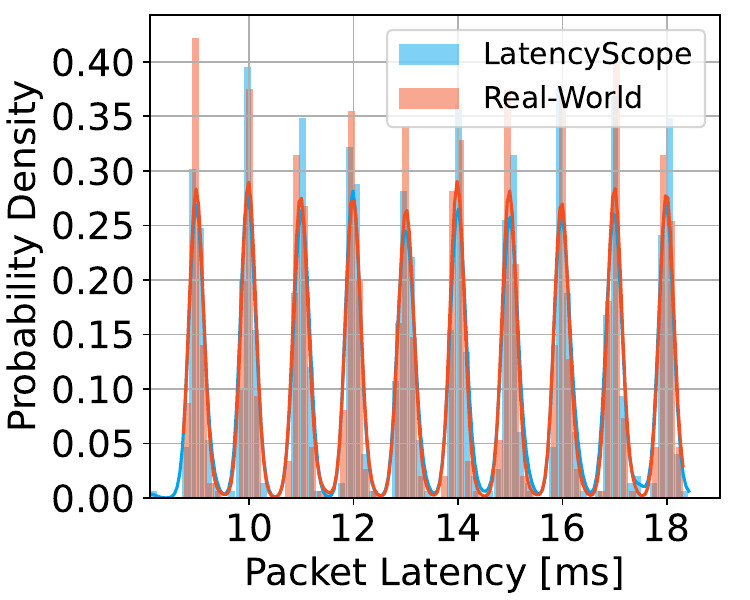}
  }
  \hfil
  \subfloat[Uplink, TDD (7\textbf{D}/3\textbf{U}), SR = 10 ms, $k_2=4$, $a_1=7$, Constant, srsRAN\label{fig:10613_sr10_bsr1_k4_mac5}]{
    \includegraphics[width=0.185\linewidth]{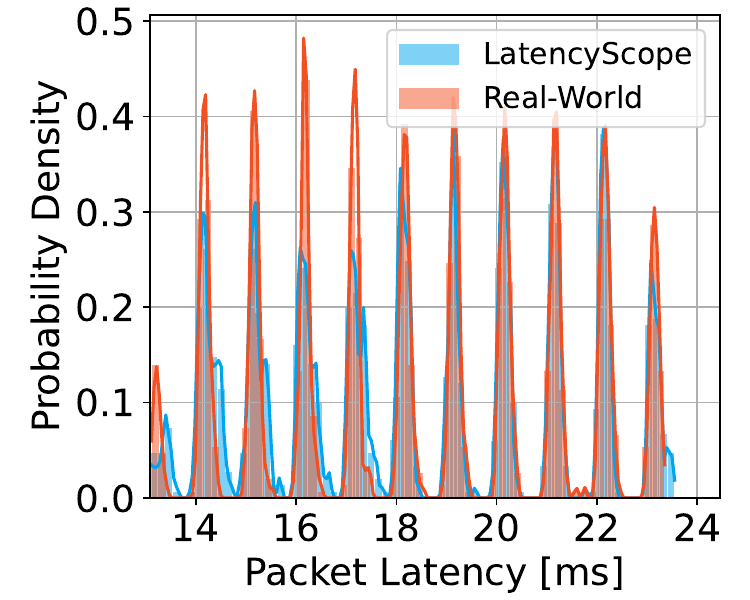}
  }
  \hfil
  \subfloat[Uplink, TDD (7\textbf{D}/3\textbf{U}), SR = 20 ms, $k_2=3$, $a_1=7$, Constant, srsRAN\label{fig:10613_sr20_bsr1_k3_mac5}]{
    \includegraphics[width=0.185\linewidth]{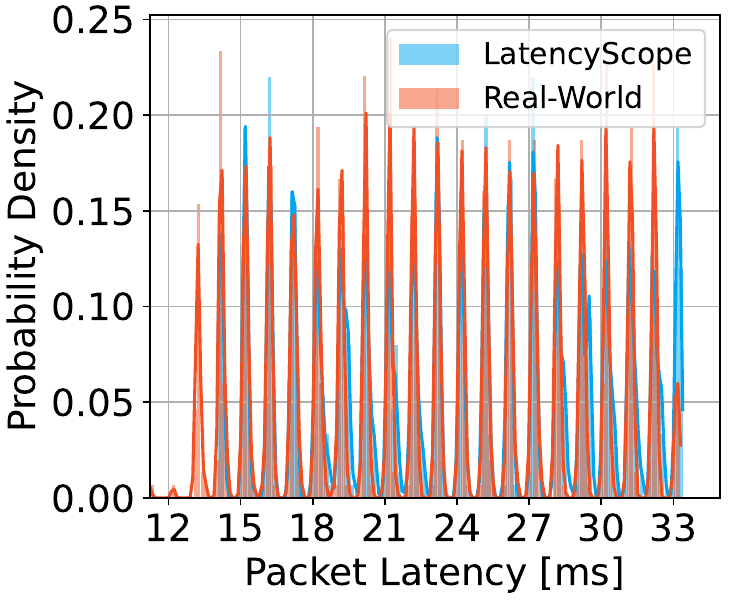}
  }
  \hfil
  \subfloat[Uplink, TDD (7\textbf{D}/3\textbf{U}), SR = 20 ms, $k_2=3$, $a_1=7$, Gaussian, srsRAN\label{fig:10613_sr20_bsr1_k3_mac5_gauss}]{
    \includegraphics[width=0.185\linewidth]{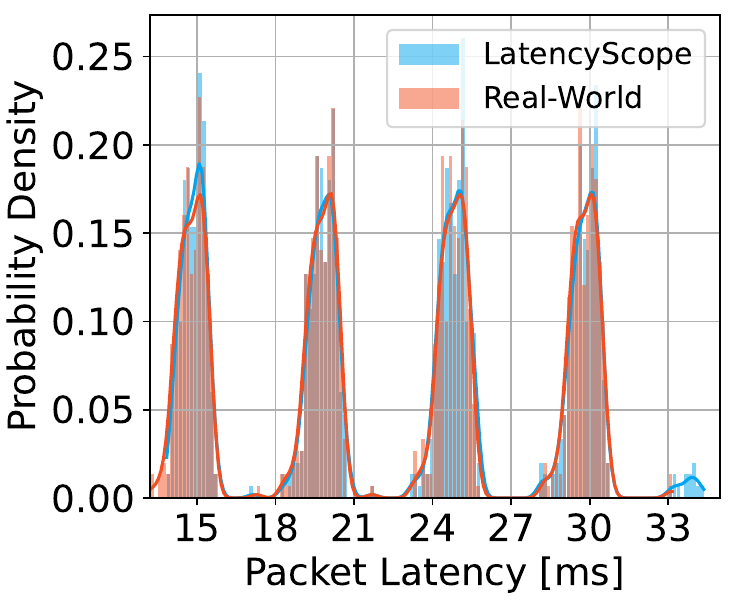}
  }

  \caption{Latency distributions for additional scenarios, shown over the 0.2nd-99th percentile latency range. We generate traffic using different inter-arrival distributions and packet sizes: 1) Constant - Packets of \SI{64}{\bytes} with a fixed inter-arrival time of \SI{101}{\milli\s}. 2) Gaussian - Packets of \SI{64}{\bytes} with a Gaussian inter-arrival distribution (mean: \SI{105}{\milli\s}, standard deviation: \SI{0.05}{\milli\s}).}
  \label{fig:latency-distribution-appendix}

   \renewcommand{\thesubfigure}{\alph{subfigure}}
\end{figure*}

\begin{figure*}[t]
  \centering
  \begin{minipage}[t]{0.65\linewidth}
    \centering
    \includegraphics[width=\linewidth]{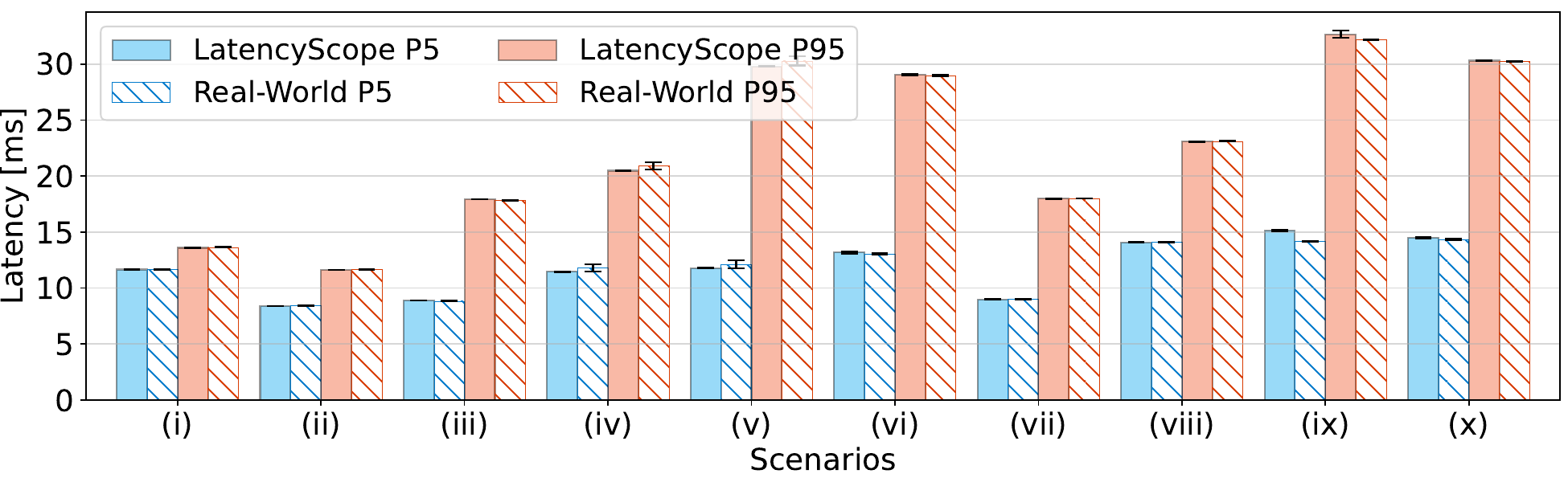}

    \caption{5th and 95th percentile latency bounds comparing \thesystem and real-world measurements under the same scenarios as \cref{fig:latency-distribution-appendix}. Error bars show one-percentile variation.}\label{fig:latency_bar_appendix}
  \end{minipage}
  \hfil
  \begin{minipage}[t]{0.33\linewidth}
    \centering
    \includegraphics[width=\linewidth]{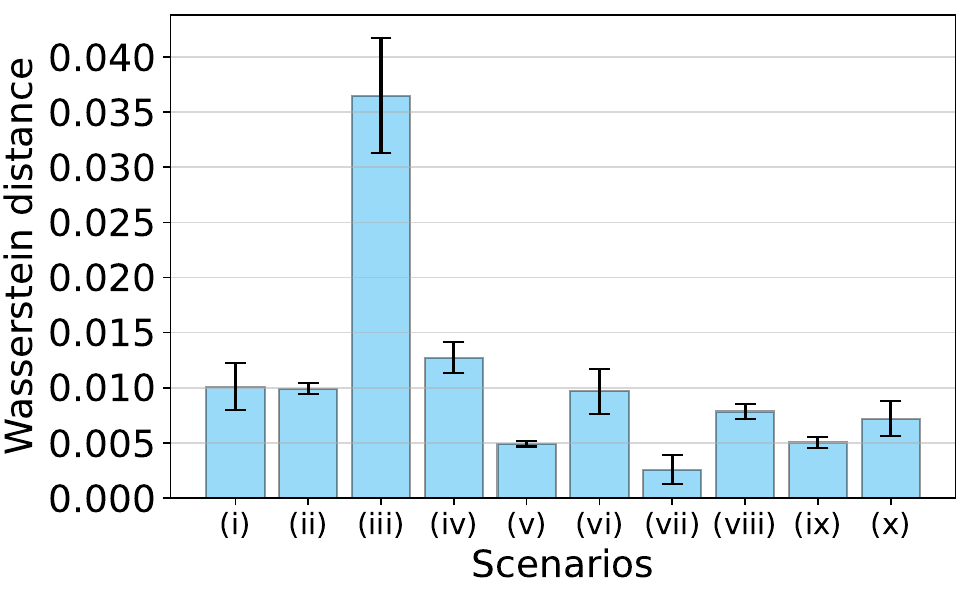}

    \caption{Wasserstein distances between \thesystem and real-world measurements over the 5th-95th percentile latency range. Error bars show 4th-96th percentiles.}\label{fig:wasserstein_distance_appendix}
  \end{minipage}

\end{figure*}

We present latency distributions for additional scenarios in Fig.~\ref{fig:latency-distribution-appendix}. Fig.~\ref{fig:latency_bar_appendix} and Fig.~\ref{fig:wasserstein_distance_appendix} show the corresponding latency bounds and Wasserstein distances for these scenarios.

\subsection{Traffic Generation}
\label{sec:traffic_generator_evaluation_app}

To validate \thesystem against latency measurements across diverse traffic patterns, we generate and send traffic using various arrival and inter-arrival distributions and packet sizes.
Existing tools such as ping~\cite{ping}, fping~\cite{fping}, and tcpreplay~\cite{tcpreplay} lack the required accuracy, so we built a custom C-based traffic generator.
Unlike general-purpose tools that trade precision for memory efficiency, our tool precomputes and stores packets with timestamps, then transmits them in a lightweight loop.
This reduces inter-arrival jitter at the cost of memory efficiency, which is acceptable as we only use this to validate the accuracy of \thesystem.
We further improve accuracy by raising process priority, pinning it to a dedicated core, and using expiration-based timers.
We evaluate the accuracy of the traffic generation process on an Intel Xeon W-2225 with \SI{64}{\giga\byte} RAM running Linux \textit{6.8.0-50-lowlatency}.
Our tool consists of two components: a \textit{generator} and a \textit{replayer}.
The generator produces packets with Gaussian, Poisson, or constant inter-arrival times, while the replayer reproduces traces with high fidelity.

\begin{figure}[t]
    \centering
    \subfloat[Constant distribution\label{fig:dist_compare}]{
        \includegraphics[width=0.97\linewidth]{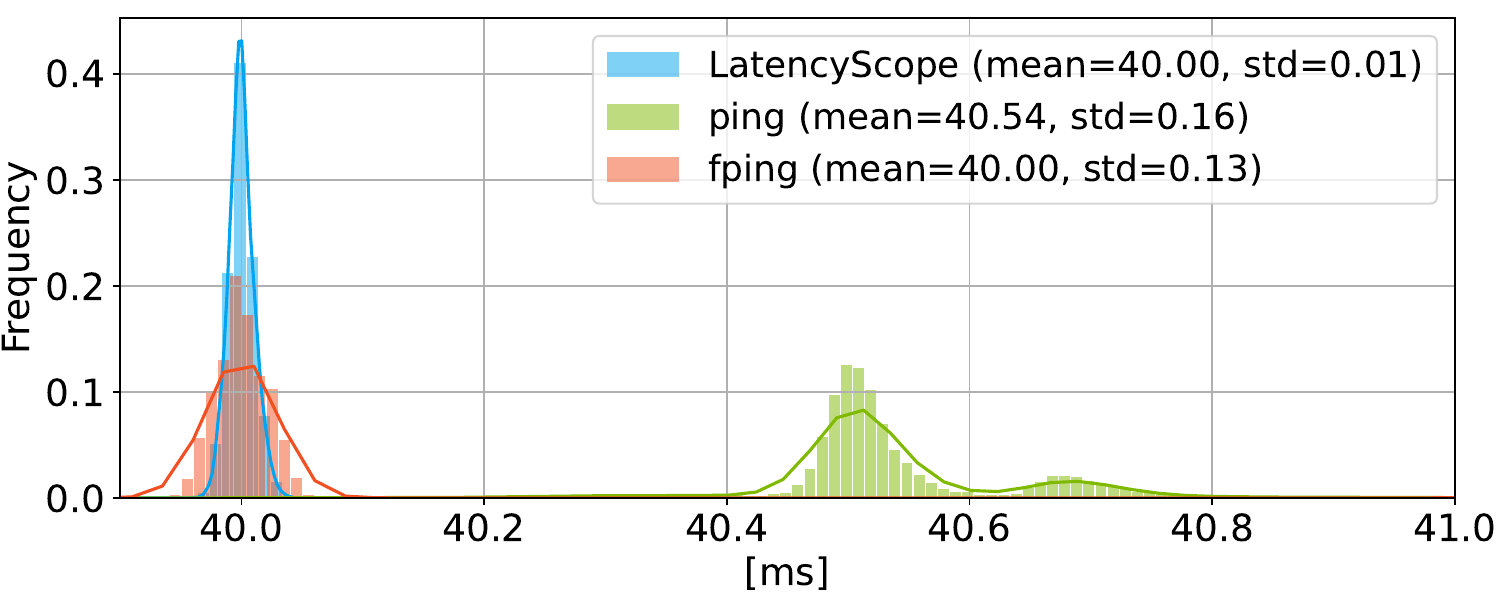}
    }\\
    \subfloat[Gaussian distribution\label{fig:normal_distribution}]{
        \includegraphics[width=0.47\linewidth]{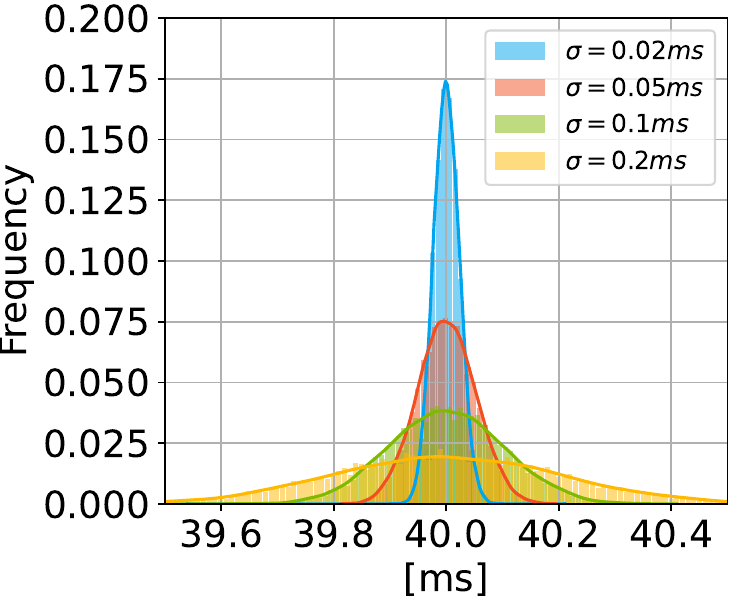}
    }
    \hfill
    \subfloat[Poisson distribution\label{fig:poisson_distribution}]{
        \includegraphics[width=0.47\linewidth]{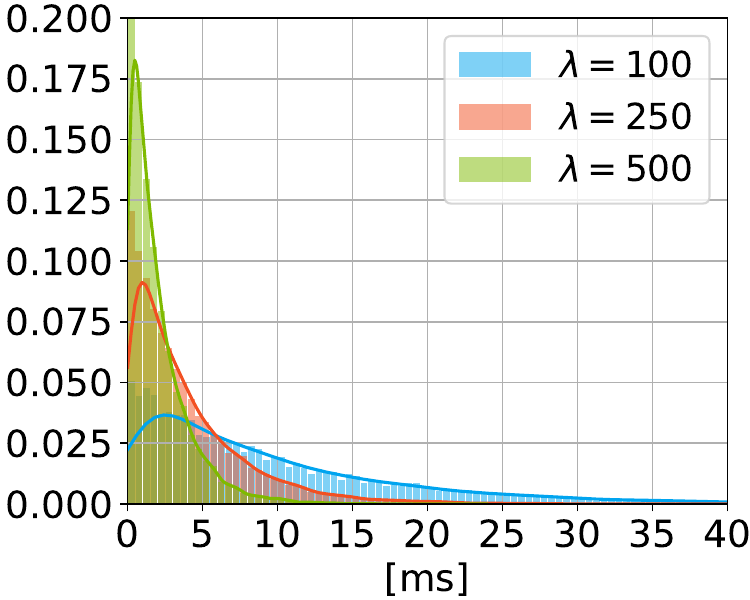}
    }
    \caption{Inter-arrival time of generated packets.}\label{fig:distributions}

\end{figure}

To assess precision, we generated \num{10000} packets with a constant \SI{40}{\milli\second} interval using our generator, \texttt{ping}~\cite{ping}, and \texttt{fping}~\cite{fping}.
As shown in \cref{fig:dist_compare}, our tool achieves the narrowest distribution, centered on the target interval.
The standard deviation is only \SI{0.01}{\milli\second}, compared to \SI{0.16}{\milli\second} and \SI{0.13}{\milli\second} for \texttt{ping} and \texttt{fping}, respectively.
Finally, \cref{fig:normal_distribution,fig:poisson_distribution} demonstrate the ability to accurately generate Gaussian and Poisson traffic with varying parameters.

\section{Configuration Analyzer} \label{sec:eval_analyzer_config}

We provide the configuration variables used in \cref{sec:config_analyzer} to evaluate the performance of \thesystem's configuration analyzer.
Taking the Cartesian product of these variables yields approximately \num{32.7} billion possible combinations.

\begin{lstlisting}
ue_preparation_time_sr_mean: {value: 1.464, optimize: false} # ms
ue_preparation_time_sr_std: {value: 0.175, optimize: false} # ms
gnb_processing_time_sr: {value: 0.1, optimize: false} # ms
mac_scheduling_time: {value: 0.1, optimize: false} # ms
gnb_phy_processing_time: {value: 0.01, optimize: false} # ms
radio_preparation_time: {value: 0.5, optimize: false} # ms
ue_l2_down_processing_time: {value: 0.3, optimize: false} # 0.6 # ms
gnb_processing_time_l1_up_mean: {value: 0.13, optimize: false} # ms
gnb_processing_time_l1_up_std: {value: 0.40, optimize: false} # ms
gnb_processing_time_l1_up_loc: {value: 0.27, optimize: false} # ms
slot_duration: {value: 0.5, optimize: true, discrete_values: [0.25, 0.5, 1.0]} # ms
dl_ul_tx_period: {value: 4, optimize: true, discrete_values: [2, 4, 5, 6, 8, 10, 12, 16, 20, 40, 80, 160, 320]} # slots
nof_dl_slots: {value: 2, optimize: true, discrete_values: [1, 2, ..., 80]} # slots
no_mixed_slot: {value: true, optimize: false}
k2: {value: 1, optimize: true, discrete_values: [1, 2, ..., 16]} # slots
sr_period: {value: 4, optimize: true, discrete_values: [1, 2, 4, 5, 8, 10, 16, 20, 40, 80, 160, 320, 640]} # slots
sr_offset: {value: 3, optimize: true, discrete_values: [0, 1, ..., 79]} # slots
pucch_st_sym: {value: 13, optimize: true, discrete_values: [0, 1, ..., 13]} # symbol
pucch_nof_sym: {value: 1, optimize: true, discrete_values: [1, 2, 3]} # number
pdcch_nof_sym: {value: 1, optimize: true, discrete_values: [1, 2, 3]} # number
in_advance_submission: {value: 1, optimize: true, discrete_values: [0, 1, 2, 3, 4]} # slots
\end{lstlisting}

\section{MNO Parameters} \label{sec:mno1-params}

We provide the most important parameters used to run \thesystem for \textit{MNO} in Tab.~\ref{table:mno1-params}.
We mark each as either directly obtained by decoding RRC messages or learned as explained in~\cref{sec:experimental-setup}.

\begin{table}[t]
\centering
\caption{MNO parameters.}
\label{table:mno1-params}
\begin{tabular}{lcc} \hline
\textbf{Parameter} & \textbf{Value} & \textbf{Source} \\ \hline
TDD Pattern & 4\textbf{D}/1\textbf{U} & RRC Message \\
Numerology ($\mu$) & 1 & RRC Message \\
SR Period & \SI{40}{\milli\s} & RRC Message \\
Radio Latency & $<$ \SI{250}{\micro\s} & Learned \\
$a_1$ & 1 & Learned \\
Wired Delay Mean & \SI{5.5}{\milli\s} & Learned \\
Wired Delay Std Dev & \SI{0.8}{\milli\s} & Learned \\
\hline
\end{tabular}
\end{table}

\section{Modeling the Latency for the Remaining Scenarios}
\label{sec:appendix_c}

As discussed in Sec.~\ref{sec:background} and~\ref{sec:latency-analysis}, and as mathematically modeled in Sec.~\ref{sec:model-foundation}, our initial analysis concentrated on small (Size 1) packets in a 5G network.
If we define the size of initial grant as $g_I$ bytes, which is determined by network configuration, the formal definition of small packet is as follows.

{\small
\begin{equation}
    P \le g_I
    \label{eq:packet-size-greater-than-initial-grant}
\end{equation}
}

This means that the packet is smaller than the initial grant size, and it can be transmitted using the initial grant.
In this appendix we thoroughly model and discuss all the remaining scenarios.
We discuss larger packets, UL latency for a train of packets, UL and DL latency with TDD mini-slot configuration, UL latency with grant-free access, DL latency in TDD, and frequency-division duplexing (FDD) configurations.

We classify the various scenarios as follows, and provide the corresponding mathematical modeling for each case.
In order to reuse the formulation in \cref{sec:model-foundation}, we summarize the section in \cref{alg:small-packet}.

\begin{algorithm}
    \caption{Total Latency of TDD UL for Size 1 Packets}
    \label{alg:small-packet}
    \begin{algorithmic}[1]
        \State Initialize $o_1 \gets t_{arrival}$ where $t_{arrival}$ is the arrival time of the packet.
        \State Calculate $w_1$ using \cref{eq:aSubsetOneT} to \cref{eq:w1}.
        \State Calculate $w_2$ using \cref{eq:combined_w2_constraint}.
        \State Calculate $w_3$ using \cref{eq:startOfMACProcessing-w3} to \cref{eq:w3}.
        \State Calculate $w_4$ using \cref{eq:combined_w4_pdcchConstraint}.
        \State Calculate $w_5$ using \cref{eq:combined_k2min_constraint} to \cref{eq:combined_k2_w5}.
        \State Calculate $w_6$ using \cref{eq:combined_w6_w7}.
        \State Calculate $w_7$ using \cref{eq:combined_w6_w7}.
        \State Calculate $Total_{Latency}$ using \cref{eq:totalUL}.
        \State \Return $w_1$, $w_2$, $w_3$, $w_4$, $w_5$, $w_6$, $w_7$, $Total_{Latency}$
    \end{algorithmic}
\end{algorithm}

\subsection{UL Latency for Size 2 packets in TDD}\label{sec:ul-packets-size2}

\begin{figure*}[!t]
    \includegraphics[width=\linewidth]{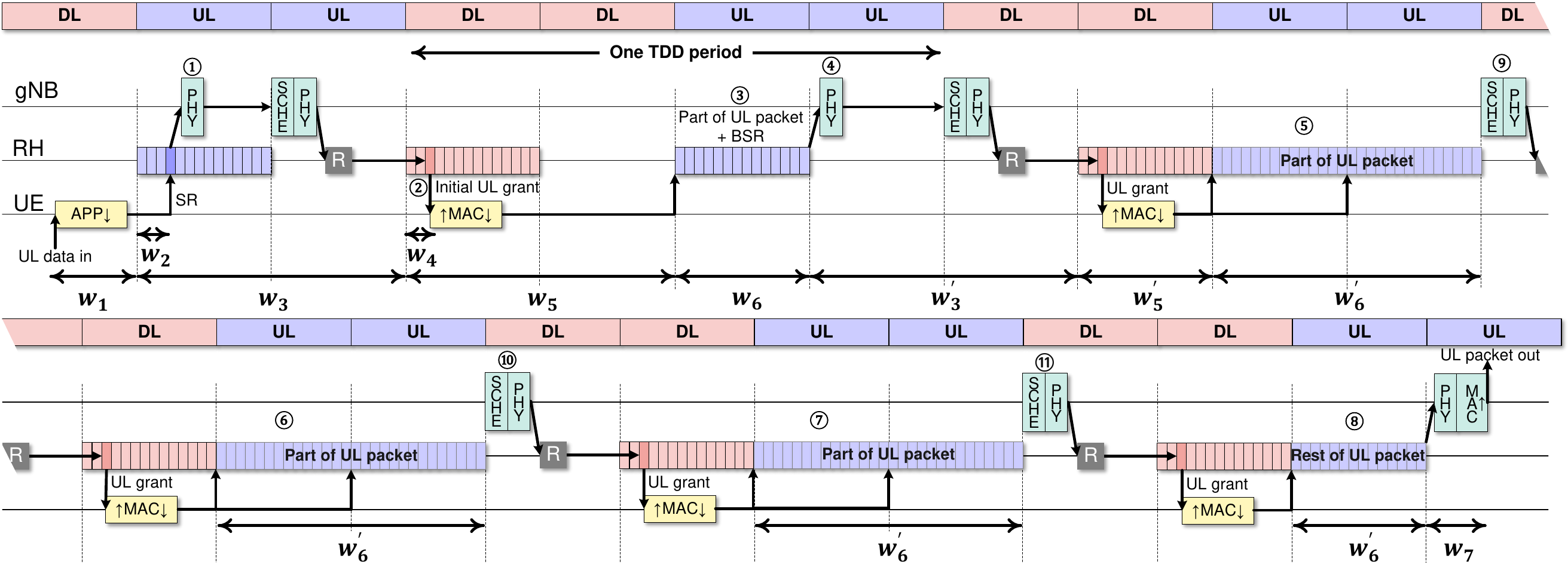}
    \caption{Journey of a packet for larger packets. A \textit{TDD Common Configuration} with the \textbf{DDUU} pattern is used.}

    \label{fig:bsr-journey-congestion}

\end{figure*}

Consider the case where a UE transmits a packet larger than the initial grant size, i.e., \cref{eq:packet-size-greater-than-initial-grant} no longer holds.
The packet's journey is shown in \cref{fig:bsr-journey-congestion} for a \textit{TDD Common Configuration} with the \textbf{DDUU} pattern.
The first steps mirror those of the ping packet in \cref{sec:latency-analysis}: the UE sends a scheduling request after $w_1$, receives the initial grant after $w_3$, and transmits data after $w_5$.
The difference arises because the packet exceeds the initial grant, which is typically small and mainly intended for buffer status reporting (\ballnumber{2}).
Using this grant, the UE sends $g_I$ bytes of data and a report showing the remaining bytes in its buffer. This report is known as the Buffer Status Report (BSR)
\footnote{The BSR is small relative to the packet sizes considered, so we neglect its contribution.} \ballnumber{3}.
The gNB uses this information to allocate additional resources to the UE in the next UL slots.
Unlike SR, which is sent on PUCCH and processed at the start of the uplink slot \ballnumber{1}, the BSR is carried on PUSCH and only processed in the subsequent slot \ballnumber{4}.

The variables $w_1$-$w_7$ are defined and computed as in \cref{sec:model-foundation}.
Specifically, we had the following.

{\small
\begin{equation}
    w_6 = S, \quad w_7 = p_4
    \label{eq:combined_w6_w7}
\end{equation}
}

Once the gNB receives the BSR, we introduce additional variables to capture the extra latency, as illustrated in \cref{fig:bsr-journey-congestion}.

\vskip 0.06in \noindent {\bf $w'_3$:} Time between the gNB receiving the BSR and sending a new grant.
Unlike the initial grant, this allocation carries the remaining packet data.

\vskip 0.06in \noindent {\bf $w'_5$:} Time from the slot carrying the new grant to the uplink slot where the UE transmits using it.

\vskip 0.06in \noindent {\bf $w'_6$:} Transmission time of data on granted uplink slots within a TDD period.

\vskip 0.06in \noindent {\bf Modeling of $w'_3$:} As in \cref{sec:model-foundation}, the gNB requires $p_1$ seconds to process UL samples, decoding symbol by symbol.
Since scheduling starts at slot boundaries, the last symbol dominates processing time, adding $p_1$.
The MAC processing and grant scheduling can be expressed as:

{\small
\begin{equation}
    \begin{aligned}
        \text{Start of MAC Processing} = o_1 + w_1 + w_3 + w_5 + w_6 + \left\lceil \tfrac{p_1}{S} \right\rceil S
        \label{eq:start-mac-processing-w-prime-3}
    \end{aligned}
\end{equation}
}
{\small
\begin{equation}
    \begin{aligned}
    \text{Scheduled Slot for Grant} = \tfrac{o_1 + w_1 + w_3 + w_5 + w_6}{S} + \left\lceil \tfrac{p_1}{S} \right\rceil + a_1 + 1
    \label{eq:scheduled-slot-for-grant-w-prime-3}
    \end{aligned}
\end{equation}
}

Thus,

{\small
\begin{equation}
    re'_1 \equiv_T \frac{o_1 + w_1 + w_3 + w_5 + w_6}{S} + \left\lceil \frac{p_1}{S} \right\rceil + a_1 + 1
    \label{eq:re_prime_1}
\end{equation}

\begin{equation}
    w'_3 =
    \begin{cases}
        \big(\left\lceil \tfrac{p_1}{S} \right\rceil + a_1 + 1\big) S, & re'_1 \leq d; \\[6pt]
        \big(\left\lceil \tfrac{p_1}{S} \right\rceil + a_1 + 1 + (T - re'_1)\big) S, & \text{otherwise.}
    \end{cases}
    \label{eq:w_prime_3}
\end{equation}
}

\vskip 0.06in \noindent {\bf Modeling of $w'_5$:}
Similar to $w_5$, but with potentially larger UL data, so the UE might need more time to prepare.
We define $l'_2$ as the time for the UE to prepare samples after the grant.
Then,

{\small
\begin{equation}
    k'_{2_{\min}} = \left\lceil \tfrac{w_4 + l'_2}{S} \right\rceil, \quad k'_{2_{\min}} \leq k_2
    \label{eq:k_prime_2_min}
\end{equation}
}

Minimizing $k_2$ gives:

\begin{equation}
    re'_2 \equiv_T \tfrac{o_1 + w_1 + w_3 + w_5 + w_6 + w'_3}{S} + k'_{2_{\min}}
    \label{eq:re_prime_2}
\end{equation}
\begin{equation}
    k_2 =
    \begin{cases}
        k'_{2_{\min}}, & re'_2 \geq d; \\
        k'_{2_{\min}} + d - re'_2, & \text{otherwise.}
    \end{cases}
    \label{eq:k_2_for_w_prime_5}
\end{equation}
\begin{equation}
    w'_5 = k_2 S
    \label{eq:w_prime_5}
\end{equation}

\vskip 0.06in \noindent {\bf Modeling of $w'_6$:} By definition, $w'_6$ is the transmission time over contiguous UL slots within one TDD period.
If the packet spans multiple periods (\ballnumber{5}--\ballnumber{8} in \cref{fig:bsr-journey-congestion}), $w'_6$ accounts for all.
In this example we assume a single UE (although we also model multi-UE in Appendix~\ref{sec:contention-model}), so all available UL resources are allocated.
The number of required UL slots is determined by the maximum bytes per slot, computed later in this section.

The bytes per uplink (UL) slot depend on (i) the bandwidth and subcarrier spacing (SCS), and (ii) the modulation and coding scheme (MCS).
In 5G, a resource block (RB) spans 12 subcarriers in frequency and 14 symbols in time.
The number of RBs for each bandwidth/SCS pair is prescribed in the standard~\cite{3GPP-TS-38.101-1-version-17.8.0}.
The MCS, determined by channel quality, specifies modulation order $Q_m$ and coding rate $R$.
The mapping between them is also prescribed in the standard~\cite{3GPP-TS-38.214-version-16.2.0}.
Since SCS is tied to slot duration $T$ (which is uniquely determined from SCS), we write $N_{RB}[B,T]$ for the RB count at bandwidth $B$ and slot duration $T$.
Given MCS index $I_{MCS}$, $Q_m[I_{MCS}]$ and $R[I_{MCS}]$ are obtained from the tables in the standard~\cite{3GPP-TS-38.214-version-16.2.0}.
The maximum payload per UL slot (excluding control) is defined and calculated as follows.

{\small
\begin{equation}
    Bi_{UL} = \frac{N_{RB}[B, T] \cdot 12 \cdot Q_m[I_{MCS}] \cdot R[I_{MCS}] \cdot (14 - UC_{no})}{8}
    \label{eq:bi_ul}
\end{equation}
}

After receiving the BSR, the gNB records the buffer status and allocates resources until the buffer is empty, updating it only upon new BSRs.
Here we assume a single big packet in the buffer, so no further BSRs are sent (multi-packet arrivals are discussed in Appendix~\ref{sec:multi-packet}).

In the first TDD period, the UE receives a grant (\ballnumber{5} in \cref{fig:bsr-journey-congestion}).
Subsequent UL slots in the same and following periods are also granted (\ballnumber{6}--\ballnumber{8}), since the gNB can allocate multiple outstanding grants until the buffer is emptied.
A more relaxed case is shown in \cref{fig:bsr-journey-congestion}, where new grants (\ballnumber{9}--\ballnumber{11}) arrive in each TDD period after the first.
The number of granted UL slots per TDD period (assuming full allocation to the UE) is:

{\small
\begin{equation}
    N_{UL, slots} = \min\!\left(\left\lceil \tfrac{P_{rem}}{Bi_{UL}} \right\rceil,\, T-d\right)
    \label{eq:n_ul_slots}
\end{equation}
}

where $P_{rem}$ is the remaining buffer, updated as data is transmitted.
Thus,

{\small
\begin{equation}
    w'_6 = N_{UL, slots} \cdot S
    \label{eq:w_prime_6}
\end{equation}
}

\begin{algorithm}
    \caption{Total Latency of UL for Size 2 Packets with a Single Scheduling Request in TDD}
    \label{alg:large-packet-with-congestion-oneFlow}
    \begin{algorithmic}[1]
        \State Initialize $o_1 \gets t_{arrival}$ \label{line:arrival-init-algo-oneFlow}
        \State Calculate $w_1, w_2, w_3, w_4, w_5, w_6$ (\cref{eq:aSubsetOneT} to \cref{eq:combined_w6_w7})
        \State Initialize $Time \gets o_1 + w_1 + w_3 + w_5 + w_6$ \label{line:time-init-algo-oneFlow}
        \State Initialize $P_{rem} \gets P - g_I$ \label{line:p_rem-init-algo-oneFlow}
        \State Calculate $w'_3$ (\cref{eq:start-mac-processing-w-prime-3} to \cref{eq:w_prime_3}) by substituting $Time \rightarrow o_1 + w_1 + w_3 + w_5 + w_6$ \label{line:w-prime-3-algo-oneFlow}
        \State Update $Time \gets Time + w'_3$
        \State Calculate $w'_5$ (\cref{eq:k_prime_2_min} to \cref{eq:w_prime_5}) by substituting $Time \rightarrow o_1 + w_1 + w_3 + w_5 + w_6 + w'_3$ \label{line:w-prime-5-algo-oneFlow}
        \State Update $Time \gets Time + w'_5$
        \State Calculate $B_{UL}$ (\cref{eq:bi_ul}) \label{line:bi-ul-algo-oneFlow}
        \State Calculate $N_{UL, slots}$ (\cref{eq:n_ul_slots}) using $P_{rem}$ \label{line:n-ul-slots-algo-oneFlow}
        \State Calculate $w'_6$ (\cref{eq:w_prime_6}) by using $N_{UL, slots}$ \label{line:w-prime-6-algo-oneFlow}
        \State Update $P_{rem} \gets P_{rem} - N_{UL, slots} \cdot B_{UL}$ \label{line:p-rem-update-algo-oneFlow}
        \State Update $Time \gets Time + w'_6$  \label{line:update-with-w-prime-6-algo-oneFlow}
        \If {$P_{rem} \leq 0$}
            \State Calculate $w_7$ (\cref{eq:combined_w6_w7})
            \State \Return $Total\_Latency \gets Time + w_7 - o_1$
        \EndIf
        \While {True} \label{line:loop-algo-oneFlow}
            \State Update $Time \gets Time + d \cdot S$ \label{line:next-tdd-period-algo-oneFlow}
            \State Calculate $N_{UL, slots}$ (\cref{eq:n_ul_slots}) using $P_{rem}$
            \State Calculate $w'_6$ (\cref{eq:w_prime_6}) by using $N_{UL, slots}$
            \State Update $P_{rem} \gets P_{rem} - N_{UL, slots} \cdot B_{UL}$
            \State Update $Time \gets Time + w'_6$
            \If {$P_{rem} \leq 0$} \label{line:check-buffer-algo-oneFlow}
                \State Calculate $w_7$ (\cref{eq:combined_w6_w7})
                \State \Return $Total\_Latency \gets Time + w_7 - o_1$ \label{line:total-latency-algo-oneFlow}
            \EndIf
        \EndWhile
    \end{algorithmic}
\end{algorithm}

Finally, to compute total latency, we iterate over TDD periods, updating $P_{rem}$ until the packet is fully transmitted.
This procedure is summarized in \cref{alg:large-packet-with-congestion-oneFlow}.

\subsection{\texorpdfstring{UL latency when radio delay $>$ TDD period}{UL latency when radio delay is greater than TDD period}}\label{sec:large-packet-with-congestion}

\begin{figure*}[!t]
    \includegraphics[width=\linewidth]{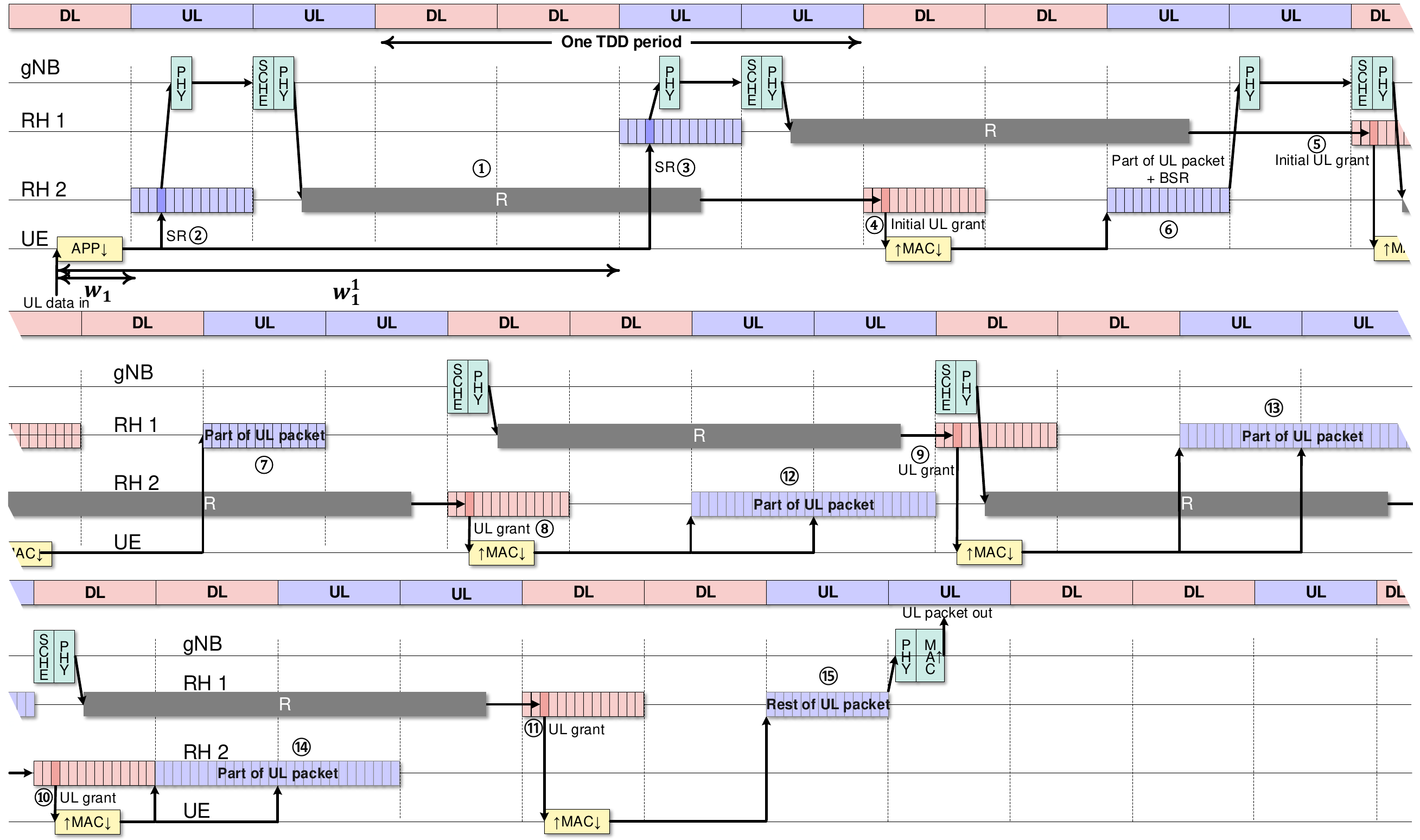}

    \caption{Journey of a packet for larger packets for the special case of extended grant latency.}

    \label{fig:bsr-journey-congestion-lagged-grant}

\end{figure*}

A special case arises when the grant ($w_3$ or $w'_3$) arrives later than the next SR opportunity, as shown in \cref{fig:bsr-journey-congestion-lagged-grant}.
Here we assume only the first UL slot per TDD period is available for SR, and a larger radio latency (\ballnumber{1}) causes the grant to be received after the next SR slot.
For clarity, two RH timelines are drawn in the figure, though they represent the same RH.

In this case, since the UE has not yet received the grant for its first SR (\ballnumber{2}), it will retransmit SRs at each new opportunity (\ballnumber{3}) until the grant (\ballnumber{4}) arrives.
This does not affect small packets (cf. \cref{sec:model-foundation}), as the initial grant is sufficient.
For large packets, however, the UE may receive multiple initial grants (\ballnumber{4}, \ballnumber{5}), sending small chunks (\ballnumber{6}, \ballnumber{7}) before larger grants (\ballnumber{8}--\ballnumber{11}) arrive.
After the first major transmission (\ballnumber{12}), the remaining packet is transmitted in subsequent TDD periods (\ballnumber{13}--\ballnumber{15}), as in Appendix~\ref{sec:ul-packets-size2}.

\begin{algorithm}
    \caption{Calculation of the Start of Repeated Scheduling Requests for TDD UL}
    \label{alg:calculation-start-logical-flows}
    \begin{algorithmic}[1]
        \State $w_1^0 \gets w_1$ \label{line:w_1_init}
        \State Initialize $j = 1$
        \While {True}
            \State Compute $SR_{ready}^j$ similar to \cref{eq:srReady}: \label{line:start-w1-calc}
            \begin{equation}
                SR_{ready}^j =
                \left\lfloor
                \frac{o_1 + \sum_{i=0}^{j-1} w_1^i}{S}
                \right\rfloor - SR_O
            \end{equation}
            \State Compute $q^j$ based on \cref{eq:q_r_w1_combined} by substituting $q^j \rightarrow q$ and $SR_{ready}^j \rightarrow SR_{ready}$
            \State Compute $r^j$ based on \cref{eq:q_r_w1_combined} by substituting $r^j \rightarrow r$ and $SR_{ready}^j \rightarrow SR_{ready}$
            \State Compute $w_1^j$ based on \cref{eq:w1} by substituting $w_1^j \rightarrow w_1$ and $r^j \rightarrow r$ \label{line:end-w1-calc}
            \If {$w_1^j > w_3$} \label{line:check-w1-w3}
                \State $SR_{num} \gets j$
                \State $W_1 \gets \{w_1^0, w_1^1, \ldots, w_1^{SR_{num} - 1}\}$ \label{line:W1-set}
                \State \Return $W_1$, $SR_{num}$
            \Else
                \State $j \gets j + 1$
            \EndIf
        \EndWhile
    \end{algorithmic}
\end{algorithm}

The procedure to compute the latency with repeated SRs is given in \cref{alg:calculation-start-logical-flows} and \cref{alg:large-packet-with-congestion-multipleFlows}.
The first algorithm identifies all SR opportunities before the first large grant, producing the set $W_1$ (e.g., $w_1, w_1^1$ in \cref{fig:bsr-journey-congestion-lagged-grant}).
The second algorithm then iterates over $W_1$, tracking the buffer and absolute times until the packet is fully transmitted.
Finally, the procedure is: run \cref{alg:calculation-start-logical-flows} to check whether there are single or multiple SRs, and then apply \cref{alg:large-packet-with-congestion-oneFlow} or \cref{alg:large-packet-with-congestion-multipleFlows}, respectively.

\begin{algorithm}
    \caption{Total Latency of UL for Size 2 Packets with Multiple Scheduling Requests in TDD}
    \label{alg:large-packet-with-congestion-multipleFlows}
    \begin{algorithmic}[1]
        \State Initialize $P_{rem} \gets P$, $o_1 \gets t_{arrival}$ \label{line:p_rem-init-algo-multipleFlows}
        \State Initialize $Time$: An array of size $SR_{num}$ with all elements set to $o_1$ \label{line:time-init-algo-multipleFlows}
        \For{$w_1^i \in W_1$} \label{line:iterate-flows-for-bsr-algo-multipleFlows}
            \State Calculate $w_2, w_3, w_4, w_5, w_6$ (\cref{eq:combined_w2_constraint} to \cref{eq:combined_w6_w7}) by substituting $w_1^i \rightarrow w_1$
            \State Update $Time[i] \gets Time[i] + w_1^i + w_3 + w_5 + w_6$
            \State Update $P_{rem} \gets P_{rem} - g_I$
            \If {$P_{rem} \leq 0$}
                \State Calculate $w_7$ (\cref{eq:combined_w6_w7})
                \State $Total\_Latency \gets Time[i] + w_7 - o_1$
                \State \Return $Total\_Latency$
            \EndIf
        \EndFor \label{line:end-iterate-flows-for-bsr-algo-multipleFlows}
        \State Calculate $w'_3$ (\cref{eq:start-mac-processing-w-prime-3} to \cref{eq:w_prime_3}) by substituting $Time[0] \rightarrow o_1 + w_1 + w_3 + w_5 + w_6$ \label{line:w-prime-3-first-sr}
        \State Update $Time[0] \gets Time[0] + w'_3$
        \State Calculate $w'_5$ (\cref{eq:k_prime_2_min} to \cref{eq:w_prime_5}) by substituting $Time[0] \rightarrow o_1 + w_1 + w_3 + w_5 + w_6 + w'_3$
        \State Update $Time[0] \gets Time[0] + w'_5$
        \State Calculate $B_{UL}$ (\cref{eq:bi_ul})
        \State Calculate $N_{UL, slots}$ (\cref{eq:n_ul_slots}) by substituting $P_{rem} \rightarrow (P_{rem} - g_I)$
        \State Update $P_{rem} \gets P_{rem} - N_{UL, slots} \cdot B_{UL}$
        \State Update $Time[0] \gets Time[0] + w'_6$ \label{line:first-big-transmission-end-algo-multipleFlows}
        \If {$P_{rem} \leq 0$}
            \State Calculate $w_7$ (\cref{eq:combined_w6_w7})
            \State $Total\_Latency \gets Time[0] + w_7 - o_1$
            \State \Return $Total\_Latency$
        \EndIf
        \While {True} \label{line:loop-algo-multipleFlows}
            \State Update $Time[0] \gets Time[0] + d \cdot S$
            \State Calculate $N_{UL, slots}$ (\cref{eq:n_ul_slots}) by substituting $P_{rem} \rightarrow (P_{rem} - g_I)$
            \State Calculate $w'_6$ (\cref{eq:w_prime_6}) by using $N_{UL, slots}$
            \State Update $P_{rem} \gets P_{rem} - N_{UL, slots} \cdot B_{UL}$
            \State Update $Time[0] \gets Time[0] + w'_6$
            \If {$P_{rem} \leq 0$}
                \State Calculate $w_7$ (\cref{eq:combined_w6_w7})
                \State \Return $Total\_Latency \gets Time[0] + w_7 - o_1$
            \EndIf
        \EndWhile \label{line:end-loop-algo-multipleFlows}
    \end{algorithmic}
\end{algorithm}

\subsection{UL latency under congestion with RLC-layer queuing}\label{sec:multi-packet}

In this section, we extend our analysis to multiple packets that arrive dynamically at the UE and are transmitted over the same TDD uplink channel.
Previously, we assumed that when a new packet arrived at the UE’s RLC buffer, no other packet was present.
Under this assumption, each packet could be analyzed independently, allowing concurrent evaluation to accelerate computation.
To capture interactions in the multi-packet case, we model the per-byte evolution in the RLC buffer as a finite-state machine with the following states.

\vskip 0.06in \noindent $\bullet~$ $S_0$ (Arrived): The byte has just entered the buffer and has not yet been requested, reported, or granted.
\vskip 0.06in \noindent $\bullet~$ $S_1$ (Requested): The byte was present when the UE sent the scheduling request (SR).
\vskip 0.06in \noindent $\bullet~$ $S_2$ (Reported): The byte was present when the UE computed and sent the buffer status report (BSR).
\vskip 0.06in \noindent $\bullet~$ $S_3$ (Granted): The byte is read for transmission based on resources granted to the UE.

\begin{figure*}[!t]
    \includegraphics[width=\linewidth]{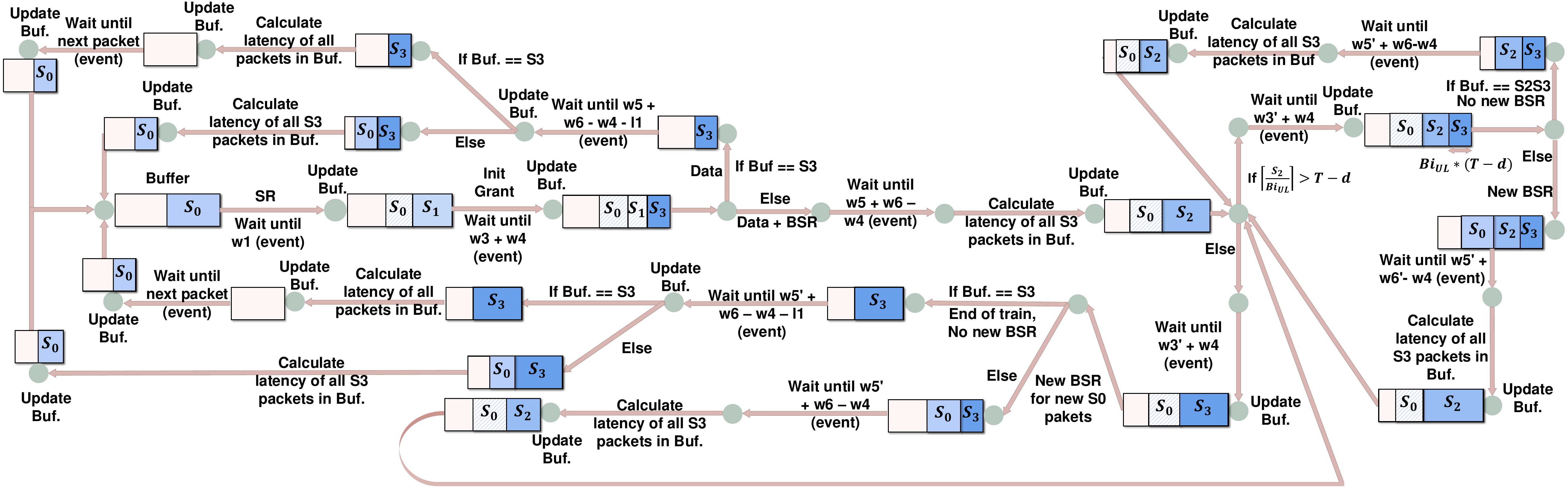}

    \caption{Finite-state machine of RLC-buffer byte states for a user.}

    \label{fig:multi-packet-states}

\end{figure*}

We derive the finite-state machine shown in~\cref{fig:multi-packet-states}, which characterizes the possible byte states in the UE's RLC buffer.
The hatched states indicate that those bytes may or may not be present.
We implement this machine in Python and use it to evaluate scenarios with multiple packet arrivals, including Zoom traffic and \textit{Dota~2} multiplayer gaming traffic, as reported in~\cref{fig:latency-distribution-2}.

\subsection{UL Grant-Free Access}
\label{sec:Grant-Free}

In uplink grant-free access, no grant message is sent by the gNB; instead, the UE uses predefined resources and slots.
Thus, $w_3$, $w_4$, and $w_5$ are not applicable, and the first transmission latency is:

{\small
\begin{equation}
    \text{First Transmission UL Latency in Grant-Free} = w_1 + w_6 + w_7
    \label{eq:totalUL_grant_free}
\end{equation}
}

Here, $w_1$ is the wait for the next pre-configured slot, modeled as in~\cref{eq:aSubsetOneT}-\cref{eq:w1} using the offset and periodicity of pre-configured grants.
If one slot is insufficient, the UE repeats the process until all data is sent, as shown in \cref{alg:Grant-Free}.
The capacity of each pre-configured slot is denoted by $Bi_{UL}$.

\begin{algorithm}
    \caption{UL Total Latency for Grant-Free Access in TDD}
    \label{alg:Grant-Free}
    \begin{algorithmic}[1]
        \State Initialize $Time \gets o_1 + w_1 + w_6$
        \State Initialize $P_{rem} \gets P - Bi_{UL}$
        \If {$P_{rem} \leq 0$}
            \State \Return $Total\_Latency \gets Time + w_7 - o_1$
        \EndIf
        \While {True}
            \State Calculate $w_1$ based on the next pre-defined slot
            \State Calculate $w_6$
            \State Update $Time \gets Time + w_1 + w_6$
            \State Calculate $Bi_{UL}$ based on pre-defined resources for the slot
            \State Update $P_{rem} \gets P_{rem} - Bi_{UL}$
            \If {$P_{rem} \leq 0$}
                \State \Return $Total\_Latency \gets Time + w_7 - o_1$
            \EndIf
        \EndWhile
    \end{algorithmic}
\end{algorithm}

\subsection{DL latency in TDD}
\label{sec:dl-model}

So far, we have modeled uplink latency.
We now extend the model to the downlink (DL) in a TDD configuration.
Since the gNB knows all DL demands, no scheduling request is needed, making DL latency lower and the model simpler.
As in \cref{sec:model-foundation}, we decompose the latency into components shown in \cref{fig:overall-latency}.

\vskip 0.06in \noindent {\bf Modeling of $u_1$:}
Time from packet arrival at the gNB to the start of scheduling.
With $p_5$ as gNB processing time,

{\small
\begin{equation}
    u_1 = \left(\left\lceil \tfrac{o_1 + p_5}{S} \right\rceil \cdot S \right) - o_1
    \label{eq:u1}
\end{equation}
}

\vskip 0.06in \noindent {\bf Modeling of $u_2$:}
Time from scheduling start to the start of DL transmission. Ensuring it falls in a DL slot, we write:

{\small
\begin{equation}
re_1 \equiv_T \tfrac{o_1 + u_1}{S} + a_1 + 1
\end{equation}
\begin{equation}
u_2 =
\begin{cases}
    (a_1 + 1)S, & re_1 \leq d; \\[6pt]
    (a_1 + 1 + (T - re_1))S, & \text{otherwise}.
    \label{eq:return}
\end{cases}
\end{equation}
}

\vskip 0.06in \noindent {\bf Modeling of $u_3$:}
DL transmission time, proportional to the number of DL slots:

{\small
\begin{equation}
    u_3 = N_{DL, slots} \cdot S
    \label{eq:u3}
\end{equation}
}

\vskip 0.06in \noindent {\bf Modeling of $u_4$:}
UE processing time after receiving DL data:

{\small
\begin{equation}
    u_4 = l_3
    \label{eq:u_4}
\end{equation}
}

\vskip 0.06in \noindent {\bf Modeling the rest of the transmissions:}
If one TDD period is insufficient, the process mirrors UL Size-2 packets (Appendix~\ref{sec:ul-packets-size2}).
The DL slot capacity and number of required slots are:

{\small
\begin{equation}
    Bi_{DL} = N_{RB}[B, T] \cdot 12 \cdot Q_m[I_{MCS}] \cdot R[I_{MCS}] \cdot (14 - DC_{no})
    \label{eq:bi_dl}
\end{equation}
\begin{equation}
    N_{DL, slots} = \min\!\left(\left\lceil \tfrac{P_{rem}}{Bi_{DL}} \right\rceil,\, d\right)
    \label{eq:n_dl_slots}
\end{equation}
}

As in Appendix~\ref{sec:ul-packets-size2}, an iterative procedure updates $P_{rem}$ across TDD periods until the packet is fully sent, as shown in \cref{alg:dl}.

\begin{algorithm}[!htbp]
    \caption{DL Total Latency in TDD}
    \label{alg:dl}
    \begin{algorithmic}[1]
        \State Initialize $Time \gets o_1 + u_1 + u_2$
        \State Initialize $P_{rem} \gets P$
        \State Calculate $B_{DL}$ (\cref{eq:bi_dl})
        \State Calculate $N_{DL, slots}$ (\cref{eq:n_dl_slots}) using $P_{rem}$
        \State Calculate $u_3$ (\cref{eq:u3}) using $N_{DL, slots}$
        \State Update $P_{rem} \gets P_{rem} - N_{DL, slots} \cdot B_{DL}$
        \State Update $Time \gets Time + u_3$
        \If {$P_{rem} \leq 0$}
            \State Calculate $u_4$ (\cref{eq:u_4})
            \State \Return $Total\_Latency \gets Time + u_4 - o_1$
        \EndIf
        \While {True}
            \State Update $Time \gets Time + (T - d) \cdot S$
            \State Calculate $N_{DL, slots}$ (\cref{eq:n_dl_slots}) using $P_{rem}$
            \State Calculate $u_3$ (\cref{eq:u3}) using $N_{DL, slots}$
            \State Update $P_{rem} \gets P_{rem} - N_{DL, slots} \cdot B_{DL}$
            \State Update $Time \gets Time + u_3$
            \If {$P_{rem} \leq 0$}
                \State Calculate $u_4$ (\cref{eq:u_4})
                \State \Return $Total\_Latency \gets Time + u_4 - o_1$
            \EndIf
        \EndWhile
    \end{algorithmic}
\end{algorithm}

\subsection{UL and DL Latency in Mini-Slot Configuration}
\label{sec:mini-slot}

\begin{figure}[t]
    \centering
    \includegraphics[width=0.95\columnwidth]{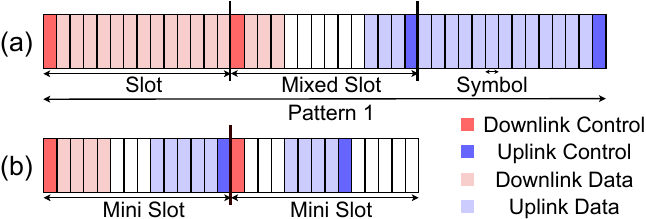}
    \caption{Comparison of (a) \textit{Common Configuration} and (b) \textit{Mini-Slot} in TDD}
    \label{fig:tdd-pattern-common-config}

\end{figure}

The 5G \textit{Mini-Slot} configuration enables finer scheduling: each slot is split into DL symbols, guard symbols, and UL symbols (\cref{fig:tdd-pattern-common-config}).
Unlike the \textit{Common Configuration}, every slot can carry both DL and UL traffic, which simplifies latency modeling.
The gNB signals the slot format to UEs via downlink control information (DCI) on the PDCCH.

\noindent \textbf{Uplink Latency:} Every slot is an UL opportunity, so modeling follows \cref{sec:model-foundation} with simpler constraints:

\vskip 0.06in \noindent {\bf Modeling of $w_1$:}
Any slot can host SR, so $A=\{1,\dots,T\}$.
$w_1$ can be calculated from \cref{eq:w1}.

\vskip 0.06in \noindent {\bf Modeling of $w_2$:}
For $w_2$, the difference from the \textit{Common Configuration} is that in \textit{Mini-Slot}, $UC_{st}$ must start no earlier than the first UL symbol $Sy_{fUL}$.
Thus, $w_2$ is as follows with this added constraint.

{\small
\begin{equation}
    w_2 = \tfrac{UC_{st}+UC_{no}}{14} S, \quad UC_{st}+UC_{no}\leq 14, \; UC_{st}\geq Sy_{fUL}
    \label{eq:combined_w2_constraint_mini_slot}
\end{equation}
}

\vskip 0.06in \noindent {\bf Modeling of $w_3$:}
Since every slot is also DL, $w_3$ is simpler than in the \textit{Common Configuration}: no DL-slot check is needed.

{\small
\begin{equation}
    w_3 = \left(\left\lceil \tfrac{w_2+p_1}{S}\right\rceil + a_1 + 1\right)S
    \label{eq:w3_first_case}
\end{equation}
}

\vskip 0.06in \noindent {\bf Modeling of $w_4$:}
$w_4$ is computed from \cref{eq:combined_w4_pdcchConstraint}, with the added requirement that the PDCCH length not exceed the DL symbols in the slot, i.e.,

{\small
\begin{equation}
    DC_{no} \leq Sy_{lDL}.
    \label{eq:combined_w4_constraint_mini_slot}
\end{equation}
}

\vskip 0.06in \noindent {\bf Modeling of $w_5$:}
With every slot valid for DL, no DL-slot check is needed.
Thus, using \cref{eq:combined_k2min_constraint}:

{\small
\begin{equation}
    k_2 = k_{2_{min}}, \quad w_5 = k_2 S
    \label{eq:w5_first_case}
\end{equation}
}

\vskip 0.06in \noindent {\bf Modeling of $w_6$:}
UL transmission may end before the slot boundary, so $w_6$ can be smaller than $S$:

{\small
\begin{equation}
    w_6 = \tfrac{Sy_{lUL}}{14} S
    \label{eq:w6_mini_slot}
\end{equation}
}

\vskip 0.06in \noindent {\bf Modeling of $w_7$:}
$ w_7 $ is calculated based on~\cref{eq:combined_w6_w7}.

\vskip 0.06in \noindent {\bf Modeling of $w_3'$:}
Since every slot is DL:

{\small
\begin{equation}
    w'_3 = \left(\left\lceil \tfrac{p_1}{S}\right\rceil + a_1 + 1\right) S
    \label{eq:w_prime_3_mini_slot}
\end{equation}
}

\vskip 0.06in \noindent {\bf Modeling of $w_5'$:}
Since every slot is UL, from \cref{eq:k_prime_2_min}:

{\small
\begin{equation}
    k_2 = k'_{2_{min}}, \quad w'_5 = k_2 S
    \label{eq:w_prime_5_mini_slot}
\end{equation}
}

\vskip 0.06in \noindent {\bf Modeling the rest of the transmissions:}
As in Appendix~\ref{sec:large-packet-with-congestion}, after $w'_5$ every slot can be used for UL.
Since each slot has both DL and UL, the same reasoning applies slot by slot.
With $Sy_{fUL}$ and $Sy_{lUL}$ as the first and last UL symbols, the per-slot payload is:

{\small
\begin{equation}
    \begin{aligned}
        Bi_{UL}
        &= \frac{N_{RB}[B,T] \cdot 12 \cdot Q_m[I_{MCS}] \cdot R[I_{MCS}]}{8} \\
        &\quad \cdot (Sy_{lUL} - Sy_{fUL} + 1 - UC_{no})
    \end{aligned}
    \label{eq:bi_ul_mini_slot}
\end{equation}
}
The remaining transmissions are then obtained iteratively as in \cref{alg:mini-slot-ul-and-fdd-ul}.

\begin{algorithm}[!htbp]
    \caption{UL Total Latency with \textit{Mini-Slot} Configuration in TDD and UL Total Latency in FDD}
    \label{alg:mini-slot-ul-and-fdd-ul}
    \begin{algorithmic}[1]
        \State Initialize $P_{rem} \gets P - g_I$
        \If {$P_{rem} \leq 0$}
            \State Calculate $w_1$, $w_2$, $w_3$, $w_4$, $w_5$, $w_7$
            \State Calculate $w_6$ according to the selected duplexing mode (mini-slot vs. FDD)
            \State \Return $Total\_Latency \gets w_1 + w_3 + w_5 + w_6 + w_7$
        \EndIf
        \State Calculate $ w_3 $, $ w_5 $, $ w_3' $, and $ w_5' $
        \State Initialize $Time \gets o_1 + w_1 + w_3 + w_5 + S$
        \State Update $Time \gets Time + w_3' + w_5'$
        \While {True}
            \State Calculate $Bi_{UL}$ based on \cref{eq:bi_ul_mini_slot} in the case of mini-slot and \cref{eq:bi_ul} in the case of FDD
            \State Update $P_{rem} \gets P_{rem} - Bi_{UL}$
            \If {$P_{rem} \leq 0$}
                \State Calculate $w_6$ according to the selected duplexing mode (mini-slot vs. FDD)
                \State Calculate $w_7$
                \State \Return $Total\_Latency \gets Time + w_6 + w_7 - o_1$
            \EndIf
            \State Update $Time \gets Time + S$
        \EndWhile
    \end{algorithmic}
\end{algorithm}

\noindent \textbf{Downlink Latency:}
DL latency in the \textit{Mini-Slot} configuration follows the same approach as Appendix~\ref{sec:dl-model}, with simplified constraints.

\vskip 0.06in \noindent {\bf Modeling of $u_1$:}
$u_1$ can be calculated using~\cref{eq:u1}.

\vskip 0.06in \noindent {\bf Modeling of $u_2$:}
Every slot includes DL, so:

{\small
\begin{equation}
    u_2 = (a_1+1)S
    \label{eq:u2_mini_slot}
\end{equation}
}

\vskip 0.06in \noindent {\bf Modeling of $u_3$:}
Transmission time depends on the number of DL symbols:

{\small
\begin{equation}
    u_3 = \tfrac{Sy_{lDL}}{14} S
    \label{eq:u_3_mini_slot}
\end{equation}
}

\vskip 0.06in \noindent {\bf Modeling of $u_4$:}
$u_4$ can be calculated using~\cref{eq:u_4}.

\vskip 0.06in \noindent {\bf Modeling the rest of the transmissions:}
Since each slot starts with a DL symbol, defining the number of DL symbols in the slot as $ {Sy_{fDL}} $, then the value of $ Bi_{DL} $ can be calculated as follows:

{\small
\begin{equation}
    Bi_{DL} = N_{RB}[B,T] \cdot 12 \cdot Q_m[I_{MCS}] \cdot R[I_{MCS}] \cdot (Sy_{fDL} - DC_{no})
    \label{eq:bi_dl_mini_slot}
\end{equation}
}

The remaining latency is then computed iteratively as in \cref{alg:mini-slot-dl-and-fdd-dl}.

\begin{algorithm}[!htbp]
    \caption{DL Total Latency with \textit{Mini-Slot} Configuration in TDD and DL Total Latency in FDD}
    \label{alg:mini-slot-dl-and-fdd-dl}
    \begin{algorithmic}[1]
        \State Calculate $u_1$ and $u_2$ based on \cref{eq:u1} and \cref{eq:u2_mini_slot}
        \State Calculate $Bi_{DL}$ based on \cref{eq:bi_dl_mini_slot}
        \State Initialize $P_{rem} \gets P - Bi_{DL}$
        \If {$P_{rem} \leq 0$}
            \State Calculate $u_3$ according to the selected duplexing mode (mini-slot vs. FDD)
            \State Calculate $u_4$ based on \cref{eq:u_4}
            \State \Return $Total\_Latency \gets u_1 + u_2 + u_3 + u_4$
        \EndIf
        \State Initialize $Time \gets u_1 + u_2 + S$
        \While {True}
            \State Calculate $Bi_{DL}$ based on \cref{eq:bi_dl_mini_slot} in the case of mini-slot and \cref{eq:bi_dl} in the case of FDD
            \State Update $P_{rem} \gets P_{rem} - Bi_{DL}$
            \If {$P_{rem} \leq 0$}
                \State Calculate $u_3$ according to the selected duplexing mode (mini-slot vs. FDD)
                \State Calculate $u_4$ based on \cref{eq:u_4}
                \State \Return $Total\_Latency \gets Time + u_3 + u_4$
            \EndIf
            \State Update $Time \gets Time + S$
        \EndWhile
    \end{algorithmic}
\end{algorithm}

\subsection{UL and DL latency in FDD}
\label{sec:fdd-model}

In FDD,\footnote{FDD is only supported in sub-\SI{2.69}{\giga\hertz} bands~\cite{3GPP-scs-fr1}, typically used by public operators. Due to the limited availability of FDD spectrum, private 5G deployments are usually allocated higher bands that only support TDD, so for URLLC applications such as industrial automation~\cite{brown2018ultra} and hospitals~\cite{9792172}, TDD is the practical option.}
uplink and downlink use separate frequency bands, so each slot is simultaneously a full DL and a full UL slot.

\noindent\textbf{Uplink Latency: }
The FDD uplink model follows the \textit{Mini-Slot} formulation. We use
$w_1$, $w_3$, $w_5$, $w'_3$, and $w'_5$ from \cref{eq:w1} with
$A=\{1,\dots,T\}$, \cref{eq:w3_first_case}, \cref{eq:w5_first_case},
\cref{eq:w_prime_3_mini_slot}, and \cref{eq:w_prime_5_mini_slot},
respectively. Since every symbol is available for uplink transmission,
$w_2$, $w_4$, $w_7$, and $Bi_{UL}$ are obtained from
\cref{eq:combined_w2_constraint}, \cref{eq:combined_w4_pdcchConstraint},
\cref{eq:combined_w6_w7}, and \cref{eq:bi_ul}, respectively, and
$w_6=S$. The total UL latency is then computed as in
\cref{alg:mini-slot-ul-and-fdd-ul}.

\noindent\textbf{Downlink Latency: }
The FDD downlink model is also inherited from the \textit{Mini-Slot}
case. We use $u_1$, $u_2$, and $u_4$ from \cref{eq:u1},
\cref{eq:u2_mini_slot}, and \cref{eq:u_4}, respectively. Since every
symbol is available for downlink transmission, $Bi_{DL}$ is obtained
from \cref{eq:bi_dl}, and $u_3=S$. The total DL latency is then computed
as in \cref{alg:mini-slot-dl-and-fdd-dl}.

\subsection{UL latency under multi-UE contention}
\label{sec:contention-model}

We extend the single-UE multi-packet model from Appendix~\ref{sec:multi-packet} to the case of multiple UEs transmitting over the same channel.
Contention occurs whenever the total requested resources exceed the available resources within a single slot.
In this case, packet latency depends on the gNB scheduling policy.
We model contention under a round-robin scheduler.
In each slot, UEs are ordered by ID and granted resources sequentially until the slot resources are exhausted.
The ordering is shifted by one UE in every slot, ensuring fairness over time.
While we focus on round-robin scheduling in this work, one can extend the model to more scheduling policies. However, we leave such extensions to future work which may incorporate other schedulers by modifying the resource allocation rule among contending UEs.

\noindent\textbf{Contention modeling procedure.}
First, we run the single-UE model for each UE independently, assuming no contention.
This yields, for every packet, its transmission start time and latency baseline.
Next, we identify overlaps across UE transmissions and partition the timeline into segments.
Within a segment where exactly $N$ UEs contend, each UE receives on average a fraction $\frac{1}{N}$ of its requested resources.
Therefore, a segment of nominal duration $L$ is stretched under contention to $N \cdot L$.

Packet latency is then recomputed by summing over all stretched segments.

\noindent\textbf{Additional TDD delay.}
In TDD configuration, there is another contention effect due to limited uplink slots.
For example, consider $N=4$ contending UEs using a TDD configuration with period $T=10$ slots, of which only $U=3$ slots are uplink.
Since only three UEs can be scheduled per period, one UE would not be scheduled in each period.
Thus, every UE experiences an additional delay of one full TDD period every four periods, increasing packet latency beyond the resource-sharing effect.
Let the TDD period be $T$ slots, containing $U$ uplink slots.

In every $N$ TDD periods, there are exactly $N-U$ periods in which a UE is not scheduled in uplink (assuming full utilization by others).
If a packet requires total transmission time $t$ slots, the additional delay due to missed UL turns is

\begin{equation}
\max\left\{
\left\lfloor \frac{t}{N T} \right\rfloor (N-U)T,\; 0
\right\}
\end{equation}

\noindent\textbf{Final latency.}
The final packet latency under contention is the sum of
(i) the baseline latency without contention,
(ii) the segment expansion due to resource sharing,
and (iii) the additional TDD delay:

\begin{equation}
\mathrm{Latency}
=
\mathrm{Latency}_{\mathrm{no\;cont.}}
+
\mathrm{Delay}_{\mathrm{resource}}
+
\Big\lfloor \frac{t}{N T} \Big\rfloor (N-U)T
\end{equation}

\end{document}